\documentclass[epj]{svjour}
\usepackage{graphicx}
\usepackage{dcolumn}
\usepackage{multirow}
\usepackage{amssymb}
\onecolumn

\begin{document}

\title{Chiral Structure of Vector and Axial-Vector Tetraquark Currents}
%
%
\author{Hua-Xing Chen}
\institute{School of Physics and Nuclear Energy Engineering and International Research Center for Nuclei and Particles in the Cosmos, Beihang University, Beijing 100191, China}
\date{Received: date / Revised version: date}
\abstract{
We investigate the chiral structure of local vector and axial-vector tetraquark currents, and study their chiral transformation properties. We consider the charge-conjugation parity and classify all the isovector vector and axial-vector local tetraquark currents of quantum numbers $I^GJ^{PC} = 1^-1^{-+}$,  $I^GJ^{PC} = 1^+1^{--}$,  $I^GJ^{PC} = 1^-1^{++}$ and  $I^GJ^{PC} = 1^+1^{+-}$. We find that there is a one to one correspondence among them. Using these currents, we perform QCD sum rule analyses. Our results suggest that there is a missing $b_1$ state having $I^G J^{PC} = 1^+1^{+-}$ and a mass around 1.47-1.66 GeV.
\PACS{
      {12.39.Mk}{Glueball and nonstandard multi-quark/gluon states}   \and
      {11.40.-q}{Currents and their properties}   \and
      {12.38.Lg}{Other nonperturbative calculations}
     } 
} 
\maketitle

\section{Introduction}

Multi-quark components always exist in the Fock space expansion of hadron states~\cite{Prelovsek:2005du,Fariborz:2009cq,Yndurain:2007qm}. Moreover, there may exist multi-quark states~\cite{Jaffe:1976ig,Weinstein:1982gc,Close:2002zu,Lipkin:1986dw,Brodsky:1977bs,Zhu:2007wz}. They are both interesting and important subjects to understand the low-energy behavior of QCD, and have been studied by lots of theoretical and experimental physicists. One of the methods to study these multi-quark components (states) is to use the group theoretical method~\cite{Weinberg:1969hw,Leinweber:1994nm,Cohen:2002st,Jido:2001nt}. This method has been used by T.~D.~Cohen and X.~D.~Ji to study the chiral structure of two-quark meson currents and three-quark baryon currents~\cite{Cohen:1996sb}. We have also used it to study the chiral structure of baryons and tetraquarks~\cite{Chen:2008qv,Chen:2012ex,Chen:2012ut}. The obtained currents (interpolating fields) can be used in QCD sum rule analyses~\cite{Matheus:2006xi,Kim:2011ut,Narison:2002pw,Chen:2010ze,Chen:2007xr} as well as Lattice QCD calculations~\cite{Okiharu:2004ve,Prelovsek:2010kg,McNeile:2006nv,Yang:2012gz,Engel:2013ig}.

Vector and axial-vector mesons are also interesting subjects~\cite{Roca:2005nm,Khemchandani:2011mf,Geng:2008ag,Cheng:2011pb,Grigoryan:2007vg,Zhao:2005vh}. In this paper we shall use the group theoretical method to study the chiral structure of local vector and axial-vector tetraquark currents. We shall study their chiral transformation properties. We shall also consider the charge-conjugation parity and classify all the local isovector vector and axial-vector tetraquark currents of quantum numbers $I^GJ^{PC} = 1^-1^{-+}$,  $I^GJ^{PC} = 1^+1^{--}$,  $I^GJ^{PC} = 1^-1^{++}$ and  $I^GJ^{PC} = 1^+1^{+-}$. We find that there is some ``symmetry'' among these currents, i.e., there is a one to one correspondence among them.

Experiments have observed three exotic mesons $\pi_1(1400)$, $\pi_1(1600)$ and $\pi(2000)$ of exotic quantum numbers $I^GJ^{PC}=1^-1^{-+}$, many $\rho$ mesons $\rho(770)$, $\rho(1450)$, $\rho(1570)$, $\rho(1700)$, $\rho(1900)$ and $\rho(2150)$ of $I^GJ^{PC}=1^+1^{--}$, two $a_1$ mesons $a_1(1260)$ and $a_1(1640)$ of $I^GJ^{PC}=1^-1^{++}$, but only one $b_1(1235)$ meson of $I^GJ^{PC}=1^+1^{+-}$~\cite{Beringer:1900zz,Abele:1998gn,Thompson:1997bs,Adams:1998ff,Akhmetshin:2001hm,Achasov:2000wy,Frabetti:2003pw,Asner:1999kj,Chung:2002pu,Baker:2003jh,Weidenauer:1993mv,Amsler:1993jz,Nozar:2002br,Ablikim:2004wn}. Particularly, in the energy region around 1.6 GeV, there are mesons of quantum numbers $I^GJ^{PC}=1^-1^{-+}$, $I^GJ^{PC}=1^+1^{--}$ and $I^GJ^{PC}=1^-1^{++}$, but there are no mesons of quantum numbers $I^GJ^{PC}=1^+1^{+-}$. To verify whether there is a missing state having these quantum numbers in this energy region, we shall perform QCD sum rule analyses using the tetraquark currents classified in this paper. We would like to note that we shall only use these tetraquark currents, but other currents representing the $\bar q q$ structure, the hybrid structure and the meson-meson structure can also contribute here (see Res.~\cite{Sugiyama:2007sg,Nakamura:2008zzc,Matheus:2009vq,Wang:2009wk,Nielsen:2010ij,Chen:2013pya} where their mixing is studied for the cases of light scalar mesons, $\Lambda(1405)$, $X(3872)$ and $X(4350)$). However, as long as the tetraquark current can couple to the physical state, it can be used to perform QCD sum rule analyses to study that physical state.

This paper is organized as follows. In Sec.~\ref{sec:currents} we classify local tetraquark currents of flavor singlet and $J^P = 1^-$, while currents of others quantum numbers are listed in Appendix.~\ref{app:othercurrents}. In Sec.~\ref{sec:chiral} we study their chiral structure, and give the chiral transformation equations for the $[({\mathbf 3},\bar{\mathbf 3}) \oplus (\bar{\mathbf 3},{\mathbf 3})]$ chiral multiplets, while equations for other chiral multiplets are given in Appendix.~\ref{app:chiraltransformation}. In Sec.~\ref{sec:vectoraxialvector} we consider the charge-conjugation parity and classify isovector vector and axial-vector tetraquark currents of quantum numbers $I^GJ^{PC} = 1^-1^{-+}$,  $I^GJ^{PC} = 1^+1^{--}$,  $I^GJ^{PC} = 1^-1^{++}$ and $I^GJ^{PC} = 1^+1^{+-}$. In Sec.~\ref{sec:sumrule} we use these currents to perform Shifman-Vainshtein-Zakharov (SVZ) sum rule analyses, and in Sec.~\ref{sec:fesr} we use them to perform finite energy sum rule (FESR) analyses. Sec.~\ref{sec:summary} is a summary.

\section{Tetraquark Currents of Flavor Singlet and $J^P = 1^-$}
\label{sec:currents}

In this section we study flavor singlet tetraquark currents of $J^P = 1^-$. There are altogether eight independent vector currents as listed in the following:
\begin{eqnarray}
\nonumber \eta_1^{\rm V, \mathbb{S}} &=& q_A^{aT} \mathbb{C} \gamma_5 q_B^b (\bar{q}_A^a \gamma_\mu \gamma_5 \mathbb{C} \bar{q}_B^{bT} - \bar{q}_A^b \gamma_\mu \gamma_5 \mathbb{C} \bar{q}_B^{aT}) \, ,
\\ \nonumber \eta_2^{\rm V, \mathbb{S}} &=& q_A^{aT} \mathbb{C} \gamma_\mu \gamma_5 q_B^b (\bar{q}_A^a \gamma_5 \mathbb{C} \bar{q}_B^{bT} - \bar{q}_A^b \gamma_5 \mathbb{C} \bar{q}_B^{aT}) \, ,
\\ \nonumber \eta_3^{\rm V, \mathbb{S}} &=& q_A^{aT} \mathbb{C} \gamma^\nu q_B^b (\bar{q}_A^a \sigma_{\mu\nu} \mathbb{C} \bar{q}_B^{bT} + \bar{q}_A^b \sigma_{\mu\nu} \mathbb{C} \bar{q}_B^{aT}) \, ,
\\ \eta_4^{\rm V, \mathbb{S}} &=& q_A^{aT} \mathbb{C} \sigma_{\mu\nu} q_B^b (\bar{q}_A^a \gamma^\nu \mathbb{C} \bar{q}_B^{bT} + \bar{q}_A^b \gamma^\nu \mathbb{C} \bar{q}_B^{aT}) \, ,
\label{eq:singletvector}
\\ \nonumber \eta_5^{\rm V, \mathbb{S}} &=& q_A^{aT} \mathbb{C} \gamma_5 q_B^b (\bar{q}_A^a \gamma_\mu \gamma_5 \mathbb{C} \bar{q}_B^{bT} + \bar{q}_A^b \gamma_\mu \gamma_5 \mathbb{C} \bar{q}_B^{aT}) \, ,
\\ \nonumber \eta_6^{\rm V, \mathbb{S}} &=& q_A^{aT} \mathbb{C} \gamma_\mu \gamma_5 q_B^b (\bar{q}_A^a \gamma_5 \mathbb{C} \bar{q}_B^{bT} + \bar{q}_A^b \gamma_5 \mathbb{C} \bar{q}_B^{aT}) \, ,
\\ \nonumber \eta_7^{\rm V, \mathbb{S}} &=& q_A^{aT} \mathbb{C} \gamma^\nu q_B^b (\bar{q}_A^a \sigma_{\mu\nu} \mathbb{C} \bar{q}_B^{bT} - \bar{q}_A^b \sigma_{\mu\nu} \mathbb{C} \bar{q}_B^{aT}) \, ,
\\ \nonumber \eta_8^{\rm V, \mathbb{S}} &=& q_A^{aT} \mathbb{C} \sigma_{\mu\nu} q_B^b (\bar{q}_A^a \gamma^\nu \mathbb{C} \bar{q}_B^{bT} - \bar{q}_A^b \gamma^\nu \mathbb{C} \bar{q}_B^{aT}) \, .
\end{eqnarray}
In these expressions the summation is taken over repeated indices ($a$, $b$, $\cdots$ for color indices, $A$, $B$, $\cdots$ for flavor indices, and $\mu$, $\nu$, $\cdots$ for Lorentz indices). The two superscripts V and $\mathbb{S}$ denote vector ($J^P = 1^-$) and flavor singlet, respectively. In this paper we also need to use the following notations: $\mathbb{C}$ is the charge-conjugation operator; $\epsilon^{ABC}$ is the totally anti-symmetric tensor; $S_P^{ABC}$ ($P=1\cdots10$) are the normalized totally symmetric matrices; $\bf \lambda_N$ ($N=1\cdots8$) are the Gell-Mann matrices; $S^{AB;CD}_{U}$ ($U=1\cdots27$) are the matrices for the $\mathbf{27}$ flavor representation, as defined in Ref.~\cite{Chen:2012ut}.

Among these eight currents, the former four currents contain diquarks and antidiquarks having the antisymmetric flavor structure $\mathbf{\bar 3} \otimes \mathbf{3}$ and the latter four currents contain diquarks and antidiquarks having the symmetric flavor structure $\mathbf{6} \otimes \mathbf{\bar6}$. To clearly see the chiral structure of Eqs.~(\ref{eq:singletvector}), we use the left-handed quark field $L_A^a \equiv q_{LA}^a = {1 - \gamma_5 \over 2} q_A^a$ and the right-handed quark field $R_A^a \equiv q_{RA}^a = {1 + \gamma_5 \over 2} q_A^a$ to rewrite these currents and then combine them properly:
\begin{eqnarray}
\nonumber \eta_1^{\rm V, \mathbb{S}} &=& 2 L_A^{aT} \mathbb{C} L_B^b (\bar{L}_A^a \gamma_\mu \mathbb{C} \bar{R}_B^{bT} - \bar{L}_A^b \gamma_\mu \mathbb{C} \bar{R}_B^{aT})
+ 2 R_A^{aT} \mathbb{C} R_B^b (\bar{R}_A^a \gamma_\mu \mathbb{C} \bar{L}_B^{bT} - \bar{R}_A^b \gamma_\mu \mathbb{C} \bar{L}_B^{aT}) \, ,
\\ \nonumber \eta_2^{\rm V, \mathbb{S}} &=& 2 L_A^{aT} \mathbb{C} \gamma_\mu R_B^b (\bar{L}_A^a \mathbb{C} \bar{L}_B^{bT} - \bar{L}_A^b \mathbb{C} \bar{L}_B^{aT})
+ 2 R_A^{aT} \mathbb{C} \gamma_\mu L_B^b (\bar{R}_A^a \mathbb{C} \bar{R}_B^{bT} - \bar{R}_A^b \mathbb{C} \bar{R}_B^{aT}) \, ,
\\ \nonumber \eta_3^{\rm V, \mathbb{S}} &=& 2 L_A^{aT} \mathbb{C} \gamma^\nu R_B^b (\bar{L}_A^a \sigma_{\mu\nu} \mathbb{C} \bar{L}_B^{bT} + \bar{L}_A^b \sigma_{\mu\nu} \mathbb{C} \bar{L}_B^{aT})
+ 2 R_A^{aT} \mathbb{C} \gamma^\nu L_B^b (\bar{R}_A^a \sigma_{\mu\nu} \mathbb{C} \bar{R}_B^{bT} + \bar{R}_A^b \sigma_{\mu\nu} \mathbb{C} \bar{R}_B^{aT}) \, ,
\\ \eta_4^{\rm V, \mathbb{S}} &=& 2 L_A^{aT} \mathbb{C} \sigma_{\mu\nu} L_B^b (\bar{L}_A^a \gamma^\nu \mathbb{C} \bar{R}_B^{bT} + \bar{L}_A^b \gamma^\nu \mathbb{C} \bar{R}_B^{aT})
+ 2 R_A^{aT} \mathbb{C} \sigma_{\mu\nu} R_B^b (\bar{R}_A^a \gamma^\nu \mathbb{C} \bar{L}_B^{bT} + \bar{R}_A^b \gamma^\nu \mathbb{C} \bar{L}_B^{aT}) \, ,
\\ \nonumber \eta_5^{\rm V, \mathbb{S}} &=& 2 L_A^{aT} \mathbb{C} L_B^b (\bar{L}_A^a \gamma_\mu \mathbb{C} \bar{R}_B^{bT} + \bar{L}_A^b \gamma_\mu \mathbb{C} \bar{R}_B^{aT})
+ 2 R_A^{aT} \mathbb{C} R_B^b (\bar{R}_A^a \gamma_\mu \mathbb{C} \bar{L}_B^{bT} + \bar{R}_A^b \gamma_\mu \mathbb{C} \bar{L}_B^{aT}) \, ,
\\ \nonumber \eta_6^{\rm V, \mathbb{S}} &=& 2 L_A^{aT} \mathbb{C} \gamma_\mu R_B^b (\bar{L}_A^a \mathbb{C} \bar{L}_B^{bT} + \bar{L}_A^b \mathbb{C} \bar{L}_B^{aT})
+ 2 R_A^{aT} \mathbb{C} \gamma_\mu L_B^b (\bar{R}_A^a \mathbb{C} \bar{R}_B^{bT} + \bar{R}_A^b \mathbb{C} \bar{R}_B^{aT}) \, ,
\\ \nonumber \eta_7^{\rm V, \mathbb{S}} &=& 2 L_A^{aT} \mathbb{C} \gamma^\nu R_B^b (\bar{L}_A^a \sigma_{\mu\nu} \mathbb{C} \bar{L}_B^{bT} - \bar{L}_A^b \sigma_{\mu\nu} \mathbb{C} \bar{L}_B^{aT})
+ 2 R_A^{aT} \mathbb{C} \gamma^\nu L_B^b (\bar{R}_A^a \sigma_{\mu\nu} \mathbb{C} \bar{R}_B^{bT} - \bar{R}_A^b \sigma_{\mu\nu} \mathbb{C} \bar{R}_B^{aT}) \, ,
\\ \nonumber \eta_8^{\rm V, \mathbb{S}} &=& 2 L_A^{aT} \mathbb{C} \sigma_{\mu\nu} L_B^b (\bar{L}_A^a \gamma^\nu \mathbb{C} \bar{R}_B^{bT} - \bar{L}_A^b \gamma^\nu \mathbb{C} \bar{R}_B^{aT})
+ 2 R_A^{aT} \mathbb{C} \sigma_{\mu\nu} R_B^b (\bar{R}_A^a \gamma^\nu \mathbb{C} \bar{L}_B^{bT} - \bar{R}_A^b \gamma^\nu \mathbb{C} \bar{L}_B^{aT}) \, ,
\end{eqnarray}
from which we can quickly find out their chiral structure (representations): $\eta^{\rm V, \mathbb{S}}_{1,4,5,8}$ belong to the chiral representation $[({\mathbf 3}, \bar {\mathbf 3}) + (\bar {\mathbf 3}, {\mathbf 3})]$ and their chirality is $L L \bar L \bar R + R R \bar R \bar L$; $\eta^{\rm V, \mathbb{S}}_{2,3,6,7}$ belong to the mirror one $[(\bar {\mathbf 3}, {\mathbf 3}) + ({\mathbf 3}, \bar {\mathbf 3})]$ and their chirality is $L R \bar L \bar L + R L \bar R \bar R$. These two chiral representations are both non-exotic. Therefore, in this case we do not find any ``exotic'' chiral structure. Their detailed structures are listed in Table~\ref{tab:singletscalar}.

\begin{table}[hbt]
\renewcommand{\arraystretch}{1.5}
\begin{center}
\caption{Flavor singlet tetraquark currents of $J^P = 1^-$, showing their chiral representations and chirality. The second and third columns show the flavor and color structures of the diquark and antidiquark inside, respectively.}
\begin{tabular}{c c c c c}
\hline\hline
Currents & Flavor & Color & Representations & Chirality
\\ \hline \hline
$\eta_{1}^{\rm V,\mathbb{S}}$ & $\mathbf{\bar 3} \otimes \mathbf{3}$ & $\mathbf{\bar 3} \otimes \mathbf{3}$ & \multirow{2}{*}{$[({\mathbf 3}, \bar {\mathbf 3}) + (\bar {\mathbf 3}, {\mathbf 3})]$} & \multirow{2}{*}{$L L \bar L \bar R + R R \bar R \bar L$}
\\ \cline{1-3} $\eta_{4}^{\rm V,\mathbb{S}}$ & $\mathbf{\bar 3} \otimes \mathbf{3}$ & $\mathbf{6} \otimes \mathbf{\bar 6}$ & &
\\ \hline
$\eta_{2}^{\rm V,\mathbb{S}}$ & $\mathbf{\bar 3} \otimes \mathbf{3}$ & $\mathbf{\bar 3} \otimes \mathbf{3}$ & \multirow{2}{*}{$[(\bar {\mathbf 3}, {\mathbf 3}) + ({\mathbf 3}, \bar {\mathbf 3})]$} & \multirow{2}{*}{$L R \bar L \bar L + R L \bar R \bar R$}
\\ \cline{1-3} $\eta_{3}^{\rm V,\mathbb{S}}$ & $\mathbf{\bar 3} \otimes \mathbf{3}$ & $\mathbf{6} \otimes \mathbf{\bar 6}$ & &
\\ \hline
$\eta_{5}^{\rm V,\mathbb{S}}$ & $\mathbf{6} \otimes \mathbf{\bar6}$ & $\mathbf{6} \otimes \mathbf{\bar 6}$ & \multirow{2}{*}{$[({\mathbf 3}, \bar {\mathbf 3}) + (\bar {\mathbf 3}, {\mathbf 3})]$} & \multirow{2}{*}{$L L \bar L \bar R+ R R \bar R \bar L$}
\\ \cline{1-3} $\eta_{8}^{\rm V,\mathbb{S}}$ & $\mathbf{6} \otimes \mathbf{\bar6}$ & $\mathbf{\bar 3} \otimes \mathbf{3}$ & &
\\ \hline
$\eta_{6}^{\rm V,\mathbb{S}}$ & $\mathbf{6} \otimes \mathbf{\bar6}$ & $\mathbf{6} \otimes \mathbf{\bar 6}$ & \multirow{2}{*}{$[(\bar {\mathbf 3}, {\mathbf 3}) + ({\mathbf 3}, \bar {\mathbf 3})]$} & \multirow{2}{*}{$L R \bar L \bar L + R L \bar R \bar R$}
\\ \cline{1-3} $\eta_{7}^{\rm V,\mathbb{S}}$ & $\mathbf{6} \otimes \mathbf{\bar6}$ & $\mathbf{\bar 3} \otimes \mathbf{3}$ & &
\\ \hline \hline
\end{tabular}
\label{tab:singletscalar}
\end{center}
\renewcommand{\arraystretch}{1}
\end{table}

To fully understand vector tetraquark currents, their chiral partners are also studied, i.e., the vector and axial-vector tetraquark currents of flavor singlet, octet, $\mathbf{10}$, $\overline\mathbf{10}$ and $\mathbf{27}$. The results are shown in Appendix.~\ref{app:othercurrents}. We can quickly find that there is some ``symmetry'' between vector and axial-vector tetraquark currents, i.e., there is a one to one correspondence between vector and axial-vector tetraquark currents: between Eqs.~(\ref{eq:singletvector}) and (\ref{eq:singletaxialvector}), between Eqs.~(\ref{eq:8vector}) and (\ref{eq:8axialvector}), between Eqs.~(\ref{eq:27vector}) and (\ref{eq:27axialvector}), and between Eqs.~(\ref{eq:10vector}) and (\ref{eq:10axialvector}).

\section{Chiral Transformations}
\label{sec:chiral}

Under the $U(1)_V$, $U(1)_A$, $SU(3)_V$ and $SU(3)_A$ chiral transformations, the quark field, $q= q_L + q_R$, transforms as
\begin{eqnarray}
\nonumber
\bf{U(1)_{V}} &:& q \to \exp(i a^0) q  = q + \delta q \, ,
\\
\bf{SU(3)_V} &:& q \to \exp (i \vec \lambda \cdot \vec a ){q} = q + \delta^{\vec{a}} q \, ,
\\ \nonumber
\bf{U(1)_{A}} &:& q \to \exp(i \gamma_5 b^0) q = q + \delta_5 q \, ,
\\ \nonumber
\bf{SU(3)_A} &:& q \to \exp (i \gamma_{5} \vec \lambda \cdot \vec b){q} = q + \delta_5^{\vec{b}} q \, ,
\end{eqnarray}
where $\vec \lambda$ are the eight Gell-Mann matrices; $a^0$ is an infinitesimal parameter for the $U(1)_V$ transformation, $\vec{a}$ are the octet of $SU(3)_V$ group parameters, $b^0$ is an infinitesimal parameter for the $U(1)_A$ transformation, and $\vec{b}$ are the octet of the $SU(3)_A$ chiral transformations.

The chiral transformation equations of tetraquark currents can be calculated straightforwardly. Here we only show the final results. Through these chiral transformation equations, we can quickly find that there are eight $[({\mathbf 3},\bar{\mathbf 3}) \oplus (\bar{\mathbf 3},{\mathbf 3})]$ chiral multiplets (including mirror multiplets):
\begin{eqnarray}
\nonumber && \big ( \eta_{1}^{\rm V, \mathbb{S}}, \eta_{1}^{\rm AV, \mathbb{S}}, \eta_{1,N}^{\rm V, \mathbb{O}} + \eta_{9,N}^{\rm V, \mathbb{O}}, \eta_{1,N}^{\rm AV, \mathbb{O}} + \eta_{9,N}^{\rm AV, \mathbb{O}} \big ) \, ,
\\ \nonumber && \big ( \eta_{2}^{\rm V, \mathbb{S}}, \eta_{2}^{\rm AV, \mathbb{S}}, \eta_{2,N}^{\rm V, \mathbb{O}} + \eta_{14,N}^{\rm V, \mathbb{O}}, \eta_{2,N}^{\rm AV, \mathbb{O}} + \eta_{14,N}^{\rm AV, \mathbb{O}} \big ) \, ,
\\ \nonumber && \big ( \eta_{3}^{\rm V, \mathbb{S}}, \eta_{3}^{\rm AV, \mathbb{S}}, \eta_{3,N}^{\rm V, \mathbb{O}} + \eta_{15,N}^{\rm V, \mathbb{O}}, \eta_{3,N}^{\rm AV, \mathbb{O}} + \eta_{15,N}^{\rm AV, \mathbb{O}} \big ) \, ,
\\ && \big ( \eta_{4}^{\rm V, \mathbb{S}}, \eta_{4}^{\rm AV, \mathbb{S}}, \eta_{4,N}^{\rm V, \mathbb{O}} + \eta_{12,N}^{\rm V, \mathbb{O}}, \eta_{4,N}^{\rm AV, \mathbb{O}} + \eta_{12,N}^{\rm AV, \mathbb{O}} \big ) \, ,
\\ \nonumber && \big ( \eta_{5}^{\rm V, \mathbb{S}}, \eta_{5}^{\rm AV, \mathbb{S}}, \eta_{5,N}^{\rm V, \mathbb{O}} + \eta_{13,N}^{\rm V, \mathbb{O}}, \eta_{5,N}^{\rm AV, \mathbb{O}} + \eta_{13,N}^{\rm AV, \mathbb{O}} \big ) \, ,
\\ \nonumber && \big ( \eta_{6}^{\rm V, \mathbb{S}}, \eta_{6}^{\rm AV, \mathbb{S}}, \eta_{6,N}^{\rm V, \mathbb{O}} + \eta_{10,N}^{\rm V, \mathbb{O}}, \eta_{6,N}^{\rm AV, \mathbb{O}} + \eta_{10,N}^{\rm AV, \mathbb{O}} \big ) \, ,
\\ \nonumber && \big ( \eta_{7}^{\rm V, \mathbb{S}}, \eta_{7}^{\rm AV, \mathbb{S}}, \eta_{7,N}^{\rm V, \mathbb{O}} + \eta_{11,N}^{\rm V, \mathbb{O}}, \eta_{7,N}^{\rm AV, \mathbb{O}} + \eta_{11,N}^{\rm AV, \mathbb{O}} \big ) \, ,
\\ \nonumber && \big ( \eta_{8}^{\rm V, \mathbb{S}}, \eta_{8}^{\rm AV, \mathbb{S}}, \eta_{8,N}^{\rm V, \mathbb{O}} + \eta_{16,N}^{\rm V, \mathbb{O}}, \eta_{8,N}^{\rm AV, \mathbb{O}} + \eta_{16,N}^{\rm AV, \mathbb{O}} \big ) \, ;
\end{eqnarray}
there are two $[(\bar{\mathbf 3},\bar{\mathbf 6}) \oplus (\bar{\mathbf 6},\bar{\mathbf 3})]$ chiral multiplets (including mirror multiplets):
\begin{eqnarray}
&& \big ( 3 \eta_{1,N}^{\rm V, \mathbb{O}} - \eta_{9,N}^{\rm V, \mathbb{O}}, 3 \eta_{1,N}^{\rm AV, \mathbb{O}} - \eta_{9,N}^{\rm AV, \mathbb{O}}, \eta_{1,P}^{\rm V, \bar\mathbb{D}}, \eta_{1,P}^{\rm AV, \bar\mathbb{D}} \big ) \, ,
\\ \nonumber && \big ( 3 \eta_{4,N}^{\rm V, \mathbb{O}} - \eta_{12,N}^{\rm V, \mathbb{O}}, 3 \eta_{4,N}^{\rm AV, \mathbb{O}} - \eta_{12,N}^{\rm AV, \mathbb{O}}, \eta_{4,P}^{\rm V, \bar\mathbb{D}}, \eta_{4,P}^{\rm AV, \bar\mathbb{D}} \big ) \, ;
\end{eqnarray}
there are two $[({\mathbf 6},{\mathbf 3}) \oplus ({\mathbf 3},{\mathbf 6})]$ chiral multiplets (including mirror multiplets):
\begin{eqnarray}
&& \big ( 3 \eta_{2,N}^{\rm V, \mathbb{O}} - \eta_{14,N}^{\rm V, \mathbb{O}}, 3 \eta_{2,N}^{\rm AV, \mathbb{O}} - \eta_{14,N}^{\rm AV, \mathbb{O}}, \eta_{2,P}^{\rm V, \mathbb{D}}, \eta_{2,P}^{\rm AV, \mathbb{D}} \big ) \, ,
\\ \nonumber && \big ( 3 \eta_{3,N}^{\rm V, \mathbb{O}} - \eta_{15,N}^{\rm V, \mathbb{O}}, 3 \eta_{3,N}^{\rm AV, \mathbb{O}} - \eta_{15,N}^{\rm AV, \mathbb{O}}, \eta_{3,P}^{\rm V, \mathbb{D}}, \eta_{3,P}^{\rm AV, \mathbb{D}} \big ) \, ;
\end{eqnarray}
there are two $[({\mathbf{15}},\bar{\mathbf 3}) \oplus (\bar{\mathbf 3},{\mathbf{15}})]$ chiral multiplets (including mirror multiplets):
\begin{eqnarray}
&& \big ( 3 \eta_{5,N}^{\rm V, \mathbb{O}} - 5 \eta_{13,N}^{\rm V, \mathbb{O}}, 3 \eta_{5,N}^{\rm AV, \mathbb{O}} - 5 \eta_{13,N}^{\rm AV, \mathbb{O}}, \eta_{1,P}^{\rm V, \mathbb{D}}, \eta_{1,P}^{\rm AV, \mathbb{D}}, \eta_{1,U}^{\rm V, \mathbb{TS}}, \eta_{1,U}^{\rm AV, \mathbb{TS}} \big ) \, ,
\\ \nonumber && \big ( 3 \eta_{8,N}^{\rm V, \mathbb{O}} - 5 \eta_{16,N}^{\rm V, \mathbb{O}}, 3 \eta_{8,N}^{\rm AV, \mathbb{O}} - 5 \eta_{16,N}^{\rm AV, \mathbb{O}}, \eta_{4,P}^{\rm V, \mathbb{D}}, \eta_{4,P}^{\rm AV, \mathbb{D}}, \eta_{4,U}^{\rm V, \mathbb{TS}}, \eta_{4,U}^{\rm AV, \mathbb{TS}} \big ) \, ;
\end{eqnarray}
there are two $[(\overline{\mathbf{15}},{\mathbf 3}) \oplus ({\mathbf 3},\overline{\mathbf{15}})]$ chiral multiplets (including mirror multiplets):
\begin{eqnarray}
&& \big ( 3 \eta_{6,N}^{\rm V, \mathbb{O}} - 5 \eta_{10,N}^{\rm V, \mathbb{O}}, 3 \eta_{6,N}^{\rm AV, \mathbb{O}} - 5 \eta_{10,N}^{\rm AV, \mathbb{O}}, \eta_{2,P}^{\rm V, \bar\mathbb{D}}, \eta_{2,P}^{\rm AV, \bar\mathbb{D}}, \eta_{2,U}^{\rm V, \mathbb{TS}}, \eta_{2,U}^{\rm AV, \mathbb{TS}} \big ) \, ,
\\ \nonumber && \big ( 3 \eta_{7,N}^{\rm V, \mathbb{O}} - 5 \eta_{11,N}^{\rm V, \mathbb{O}}, 3 \eta_{7,N}^{\rm AV, \mathbb{O}} - 5 \eta_{11,N}^{\rm AV, \mathbb{O}}, \eta_{3,P}^{\rm V, \bar\mathbb{D}}, \eta_{3,P}^{\rm AV, \bar\mathbb{D}}, \eta_{3,U}^{\rm V, \mathbb{TS}}, \eta_{3,U}^{\rm AV, \mathbb{TS}} \big ) \, .
\end{eqnarray}

Here we only show the chiral transformation equations of the $[({\mathbf 3},\bar{\mathbf 3}) \oplus (\bar{\mathbf 3},{\mathbf 3})]$ chiral multiplet, and others are shown in Appendix~\ref{app:chiraltransformation}. We use $\big ( \eta_{(\bar \mathbf{3},\mathbf{3})}^{\rm V, \mathbb{S}}, \eta_{(\bar \mathbf{3},\mathbf{3})}^{\rm AV, \mathbb{S}}, \eta_{(\bar \mathbf{3},\mathbf{3}),N}^{\rm V, \mathbb{O}}, \eta_{(\bar \mathbf{3},\mathbf{3}),N}^{\rm AV, \mathbb{O}} \big )$ to denote it, and its chiral transformation properties are
\begin{eqnarray}
\nonumber \delta_5 \eta_{(\bar \mathbf{3},\mathbf{3})}^{\rm V, \mathbb{S}} &=& 2 i b \eta_{(\bar \mathbf{3},\mathbf{3})}^{\rm AV, \mathbb{S}} \, ,
\\ \nonumber \delta^{\vec a} \eta_{(\bar \mathbf{3},\mathbf{3})}^{\rm V, \mathbb{S}} &=& 0 \, ,
\\ \nonumber \delta_5^{\vec b} \eta_{(\bar \mathbf{3},\mathbf{3})}^{\rm V, \mathbb{S}} &=& 2 i b^N \eta_{(\bar \mathbf{3},\mathbf{3}),N}^{\rm AV, \mathbb{O}} \, ,
\\ \nonumber \delta_5 \eta_{(\bar \mathbf{3},\mathbf{3})}^{\rm AV, \mathbb{S}} &=& 2 i b \eta_{(\bar \mathbf{3},\mathbf{3})}^{\rm V, \mathbb{S}} \, ,
\\ \nonumber \delta^{\vec a} \eta_{(\bar \mathbf{3},\mathbf{3})}^{\rm AV, \mathbb{S}} &=& 0 \, ,
\\ \nonumber \delta_5^{\vec b} \eta_{(\bar \mathbf{3},\mathbf{3})}^{\rm AV, \mathbb{S}} &=& 2 i b^N \eta_{(\bar \mathbf{3},\mathbf{3}),N}^{\rm V, \mathbb{O}} \, ,
\\ \nonumber \delta_5 \eta_{(\bar \mathbf{3},\mathbf{3}),N}^{\rm V, \mathbb{O}} &=& 2 i b \eta_{(\bar \mathbf{3},\mathbf{3}),N}^{\rm AV, \mathbb{O}} \, ,
\\ \nonumber \delta^{\vec a} \eta_{(\bar \mathbf{3},\mathbf{3}),N}^{\rm V, \mathbb{O}} &=& 2 a^N f_{NMO} \eta_{(\bar \mathbf{3},\mathbf{3}),O}^{\rm V, \mathbb{O}} \, ,
\\ \nonumber \delta_5^{\vec b} \eta_{(\bar \mathbf{3},\mathbf{3}),N}^{\rm V, \mathbb{O}} &=& {4\over3} i b^M \eta_{(\bar \mathbf{3},\mathbf{3})}^{\rm AV, \mathbb{S}} + 2 i b^N d_{NMO} \eta_{(\bar \mathbf{3},\mathbf{3}),O}^{\rm AV, \mathbb{O}} \, ,
\\ \nonumber \delta_5 \eta_{(\bar \mathbf{3},\mathbf{3}),N}^{\rm AV, \mathbb{O}} &=& 2 i b \eta_{(\bar \mathbf{3},\mathbf{3}),N}^{\rm V, \mathbb{O}} \, ,
\\ \nonumber \delta^{\vec a} \eta_{(\bar \mathbf{3},\mathbf{3}),N}^{\rm AV, \mathbb{O}} &=& 2 a^N f_{NMO} \eta_{(\bar \mathbf{3},\mathbf{3}),O}^{\rm AV, \mathbb{O}} \, ,
\\ \nonumber \delta_5^{\vec b} \eta_{(\bar \mathbf{3},\mathbf{3}),N}^{\rm AV, \mathbb{O}} &=& {4\over3} i b^M \eta_{(\bar \mathbf{3},\mathbf{3})}^{\rm V, \mathbb{S}} + 2 i b^N d_{NMO} \eta_{(\bar \mathbf{3},\mathbf{3}),O}^{\rm V, \mathbb{O}} \, .
\end{eqnarray}
From these equations and those listed in Appendix.~\ref{app:chiraltransformation}, we can quickly find that vector and axial-vector tetraquark currents are closely related by chiral transformations, and confirm the one to one correspondence.

\section{Charge-Conjugation Parity}
\label{sec:vectoraxialvector}

In this paper we shall concentrate on isovector tetraquark currents since there are more experimental results, including three exotic mesons $\pi_1(1400)$, $\pi_1(1600)$ and $\pi(2000)$~\cite{Beringer:1900zz,Abele:1998gn,Thompson:1997bs,Adams:1998ff}. They have exotic quantum numbers $I^GJ^{PC}=1^-1^{-+}$. Other observed isovector vector and axial-vector mesons are $\rho(770)$, $\rho(1450)$, $\rho(1570)$, $\rho(1700)$, $\rho(1900)$ and $\rho(2150)$ of $I^GJ^{PC}=1^+1^{--}$, $a_1(1260)$ and $a_1(1640)$ of $I^GJ^{PC}=1^-1^{++}$, and $b_1(1235)$ of $I^GJ^{PC}=1^+1^{+-}$~\cite{Beringer:1900zz,Akhmetshin:2001hm,Achasov:2000wy,Frabetti:2003pw,Asner:1999kj,Chung:2002pu,Baker:2003jh,Weidenauer:1993mv,Amsler:1993jz,Nozar:2002br,Ablikim:2004wn}. Particularly, in the energy region around 1.6 GeV, there are mesons of quantum numbers $I^GJ^{PC}=1^-1^{-+}$, $I^GJ^{PC}=1^+1^{--}$ and $I^GJ^{PC}=1^-1^{++}$, but there are no mesons of quantum numbers $I^GJ^{PC}=1^+1^{+-}$.

In Sec.~\ref{sec:currents} and Appendix.~\ref{app:othercurrents} we have found there is a one to one correspondence between vector and axial-vector tetraquark currents, and in Sec.~\ref{sec:chiral} and Appendix.~\ref{app:chiraltransformation} we have found that they are closely related by chiral transformations. In this section we shall consider the charge-conjugation parity, and study the isovector tetraquark currents of $I^G J^{PC} = 1^+1^{--}$, $1^+1^{+-}$, $1^-1^{++}$ and $1^-1^{-+}$. We shall also find that there is a similar ``symmetry'' among tetraquark currents of these quantum numbers. Accordingly, we propose that there might be a missing state having $I^GJ^{PC}=1^+1^{+-}$ and a mass around 1.6 GeV. We shall use the method of QCD sum rules to verify this theoretically in the following sections.

The charge-conjugation transformation changes the diquark to antidiquark, and antidiquark to diquark, while it maintains their flavor structure. If the tetraquark state has a definite charge-conjugation parity, either positive or negative, the constituent diquark ($qq$) and antidiquark ($\bar q \bar q$) must
have the same flavor symmetry, which is either symmetric $\mathbf{6_f} \otimes \mathbf{\bar 6_f}$ ($\mathbf{S}$) or antisymmetric $\mathbf{\bar 3_f} \otimes \mathbf{3_f}$ ($\mathbf{A}$). We note that a proper mixture of $\mathbf{\bar 3_f} \otimes \mathbf{\bar 6_f}$ and $\mathbf{6_f} \otimes
\mathbf{3_f}$ can also have a definite charge-conjugation parity. However, to simplify our calculation, we shall not study them in this paper.

The tetraquark currents of quantum numbers $J^{PC}=1^{-+}$ have been studied in Ref.~\cite{Chen:2008qw}. There are two independent tetraquark currents of quantum numbers $J^{PC}=1^{-+}$, where the diquark and antidiquark inside have a symmetric flavor structure $\mathbf{6_f}(qq) \otimes \mathbf{\bar 6_f}(\bar q \bar q)$ ($\mathbf{S}$):
%
\begin{eqnarray}
\label{eq:mpS}
\eta^{-+}_{S,1} &=& q_A^{aT} C \gamma_5 q_B^b (\bar{q}_C^a \gamma_\mu \gamma_5 C \bar{q}_D^{bT} + \bar{q}_C^b \gamma_\mu \gamma_5 C \bar{q}_D^{aT}) + q_A^{aT} C \gamma_\mu \gamma_5 q_B^b (\bar{q}_C^a \gamma_5 C \bar{q}_D^{bT} + \bar{q}_C^b \gamma_5 C \bar{q}_D^{aT}) \, ,
\\ \nonumber \eta^{-+}_{S,2} &=& q_A^{aT} C \gamma^\nu q_B^b (\bar{q}_C^a \sigma_{\mu\nu} C \bar{q}_D^{bT} - \bar{q}_C^b \sigma_{\mu\nu} C \bar{q}_D^{aT}) + q_A^{aT} C \sigma_{\mu\nu} q_B^b (\bar{q}_C^a \gamma^\nu C \bar{q}_D^{bT} - \bar{q}_C^b \gamma^\nu C \bar{q}_D^{aT}) \, .
\end{eqnarray}
%
Due to this symmetric flavor structure $\mathbf{6_f}(qq) \otimes \mathbf{\bar 6_f}(\bar q \bar q)$ ($\mathbf{S}$), there are two sets of isovector currents of $I^GJ^{PC}=1^-1^{-+}$: one set with the quark contents $\bar q q \bar q q$ (such as $\bar u d (\bar u u + \bar d d)$; $q$ represents $up$ or $down$ quarks)
\begin{eqnarray}
\nonumber && \left \{
\begin{array}{l}
\eta^{-+}_{S,1,1} \equiv \eta^{-+}_{S,1}(qq\bar q \bar q) \sim u_a^T C
\gamma_5 d_b (\bar{u}_a \gamma_\mu \gamma_5 C \bar{d}_b^T +
\bar{u}_b \gamma_\mu \gamma_5 C \bar{d}_a^T) + u_a^T C \gamma_\mu
\gamma_5 d_b (\bar{u}_a \gamma_5 C \bar{d}_b^T + \bar{u}_b \gamma_5
C \bar{d}_a^T) \, ,
\\ \eta^{-+}_{S,2,1} \equiv \eta^{-+}_{S,2}(qq\bar q \bar q)
\sim u_a^T C \gamma^\nu d_b (\bar{u}_a \sigma_{\mu\nu} C \bar{d}_b^T
- \bar{u}_b \sigma_{\mu\nu} C \bar{d}_a^T) + u_a^T C \sigma_{\mu\nu}
d_b (\bar{u}_a \gamma^\nu C \bar{d}_b^T - \bar{u}_b \gamma^\nu C
\bar{d}_a^T) \, ,
\end{array} \right.
\end{eqnarray}
and the other with $\bar q q \bar s s$ (such as $\bar u d \bar s s$)
\begin{eqnarray}
\nonumber && \left \{
\begin{array}{l} \eta^{-+}_{S,1,2} \equiv \eta^{-+}_{S,1}(q s \bar q \bar s) \sim u_a^T C \gamma_5 s_b
(\bar{u}_a \gamma_\mu \gamma_5 C \bar{s}_b^T + \bar{u}_b \gamma_\mu
\gamma_5 C \bar{s}_a^T) + u_a^T C \gamma_\mu \gamma_5 s_b (\bar{u}_a
\gamma_5 C \bar{s}_b^T + \bar{u}_b \gamma_5 C \bar{s}_a^T) \, ,
\\ \eta^{-+}_{S,2,2} \equiv \eta^{-+}_{S,2}(q s \bar q \bar s)
\sim u_a^T C \gamma^\nu s_b (\bar{u}_a \sigma_{\mu\nu} C \bar{s}_b^T
- \bar{u}_b \sigma_{\mu\nu} C \bar{s}_a^T) + u_a^T C \sigma_{\mu\nu}
s_b (\bar{u}_a \gamma^\nu C \bar{s}_b^T - \bar{u}_b \gamma^\nu C
\bar{s}_a^T) \, .
\end{array} \right.
\end{eqnarray}
Similarly we find the following two independent tetraquark currents of quantum numbers $J^{PC}=1^{-+}$, where the diquark and antidiquark inside have an antisymmetric flavor structure $\mathbf{\bar 3_f} (qq) \otimes \mathbf{3_f} (\bar q \bar q)$ ($\mathbf{A}$):
%
\begin{eqnarray}
\eta^{-+}_{A,1} &=& q_A^{aT} C \gamma_5 q_B^b (\bar{q}_C^a \gamma_\mu \gamma_5 C \bar{q}_D^{bT} - \bar{q}_C^b \gamma_\mu \gamma_5 C \bar{q}_D^{aT}) + q_A^{aT} C
\gamma_\mu \gamma_5 q_B^b (\bar{q}_C^a \gamma_5 C \bar{q}_D^{bT} - \bar{q}_C^b \gamma_5 C \bar{q}_D^{aT}) \, ,
\\ \nonumber \eta^{-+}_{A,2} &=&  q_A^{aT} C \gamma^\nu q_B^b (\bar{q}_C^a \sigma_{\mu\nu} C \bar{q}_D^{bT} + \bar{q}_C^b \sigma_{\mu\nu} C \bar{q}_D^{aT}) + q_A^{aT} C \sigma_{\mu\nu} q_B^b (\bar{q}_C^a \gamma^\nu C \bar{q}_D^{bT} + \bar{q}_C^b \gamma^\nu C \bar{q}_D^{aT})\, ,
\end{eqnarray}
%
Due to this antisymmetric flavor structure $\mathbf{\bar 3_f} (qq) \otimes \mathbf{3_f} (\bar q \bar q)$ ($\mathbf{A}$), there is one set of isovector currents with the quark contents $\bar q q \bar s s$ (such as $\bar u d \bar s s$):
\begin{eqnarray}
&& \nonumber \left \{
\begin{array}{l}
\eta^{-+}_{A,1,1} \equiv \eta^{-+}_{A,1}(q s\bar q \bar s) \sim u_a^T C
\gamma_5 s_b (\bar{u}_a \gamma_\mu \gamma_5 C \bar{s}_b^T -\bar{u}_b
\gamma_\mu \gamma_5 C \bar{s}_a^T) + u_a^T C \gamma_\mu \gamma_5 s_b
(\bar{u}_a \gamma_5 C \bar{s}_b^T - \bar{u}_b \gamma_5 C
\bar{s}_a^T) \, ,
\\ \eta^{-+}_{A,1,2} \equiv \eta^{-+}_{A,2}(q s\bar q \bar s)
\sim u_a^T C \gamma^\nu s_b (\bar{u}_a \sigma_{\mu\nu} C \bar{s}_b^T
+ \bar{u}_b \sigma_{\mu\nu} C \bar{s}_a^T) + u_a^T C \sigma_{\mu\nu}
s_b (\bar{u}_a \gamma^\nu C \bar{s}_b^T + \bar{u}_b \gamma^\nu C
\bar{s}_a^T) \, ,
\end{array} \right.
\end{eqnarray}
We use $\sim$ to make clear that the quark contents here are not
exactly correct. For instance, in the current $\eta^{-+}_{A,1,1}$, the
state $us\bar u \bar s$ does not have isospin one. The correct quark
contents should be $(u s \bar u \bar s - d s \bar d \bar s)$.
However, in the following QCD sum rule analyses, we shall not
include the masses of $up$ and $down$ quarks and we shall choose the same value
for $\langle \bar u u \rangle$ and $\langle \bar d d \rangle$.
Therefore, the QCD sum rule results are the same.

We also construct tetraquark currents of other quantum numbers. We find the following two independent tetraquark currents of quantum numbers $J^{PC}=1^{--}$, where the diquark and antiquark inside have a symmetric flavor structure $\mathbf{6_f}(qq) \otimes \mathbf{\bar 6_f}(\bar q \bar q)$ ($\mathbf{S}$):
%
\begin{eqnarray}
\label{eq:mmS}
\eta^{--}_{S,1} &=& q_A^{aT} C \gamma_5 q_B^b (\bar{q}_C^a \gamma_\mu \gamma_5 C \bar{q}_D^{bT} + \bar{q}_C^b \gamma_\mu \gamma_5 C \bar{q}_D^{aT}) - q_A^{aT} C
\gamma_\mu \gamma_5 q_B^b (\bar{q}_C^a \gamma_5 C \bar{q}_D^{bT} + \bar{q}_C^b \gamma_5 C \bar{q}_D^{aT}) \, ,
\\ \nonumber \eta^{--}_{S,2} &=& A_a^T C \gamma^\nu q_B^b (\bar{q}_C^a \sigma_{\mu\nu} C \bar{q}_D^{bT} - \bar{q}_C^b \sigma_{\mu\nu} C \bar{q}_D^{aT}) - q_A^{aT} C \sigma_{\mu\nu} q_B^b (\bar{q}_C^a \gamma^\nu C \bar{q}_D^{bT} - \bar{q}_C^b \gamma^\nu C \bar{q}_D^{aT}) \, .
\end{eqnarray}
%
Similarly we find the following two independent tetraquark currents of quantum numbers $J^{PC}=1^{--}$, where the diquark and antidiquark inside have an antisymmetric flavor structure $\mathbf{\bar 3_f} (qq) \otimes \mathbf{3_f} (\bar q \bar q)$ ($\mathbf{A}$):
%
\begin{eqnarray}
\eta^{--}_{A,1} &=& q_A^{aT} C \gamma_5 q_B^b (\bar{q}_C^a \gamma_\mu \gamma_5 C \bar{q}_D^{bT} - \bar{q}_C^b \gamma_\mu \gamma_5 C \bar{q}_D^{aT}) - q_A^{aT} C
\gamma_\mu \gamma_5 q_B^b (\bar{q}_C^a \gamma_5 C \bar{q}_D^{bT} - \bar{q}_C^b \gamma_5 C \bar{q}_D^{aT}) \, ,
\\ \nonumber \eta^{--}_{A,2} &=& q_A^{aT} C \gamma^\nu q_B^b (\bar{q}_C^a \sigma_{\mu\nu} C \bar{q}_D^{bT} + \bar{q}_C^b \sigma_{\mu\nu} C \bar{q}_D^{aT}) - q_A^{aT} C \sigma_{\mu\nu} q_B^b (\bar{q}_C^a \gamma^\nu C \bar{q}_D^{bT} + \bar{q}_C^b \gamma^\nu C \bar{q}_D^{aT}) \, .
\end{eqnarray}
%

We find the following two independent tetraquark currents of quantum numbers $J^{PC}=1^{++}$, where the diquark and antidiquark inside have a symmetric flavor structure $\mathbf{6_f}(qq) \otimes \mathbf{\bar 6_f}(\bar q \bar q)$ ($\mathbf{S}$):
%
\begin{eqnarray}
\label{eq:ppS}
\eta^{++}_{S,1} &=& q_A^{aT} C q_B^b (\bar{q}_C^a \gamma_\mu \gamma_5 C \bar{q}_D^{bT} + \bar{q}_C^b \gamma_\mu \gamma_5 C \bar{q}_D^{aT}) + q_A^{aT} C \gamma_\mu \gamma_5 q_B^b (\bar{q}_C^a C \bar{q}_D^{bT} + \bar{q}_C^b C \bar{q}_D^{aT}) \, ,
\\ \nonumber \eta^{++}_{S,2} &=& q_A^{aT} C \gamma^\nu q_B^b (\bar{q}_C^a \sigma_{\mu\nu} \gamma_5 C \bar{q}_D^{bT} - \bar{q}_C^b \sigma_{\mu\nu} \gamma_5 C \bar{q}_D^{aT}) + q_A^{aT} C \sigma_{\mu\nu} \gamma_5 q_B^b (\bar{q}_C^a \gamma^\nu C \bar{q}_D^{bT} - \bar{q}_C^b \gamma^\nu C \bar{q}_D^{aT})\, .
\end{eqnarray}
%
Similarly we find the following two independent tetraquark currents of quantum numbers $J^{PC}=1^{++}$, where the diquark and antidiquark inside have an antisymmetric flavor structure $\mathbf{\bar 3_f} (qq) \otimes \mathbf{3_f} (\bar q \bar q)$ ($\mathbf{A}$):
%
\begin{eqnarray}
\eta^{++}_{A,1} &=& q_A^{aT} C q_B^b (\bar{q}_C^a \gamma_\mu \gamma_5 C \bar{q}_D^{bT} - \bar{q}_C^b \gamma_\mu \gamma_5 C \bar{q}_D^{aT}) + q_A^{aT} C \gamma_\mu \gamma_5 q_B^b (\bar{q}_C^a C \bar{q}_D^{bT} - \bar{q}_C^b C \bar{q}_D^{aT}) \, ,
\\ \nonumber \eta^{++}_{A,2} &=& q_A^{aT} C \gamma^\nu q_B^b (\bar{q}_C^a \sigma_{\mu\nu} \gamma_5 C \bar{q}_D^{bT} + \bar{q}_C^b \sigma_{\mu\nu} \gamma_5 C \bar{q}_D^{aT}) + q_A^{aT} C \sigma_{\mu\nu} \gamma_5 q_B^b (\bar{q}_C^a \gamma^\nu C \bar{q}_D^{bT} + \bar{q}_C^b \gamma^\nu C \bar{q}_D^{aT}) \, .
\end{eqnarray}
%

We find the following two independent tetraquark currents of quantum numbers $J^{PC}=1^{+-}$, where the diquark and antidiquark inside have a symmetric flavor structure $\mathbf{6_f}(qq) \otimes \mathbf{\bar 6_f}(\bar q \bar q)$ ($\mathbf{S}$):
%
\begin{eqnarray}
\label{eq:pmS}
\eta^{+-}_{S,1} &=& q_A^{aT} C q_B^b (\bar{q}_C^a \gamma_\mu \gamma_5 C \bar{q}_D^{bT} + \bar{q}_C^b \gamma_\mu \gamma_5 C \bar{q}_D^{aT}) - q_A^{aT} C \gamma_\mu \gamma_5 q_B^b (\bar{q}_C^a C \bar{q}_D^{bT} + \bar{q}_C^b C \bar{q}_D^{aT}) \, ,
\\ \nonumber \eta^{+-}_{S,2} &=& q_A^{aT} C \gamma^\nu q_B^b (\bar{q}_C^a \sigma_{\mu\nu} \gamma_5 C \bar{q}_D^{bT} - \bar{q}_C^b \sigma_{\mu\nu} \gamma_5 C \bar{q}_D^{aT}) - q_A^{aT} C \sigma_{\mu\nu} \gamma_5 q_B^b (\bar{q}_C^a \gamma^\nu C \bar{q}_D^{bT} - \bar{q}_C^b \gamma^\nu C \bar{q}_D^{aT})\, .
\end{eqnarray}
%
Similarly we find the following two independent tetraquark currents of quantum numbers $J^{PC}=1^{+-}$, where the diquark and antidiquark inside have an antisymmetric flavor structure $\mathbf{\bar 3_f} (qq) \otimes \mathbf{3_f} (\bar q \bar q)$ ($\mathbf{A}$):
%
\begin{eqnarray}
\eta^{+-}_{A,1} &=& q_A^{aT} C q_B^b (\bar{q}_C^a \gamma_\mu \gamma_5 C \bar{q}_D^{bT} - \bar{q}_C^b \gamma_\mu \gamma_5 C \bar{q}_D^{aT}) - q_A^{aT} C \gamma_\mu \gamma_5 q_B^b (\bar{q}_C^a C \bar{q}_D^{bT} - \bar{q}_C^b C \bar{q}_D^{aT}) \, ,
\\ \nonumber \eta^{+-}_{A,2} &=& q_A^{aT} C \gamma^\nu q_B^b (\bar{q}_C^a \sigma_{\mu\nu} \gamma_5 C \bar{q}_D^{bT} + \bar{q}_C^b \sigma_{\mu\nu} \gamma_5 C \bar{q}_D^{aT}) - q_A^{aT} C \sigma_{\mu\nu} \gamma_5 q_B^b (\bar{q}_C^a \gamma^\nu C \bar{q}_D^{bT} + \bar{q}_C^b \gamma^\nu C \bar{q}_D^{aT}) \, .
\end{eqnarray}
%

From these expressions, we can clearly see that there is a one to one correspondence among local tetraquark currents of $J^{PC}=1^{-+}$, $J^{PC}=1^{--}$, $J^{PC}=1^{++}$ and $J^{PC}=1^{+-}$, such as among Eq.(\ref{eq:mpS}), (\ref{eq:mmS}), (\ref{eq:ppS}) and (\ref{eq:pmS}). According to these tetraquark currents, we can construct isovector currents whose quark contents are similar to those of quantum numbers $I^GJ^{PC}=1^-1^{-+}$. We use $\eta^{\pm\pm}_{S,i,1} \equiv \eta^{\pm\pm}_{S,i}(qq\bar q \bar q)$, $\eta^{\pm\pm}_{S,i,2} \equiv \eta^{\pm\pm}_{S,i}(qs\bar q \bar s)$ and $\eta^{\pm\pm}_{A,i,1} \equiv \eta^{\pm\pm}_{A,i}(qs\bar q \bar s)$ to denote them, while we do not show their explicit forms here. Here the first $\pm$ symbol denotes the $P$-parity, and the second $\pm$ symbol denotes the $C$-parity. Again, we can quickly find that there is a one to one correspondence among them.

We can also study their chiral structure and their chiral transformation properties. For example, the current $\eta^{-+}_{A,1,1} \equiv \eta^{-+}_{A,1}(qs\bar q \bar s)$ can be written as a combination of $\eta_{1}^{\rm V, \mathbb{S}} + \eta_{2}^{\rm V, \mathbb{S}}$ and $\eta_{1,N}^{\rm V, \mathbb{O}} + \eta_{2,N}^{\rm V, \mathbb{O}}$ with the quark contents $qs\bar q \bar s$, and so it contains $[({\mathbf 3}, \bar {\mathbf 3}) + (\bar {\mathbf 3}, {\mathbf 3})]$, $[({\mathbf {15}}, \bar {\mathbf 3}) + (\bar {\mathbf 3}, {\mathbf {15}})]$ and $[({\mathbf 3}, \overline{\mathbf {15}}) + (\overline{\mathbf {15}}, {\mathbf 3})]$ components. Among these components, we are interested in the $[({\mathbf 3}, \bar {\mathbf 3}) + (\bar {\mathbf 3}, {\mathbf 3})]$ one, since the physical states probably belong to it, or partly belong to it. In principle this $[({\mathbf 3}, \bar {\mathbf 3}) + (\bar {\mathbf 3}, {\mathbf 3})]$ component can be projected out, which is, however, technically difficult. Fortunately, in QCD sum rules we usually calculate the mass of the lowest-lying state which the hadronic current couples to, and this lowest-lying state probably belongs to the $[({\mathbf 3}, \bar {\mathbf 3}) + (\bar {\mathbf 3}, {\mathbf 3})]$ chiral multiplet.

\section{SVZ Sum Rules}
\label{sec:sumrule}

For the past decades QCD sum rules have proven to be a very powerful and successful non-perturbative method~\cite{Shifman:1978bx,Reinders:1984sr}. In sum rule analyses, we consider two-point correlation functions:
%
\begin{equation}
\Pi_{\mu\nu}(q^2) \, \equiv \, i \int d^4x e^{iqx} \langle 0 | T
\eta_\mu(x) { \eta_\nu^\dagger } (0) | 0 \rangle \, , \label{def:pi}
\end{equation}
%
where $\eta_\mu$ is an interpolating current for the tetraquark. The
Lorentz structure can be simplified to be:
%
\begin{equation}
\Pi_{\mu\nu}(q^2) = ( {q_\mu q_\nu \over q^2} - g_{\mu\nu} )
\Pi(q^2) + {q_\mu q_\nu \over q^2} \Pi^{(0)}(q^2) \, .
\label{def:pi1}
\end{equation}
%
We compute $\Pi(q^2)$ in the operator product expansion (OPE) of QCD up to certain order in the expansion, which is then matched with a hadronic parametrization to extract information about hadron properties. At the hadron level, we express the correlation function in the form of the dispersion relation with a spectral function:
%
\begin{equation}
\Pi(q^2)={1\over\pi}\int^\infty_{s_<}\frac{{\rm Im} \Pi(s)}{s-q^2-i\varepsilon}ds \, ,
\label{eq:disper}
\end{equation}
%
where the integration starts from the mass square of all current quarks. The imaginary part of the two-point correlation function is
%
\begin{eqnarray}
{\rm Im} \Pi(s) & \equiv & \pi \sum_n\delta(s-M^2_n)\langle 0|\eta|n\rangle\langle n|{\eta^\dagger}|0\rangle \, .
\label{eq:rho}
\end{eqnarray}
%
For the second equation, as usual, we adopt a parametrization of one pole dominance for the ground state $Y$ and a continuum contribution. The sum rule analysis is then performed after the Borel transformation of the two expressions of the correlation function, (\ref{def:pi}) and (\ref{eq:disper}):
%
\begin{equation}
\Pi^{(all)}(M_B^2)\equiv\mathcal{B}_{M_B^2}\Pi(p^2) = {1\over\pi} \int^\infty_{s_<} e^{-s/M_B^2} {\rm Im} \Pi(s) ds \, .
\label{eq:borel}
\end{equation}
%
Assuming the contribution from the continuum states can be approximated well by the spectral density of OPE above a threshold value $s_0$ (duality), we arrive at the sum rule equation
%
\begin{eqnarray}
f^2_Y e^{-M_Y^2/M_B^2} &=& \Pi(s_0, M_B^2) \equiv {1\over\pi} \int^{s_0}_{s_<} e^{-s/M_B^2} {\rm Im} \Pi(s) ds
\label{eq:fin} \, .
\end{eqnarray}
%
Differentiating Eq.~(\ref{eq:fin}) with respect to $1 / M_B^2$ and dividing it by Eq. (\ref{eq:fin}), finally we obtain
%
\begin{equation}
\label{eq:sumrule}
M^2_Y = \frac{\frac{\partial}{\partial(-1/M_B^2)}\Pi(s_0, M_B^2)}{\Pi(s_0, M_B^2)} \, .
\end{equation}

In this section we use the method of QCD sum rules to calculate the masses of vector and axial-vector mesons. We shall use the tetraquark currents $\eta^{\pm\pm}_{S,i,1} \equiv \eta^{\pm\pm}_{S,i}(qq\bar q \bar q)$, $\eta^{\pm\pm}_{S,i,2} \equiv \eta^{\pm\pm}_{S,i}(qs\bar q \bar s)$ and $\eta^{\pm\pm}_{A,i,1} \equiv \eta^{\pm\pm}_{A,i}(qs\bar q \bar s)$, which have been classified in Sec.~\ref{sec:vectoraxialvector}. They have quantum numbers $I^G J^{PC} = 1^+1^{--}$, $1^+1^{+-}$, $1^-1^{++}$ and $1^-1^{-+}$. We would like to note again that we shall only use these tetraquark currents for simplicity, but other currents representing the $\bar q q$ structure, the hybrid structure and the meson-meson structure can also contribute here.

In Ref.~\cite{Chen:2008qw} we have studied the tetraquark currents $\eta^{-+}_{S,i,1} \equiv \eta^{-+}_{S,i}(qq\bar q \bar q)$, $\eta^{-+}_{S,i,2} \equiv \eta^{-+}_{S,i}(qs\bar q \bar s)$ and $\eta^{-+}_{A,i,1} \equiv \eta^{-+}_{A,i}(qs\bar q \bar s)$ having exotic quantum numbers $1^-1^{-+}$. The first set $\eta^{-+}_{S,i,1}$ only contains $up$ and $down$ quarks. They both lead to the masses around 1.6 GeV, and so they can couple to the exotic meson $\pi_1(1600)$. The last two sets $\eta^{-+}_{S,i,2}$ and $\eta^{-+}_{A,i,1}$ contain one $s \bar s$ pair. They all lead to the masses around 2.0 GeV, and so they can couple to the exotic meson $\pi_1(2000)$.

In the following subsections we shall separately study tetraquark currents of other quantum numbers $I^G J^{PC} = 1^+1^{--}$, $1^-1^{++}$ and $1^+1^{+-}$. Our procedures are quite similar to those used in Ref.~\cite{Chen:2008qw}. In our numerical analysis, we use the following values for various condensates and $m_s$ at 1 GeV and $\alpha_s$ at 1.7 GeV~\cite{Narison:2002pw,Yang:1993bp,Gimenez:2005nt,Jamin:2002ev,Ioffe:2002be,Ovchinnikov:1988gk}:
%
\begin{eqnarray}
\nonumber &&\langle\bar qq \rangle=-(0.240 \mbox{ GeV})^3\, ,
\\
\nonumber &&\langle\bar ss\rangle=-(0.8\pm 0.1)\times(0.240 \mbox{
GeV})^3\, ,
\\
\nonumber &&\langle g_s^2GG\rangle =(0.48\pm 0.14) \mbox{ GeV}^4\, ,
\\
\label{condensates} && \langle g_s\bar q\sigma G
q\rangle=-M_0^2\times\langle\bar qq\rangle\, ,
\\
\nonumber && M_0^2=(0.8\pm0.2)\mbox{ GeV}^2\, ,
\\
\nonumber &&m_s(1\mbox{ GeV})=125 \pm 20 \mbox{ MeV}\, ,
\\
\nonumber && \alpha_s(1.7\mbox{GeV}) = 0.328 \pm 0.03 \pm 0.025 \, .
\end{eqnarray}
%

\subsection{Tetraquark currents of $I^G J^{PC} = 1^+1^{--}$}
\label{sec:sumruleNN}

In this subsection we use the tetraquark currents $\eta^{--}_{S,i,1} \equiv \eta^{--}_{S,i}(qq\bar q \bar q)$, $\eta^{--}_{S,i,2} \equiv \eta^{--}_{S,i}(qs\bar q \bar s)$ and $\eta^{--}_{A,i,1} \equiv \eta^{--}_{A,i}(qs\bar q \bar s)$ to perform QCD sum rule analyses. We calculate the OPE up to dimension twelve. Here we only show the result for the current $\eta^{--}_{A,1,1} \equiv \eta^{--}_{A,1}(q s \bar q \bar s)$, which has quark contents $q s \bar q \bar s$. Others are shown in Appendix.~\ref{app:ope}.
%
\begin{eqnarray}
\nonumber \Pi^{--}_{A,1,1}(M_B^2) &=& \int^{s_0}_{4 m_s^2} \Bigg [ {1 \over 36864
\pi^6} s^4 - { m_s^2 \over 1024 \pi^6 } s^3 + \Big ( { \langle
g_s^2 G G \rangle \over 18432 \pi^6 } + {m_s \langle \bar s s
\rangle \over 192 \pi^4} \Big ) s^2 + \Big ( { \langle \bar q q \rangle^2 \over 72
\pi^2 } + { \langle \bar s s \rangle^2 \over 72 \pi^2 }
\\ \nonumber && - { m_s
\langle g_s \bar s \sigma G s \rangle \over 192 \pi^4 }
- { m_s^2
\langle g_s^2 G G \rangle \over 4608 \pi^6 } \Big ) s + { \langle
\bar q q \rangle \langle g_s \bar q \sigma G q \rangle \over 48
\pi^2 }
+ { \langle \bar s s \rangle \langle g_s \bar s
\sigma G s \rangle \over 48 \pi^2 }
\\ \label{eq:ope} && + { 7 m_s \langle g_s^2 G G
\rangle \langle \bar q q \rangle \over 2304 \pi^4} - { m_s^2 \langle
\bar q q \rangle^2 \over 4 \pi^2 } + { m_s^2 \langle
\bar s s \rangle^2 \over 48 \pi^2 } \Bigg ] e^{-s/M_B^2} ds
\\ \nonumber && + \Big ( { \langle g_s \bar q \sigma G q \rangle^2 \over 192 \pi^2
} + { \langle g_s \bar s \sigma G
s \rangle^2 \over 192 \pi^2 } - { 5 \langle g_s^2 GG \rangle \langle
\bar q q \rangle \langle \bar s s \rangle \over 864 \pi^2 } + { 5m_s
\langle \bar q q \rangle^2 \langle \bar s s \rangle \over 9 }
+ { 5 m_s \langle g_s^2 GG \rangle \langle g_s \bar q
\sigma G q \rangle \over 4608 \pi^4 }
\\ \nonumber &&  - { m_s^2 \langle \bar q q
\rangle \langle g_s \bar q \sigma G q \rangle \over 6 \pi^2 }\Big) + {1 \over M_B^2} \Big( - {16 g_s^2 \langle
\bar q q \rangle ^2 \langle \bar s s \rangle^2 \over 81 } + { \langle g_s^2 GG \rangle \langle \bar q q \rangle
\langle g_s \bar s \sigma G s \rangle \over 1152 \pi^2 }
\\ \nonumber && + { \langle
g_s^2 GG \rangle \langle \bar s s \rangle \langle g_s \bar q \sigma
G q \rangle \over 1152 \pi^2 } - { m_s \langle \bar q q \rangle^2
\langle g_s \bar s \sigma G s \rangle \over 9} - { m_s \langle
\bar q q \rangle \langle \bar s s \rangle \langle g_s \bar q \sigma
G q \rangle \over 6} \Big)\, .
\end{eqnarray}
%
In the above equations, $\langle \bar{q}q \rangle$ and $\langle \bar{s}s \rangle$ are the dimension
$D=3$ quark condensates; $\langle g^2 GG \rangle$ is a $D=4$
gluon condensate; $\langle g\bar{q}\sigma Gq \rangle$ and $\langle g\bar{s}\sigma Gs \rangle$ are $D=5$ mixed
condensates. As usual, we assume the vacuum saturation for higher dimensional condensates such
as $\langle 0 | \bar q q \bar q q |0\rangle \sim \langle 0 | \bar q
q |0\rangle \langle 0|\bar q q |0\rangle$. We note that we have calculated the tree-level $\mathcal{O}(\alpha_s)$ correction, i.e., the condensate
$g_s^2 \langle \bar q q \rangle^2 \langle \bar s s \rangle^2$, but omitted other $\mathcal{O}(\alpha_s)$ corrections.
To obtain these results, we keep terms of order $\mathcal{O}(m_q^2)$ in the propagator of a
massive quark in the presence of quark and gluon condensates:
%
\begin{eqnarray} \nonumber
i S^{ab} & \equiv & \langle 0 | T [ q^a(x) q^b(0) ] | 0 \rangle
\\ \nonumber &=& { i \delta^{ab} \over 2 \pi^2 x^4 } \hat{x} + {i \over
32\pi^2} { \lambda^n_{ab} \over 2 } g_c G^n_{\mu\nu} {1 \over x^2}
(\sigma^{\mu\nu} \hat{x} + \hat{x} \sigma^{\mu\nu}) - { \delta^{ab}
\over 12 } \langle \bar q q \rangle \\ && + { \delta^{ab} x^2 \over
192 } \langle g_c \bar q \sigma G q \rangle - { m_q \delta^{ab}
\over 4 \pi^2 x^2 } + { i \delta^{ab} m_q \langle \bar q q \rangle
\over 48 }  \hat x + { i \delta^{ab} m_q^2 \over 8 \pi^2 x^2 }
\hat{x} \, .
\end{eqnarray}
%

\begin{figure}[hbt]
\begin{center}
\scalebox{0.6}{\includegraphics{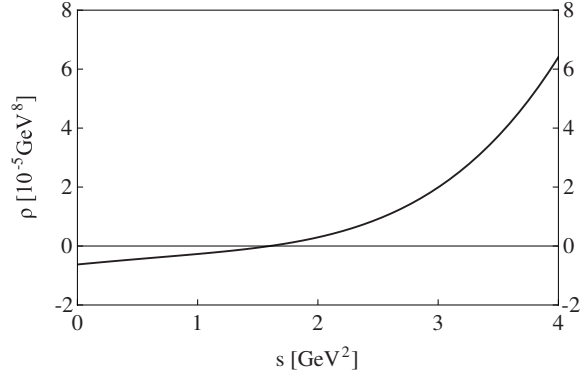}}
\caption{Spectral density for the current $\eta^{--}_{A,1,1} \equiv \eta^{--}_{A,1}(q s \bar q \bar s)$.}
\label{fig:rhoA1}
\end{center}
\end{figure}

\begin{figure}[h!t]
\begin{center}
\scalebox{0.7}{\includegraphics{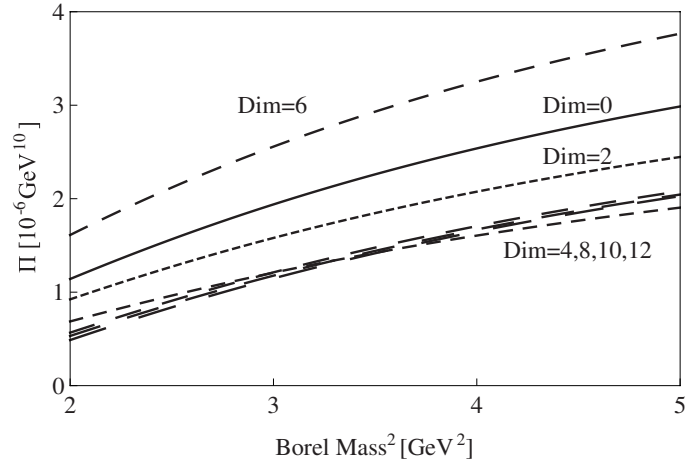}} \caption{Various
contributions to the correlation function of the current $\eta^{--}_{A,1,1} \equiv \eta^{--}_{A,1}(q s \bar q \bar s)$ as functions of the Borel mass $M_B$ in units of
GeV$^{10}$ at $s_0$ = 4 GeV$^2$. The labels indicate the dimension up
to which the OPE terms are included.} \label{fig:pi}
\end{center}
\end{figure}

The spectral density of the current $\eta^{--}_{A,1,1} \equiv \eta^{--}_{A,1}(q s \bar q \bar s)$ can be easily extracted from Eq.~(\ref{eq:ope}):
%
\begin{eqnarray}
\nonumber \rho^{--}_{A,1,1}(M_B^2) &=& {1 \over 36864
\pi^6} s^4 - { m_s^2 \over 1024 \pi^6 } s^3 + \Big ( { \langle
g_s^2 G G \rangle \over 18432 \pi^6 } + {m_s \langle \bar s s
\rangle \over 192 \pi^4} \Big ) s^2 + \Big ( { \langle \bar q q \rangle^2 \over 72
\pi^2 } + { \langle \bar s s \rangle^2 \over 72 \pi^2 }
- { m_s
\langle g_s \bar s \sigma G s \rangle \over 192 \pi^4 }
- { m_s^2
\langle g_s^2 G G \rangle \over 4608 \pi^6 } \Big ) s
\\ && + { \langle
\bar q q \rangle \langle g_s \bar q \sigma G q \rangle \over 48
\pi^2 }
+ { \langle \bar s s \rangle \langle g_s \bar s
\sigma G s \rangle \over 48 \pi^2 }
+ { 7 m_s \langle g_s^2 G G
\rangle \langle \bar q q \rangle \over 2304 \pi^4} - { m_s^2 \langle
\bar q q \rangle^2 \over 4 \pi^2 } + { m_s^2 \langle
\bar s s \rangle^2 \over 48 \pi^2 } \, .
\end{eqnarray}
%
We show it in Fig.~\ref{fig:rhoA1} as a function of the energy $s$. It is positive when $s>1.6$ GeV$^2$, and so our working
regions should be inside this. To perform QCD sum rule analyses, first we need
to study the convergence of the OPE. The Borel transformed correlation function of the current $\eta^{--}_{A,1,1} \equiv \eta^{--}_{A,1}(q s \bar q \bar s)$
is shown in Fig.~\ref{fig:pi}, when we take $s_0=4$ GeV$^2$. We can clearly see that besides the leading continuum term, the $D=6$ and $D=8$ terms
give large contributions, but the $D=10$ and $D=12$ terms are negligible. Therefore, the convergence is very good in the region of $3$ GeV$^2<M_B^2 < 4$ GeV$^2$, where OPEs are reliable.
%
%

\begin{figure}[hbt]
\begin{center}
\scalebox{1.3}{\includegraphics{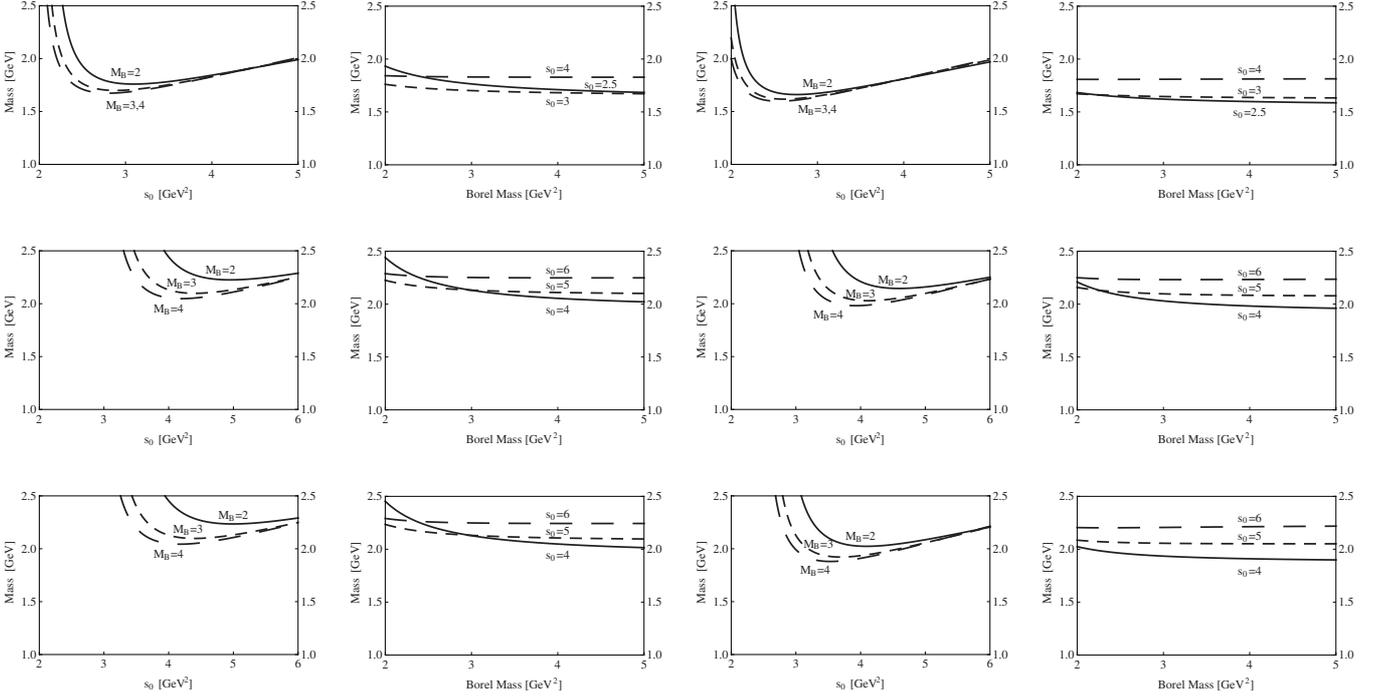}}
\caption{Masses as functions of the threshold value $s_0$ and the Borel Mass $M_B^2$. The upper four figures are obtained using the tetraquark currents $\eta^{--}_{S,1}(qq\bar q \bar q)$ (the two on the left) and $\eta^{--}_{S,2}(qq\bar q \bar q)$ (the two on the right). The threshold value is set to be $s_0 = 2.5$ GeV$^2$ (solid line), 3 GeV$^2$ (short-dashed line) and 4 GeV$^2$ (long-dashed line). Similarly, the middle four figures are obtained using the tetraquark currents $\eta^{--}_{S,i}(qs\bar q \bar s)$, and the lower four figures are obtained using the tetraquark currents $\eta^{--}_{A,i}(qs\bar q \bar s)$. The threshold value is set to be $s_0 = 4$ GeV$^2$ (solid line), 5 GeV$^2$ (short-dashed line) and 6 GeV$^2$ (long-dashed line). For all these twelve figures, the Borel Mass is set to be $M_B^2 = 2$ GeV$^2$ (solid line), 3 GeV$^2$ (short-dashed line) and 4 GeV$^2$ (long-dashed line).}
\label{fig:MM}
\end{center}
\end{figure}

The masses are calculated using Eq.~(\ref{eq:sumrule}), and the results are shown in Figs.~\ref{fig:MM} as functions of the threshold value $s_0$ and the Borel Mass $M_B^2$ for all tetraquark currents of $I^G J^{PC} = 1^+1^{--}$. The upper four figures are obtained using the tetraquark currents $\eta^{--}_{S,1}(qq\bar q \bar q)$ and $\eta^{--}_{S,2}(qq\bar q \bar q)$, whose quark contents are $qq \bar q \bar q$. These two currents lead to similar results. From those figures where masses are shown as functions of $M_B$, we find that the dependence on the Borel mass is weak when $M_B^2> 3$ GeV$^2$. From those figures where masses are shown as functions of $s_0$, we find that there is a mass minimum around 1.6 GeV for both curves where the stability is the best. Accordingly, we fix our working regions to be $3$ GeV$^2<M_B^2<4$ GeV$^2$ and $2.6$ GeV$^2<s_0<3.0$ GeV$^2$, and obtain the similar results 1.68-1.73 GeV and 1.60-1.65 GeV for $\eta^{--}_{S,1}(qq\bar q \bar q)$ and $\eta^{--}_{S,2}(qq\bar q \bar q)$, respectively. Our results suggest that these two currents can couple to $\rho(1570)$ or $\rho(1700)$.

Following the same procedures we perform QCD sum rule analyses to study other tetraquark currents of $I^G J^{PC} = 1^+1^{--}$. The middle four figures of Figs.~\ref{fig:MM} are obtained using the tetraquark currents $\eta^{--}_{S,i}(qs\bar q \bar s)$, and the lower four figures of Figs.~\ref{fig:MM} are obtained using the tetraquark currents $\eta^{--}_{A,i}(qs\bar q \bar s)$. Their quark contents are $qs \bar q \bar s$. We find that the obtained masses all have a mass minimum around 2.0 GeV against the threshold value $s_0$. Consequently, we fix our working regions to be $3$ GeV$^2<M_B^2<4$ GeV$^2$ and $4.0$ GeV$^2<s_0<4.5$ GeV$^2$ where the stability is good. Using these currents, we obtain the similar masses 2.06-2.13 GeV, 1.98-2.05 GeV, 2.05-2.13 GeV, and 1.91-1.99 GeV. Our results suggest that they can couple to $\rho(1900)$ or $\rho(2150)$.

The pole contribution
\begin{equation}
{ \int^{s_0}_{s_<} e^{-s/M_B^2}\rho(s)ds \over \int^{\infty}_{s_<}
e^{-s/M_B^2}\rho(s)ds } \, , \label{def:poleLSR}
\end{equation}
is another important criterion to fix the Borel window and perform reliable QCD sum rule analyses. However, as we have found in Fig.~\ref{fig:rhoA1}, the spectral density of the current $\eta^{--}_{A,1,1} \equiv \eta^{--}_{A,1}(q s \bar q \bar s)$ has some negative parts when $s<1.6$ GeV$^2$, which makes the pole contribution not well defined. The situation is the same for other currents of $I^G J^{PC} = 1^+1^{--}$. Therefore, our results are stable, but have a small pole contribution. To make our analyses more reliable, we have also used the method of finite energy sum rules. We shall find that the obtained results are almost the same. The details are shown in Sec.~\ref{sec:fesr}.

\subsection{Tetraquark currents of $I^G J^{PC} = 1^-1^{++}$}

\begin{figure}[hbt]
\begin{center}
\scalebox{1.3}{\includegraphics{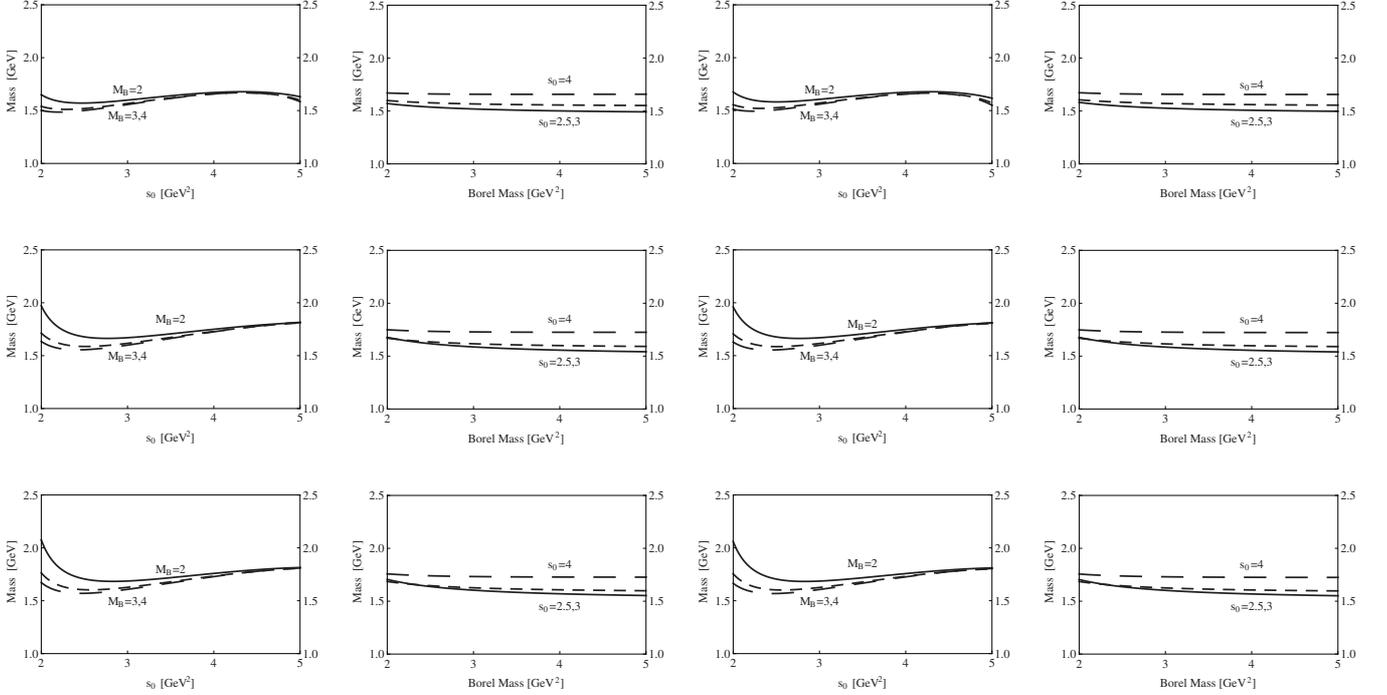}}
\caption{Masses as functions of the threshold value $s_0$ and the Borel Mass $M_B^2$. The upper four figures are obtained using the tetraquark currents $\eta^{++}_{S,1}(qq\bar q \bar q)$ (the two on the left) and $\eta^{++}_{S,2}(qq\bar q \bar q)$ (the two on the right). Similarly, the middle four figures are obtained using the tetraquark currents $\eta^{++}_{S,i}(qs\bar q \bar s)$, and the lower four figures are obtained using the tetraquark currents $\eta^{++}_{A,i}(qs\bar q \bar s)$. For all these twelve figures, the Borel Mass is set to be $M_B^2 = 2$ GeV$^2$ (solid line), 3 GeV$^2$ (short-dashed line) and 4 GeV$^2$ (long-dashed line), and the threshold value is set to be $s_0 = 2.5$ GeV$^2$ (solid line), 3 GeV$^2$ (short-dashed line) and 4 GeV$^2$ (long-dashed line).}
\label{fig:PP}
\end{center}
\end{figure}

In this subsection we follow the same procedures and use the tetraquark currents $\eta^{++}_{S,i,1} \equiv \eta^{++}_{S,i}(qq\bar q \bar q)$, $\eta^{++}_{S,i,2} \equiv \eta^{++}_{S,i}(qs\bar q \bar s)$ and $\eta^{++}_{A,i,1} \equiv \eta^{++}_{A,i}(qs\bar q \bar s)$ to perform QCD sum rule calculations. The masses are calculated using Eq.~(\ref{eq:sumrule}), and the results are shown in Figs.~\ref{fig:PP} as functions of the threshold value $s_0$ and the Borel Mass $M_B^2$. The upper four figures are obtained using the tetraquark currents $\eta^{++}_{S,i}(qq\bar q \bar q)$, the middle four figures are obtained using the tetraquark currents $\eta^{++}_{S,i}(qs\bar q \bar s)$, and the lower four figures are obtained using the tetraquark currents $\eta^{++}_{A,i}(qs\bar q \bar s)$.

We find that the obtained masses all have a mass minimum around 1.5-1.6 GeV against the threshold value $s_0$. Consequently, we fix the working regions to be $3$ GeV$^2<M_B^2<4$ GeV$^2$ and 2.6 GeV$^2 <s_0<$ 3.0 GeV$^2$ where the stability is good. Using these currents, we obtain the similar masses 1.51-1.57 GeV, 1.52-1.57 GeV, 1.56-1.62 GeV, 1.56-1.62 GeV, 1.57-1.63 GeV, and 1.57-1.63 GeV. Our results suggest they can couple to $a_1(1640)$.

\subsection{Tetraquark currents of $I^G J^{PC} = 1^+1^{+-}$}

\begin{figure}[hbt]
\begin{center}
\scalebox{1.3}{\includegraphics{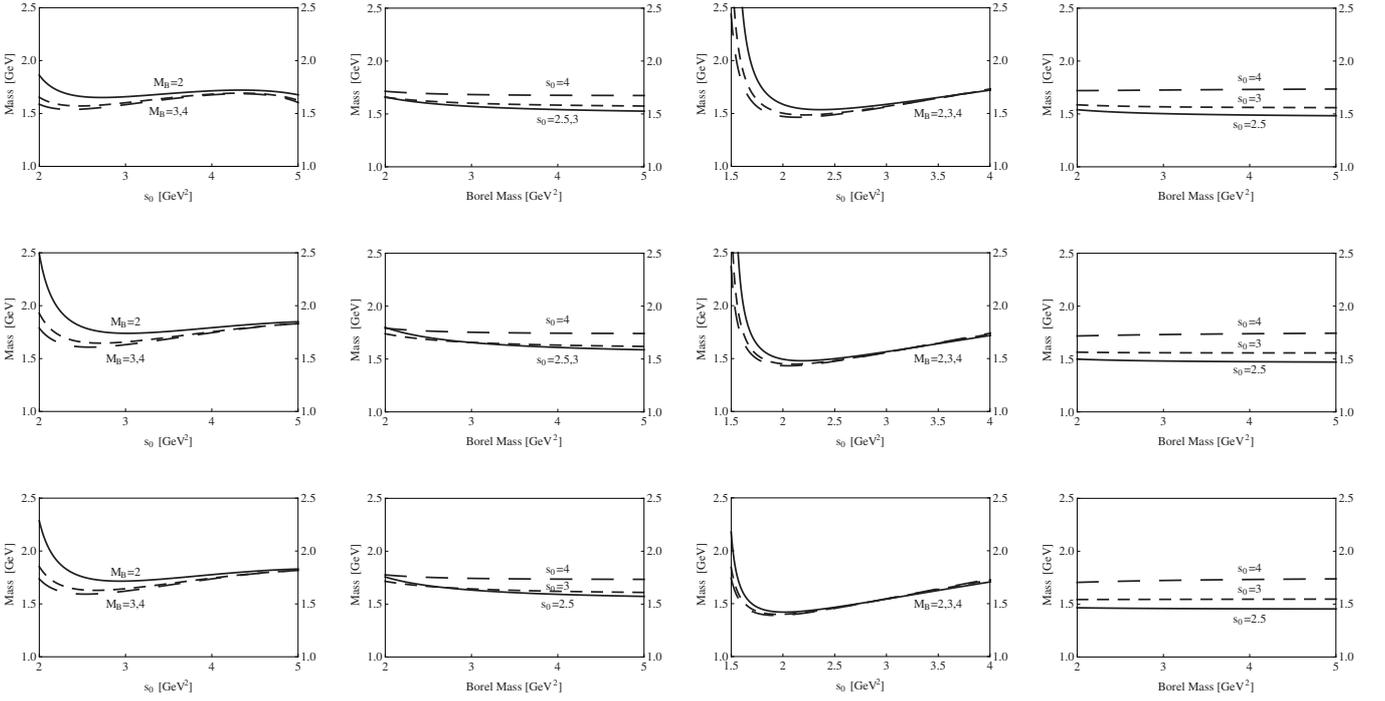}}
\caption{Masses as functions of the threshold value $s_0$ and the Borel Mass $M_B^2$. The upper four figures are obtained using the tetraquark currents $\eta^{+-}_{S,1}(qq\bar q \bar q)$ (the two on the left) and $\eta^{+-}_{S,2}(qq\bar q \bar q)$ (the two on the right). Similarly, the middle four figures are obtained using the tetraquark currents $\eta^{+-}_{S,i}(qs\bar q \bar s)$, and the lower four figures are obtained using the tetraquark currents $\eta^{+-}_{A,i}(qs\bar q \bar s)$. For all these twelve figures, the Borel Mass is set to be $M_B^2 = 2$ GeV$^2$ (solid line), 3 GeV$^2$ (short-dashed line) and 4 GeV$^2$ (long-dashed line), and the threshold value is set to be $s_0 = 2.5$ GeV$^2$ (solid line), 3 GeV$^2$ (short-dashed line) and 4 GeV$^2$ (long-dashed line).}
\label{fig:PM}
\end{center}
\end{figure}

In this subsection we follow the same procedures and use the tetraquark currents $\eta^{+-}_{S,i,1} \equiv \eta^{+-}_{S,i}(qq\bar q \bar q)$, $\eta^{+-}_{S,i,2} \equiv \eta^{+-}_{S,i}(qs\bar q \bar s)$ and $\eta^{+-}_{A,i,1} \equiv \eta^{+-}_{A,i}(qs\bar q \bar s)$ to perform QCD sum rule calculations. The masses are calculated using Eq.~(\ref{eq:sumrule}), and the results are shown in Figs.~\ref{fig:PM} as functions of the threshold value $s_0$ and the Borel Mass $M_B^2$. The upper four figures are obtained using the tetraquark currents $\eta^{+-}_{S,i}(qq\bar q \bar q)$, the middle four figures are obtained using the tetraquark currents $\eta^{+-}_{S,i}(qs\bar q \bar s)$, and the lower four figures are obtained using the tetraquark currents $\eta^{+-}_{A,i}(qs\bar q \bar s)$.

The masses obtained using all these six currents have a minimum around 1.5-1.6 GeV against the threshold value $s_0$. Consequently, we fix the working regions to be $3$ GeV$^2<M_B^2<4$ GeV$^2$ and 2.6 GeV$^2 <s_0<$ 3.0 GeV$^2$ where the stability is good. Using these currents, we obtain the similar masses 1.55-1.60 GeV, 1.50-1.57 GeV, 1.61-1.66 GeV, 1.49-1.56 GeV, 1.59-1.64 GeV, and 1.47-1.55 GeV. Therefore, our results suggest that there is a missing $b_1$ state having $I^G J^{PC} = 1^+1^{+-}$ and a mass around 1.47-1.66 GeV.

We also use them to predict decay patterns of this missing $b_1$ state. These diquark-antidiquark currents can be transformed into the meson-meson currents:
\begin{eqnarray}
\eta^{+-}_{S,1} \& \eta^{+-}_{A,1} &\rightarrow& (\bar q q) (\bar q \gamma_\mu \gamma_5 q) \& (\bar q \gamma^\nu q) (\bar q \sigma_{\mu\nu} \gamma_5 q)\, ,
\\ \nonumber \eta^{+-}_{S,2} \& \eta^{+-}_{A,2} &\rightarrow& (\bar q \gamma_\mu q) (\bar q \gamma_5 q) \& (\bar q \gamma^\nu \gamma_5 q) (\bar q \sigma_{\mu\nu} q)\, ,
\end{eqnarray}
and so the possible $S$-wave decay patterns are
\begin{eqnarray}
b_1(1600) &\rightarrow& 0^+ (\sigma_0(500), \kappa(800), f_0(980) \cdots) + 1^+(a_1(1260),b_1(1235),K_1(1270) \cdots) \, ,
\\ \nonumber b_1(1600) &\rightarrow& 1^- (\rho(770), \omega(782), K^\star(892) \cdots) + 1^-(\rho(770),\omega(782),K^\star(892) \cdots) \, ,
\\ \nonumber b_1(1600) &\rightarrow& 1^- (\rho(770), \omega(782), K^\star(892) \cdots) + 0^-(\pi, \eta, \eta^\prime, K \cdots) \, ,
\\ \nonumber b_1(1600) &\rightarrow& 1^+ (a_1(1260),b_1(1235),K_1(1270) \cdots) + 1^+(a_1(1260),b_1(1235),K_1(1270) \cdots) \, ,
\end{eqnarray}
and the possible $P$-wave decay patterns are
\begin{eqnarray}
b_1(1600) &\rightarrow& 0^+ (\sigma_0(500), \kappa(800), f_0(980) \cdots) + 0^-(\pi, \eta, \eta^\prime, K \cdots) \, ,
\\ \nonumber b_1(1600) &\rightarrow& 0^-(\pi, \eta, \eta^\prime, K \cdots) + 1^+(a_1(1260),b_1(1235),K_1(1270) \cdots) \, .
\end{eqnarray}

%
\section{Finite Energy Sum Rules}\label{sec:fesr}
%

In this section, we use the method of finite energy sum rules (FESR). Compared with the method of SVZ sum rules, it does not use the Borel transformation, and so its convergence is slower. In order to calculate the mass in the FESR, we first define the $n$th moment by using the spectral function $\rho(s)$ obtained in Eq.~(\ref{eq:rho}):
%
\begin{equation}
W(n, s_0) = \int^{s_0}_0 \rho(s) s^n ds \, . \label{eq:moment}
\end{equation}
%
This integral is used for the phenomenological side, while the
integral along the circular contour of radius $s_0$ on the $q^2$
complex plain should be performed for the theoretical side.

With the assumption of quark-hadron duality, we obtain
%
\begin{equation}
W(n, s_0)\Big |_{\rm Hadron} = W(n, s_0)\Big |_{\rm OPE} \, .
\end{equation}
%
The mass of the ground state can be obtained as
%
\begin{equation}
M^2_Y(n, s_0)= { W(n+1, s_0) \over W(n, s_0) } \, . \label{eq:FESR}
\end{equation}
%
The spectral function $\rho(s)$ can be drawn from the Borel
transformed correlation functions shown in Sec.~\ref{sec:sumrule}.
We shall omit the $D=12$ terms which are proportional to $1 / (q^2)^2$. They do not
contribute to the function $W(n,s_0)$ of Eq.~(\ref{eq:moment}) for
$n=0$, and they have a very small contribution for $n=1$.

\begin{figure}[h!t]
\begin{center}
\scalebox{1.2}{\includegraphics{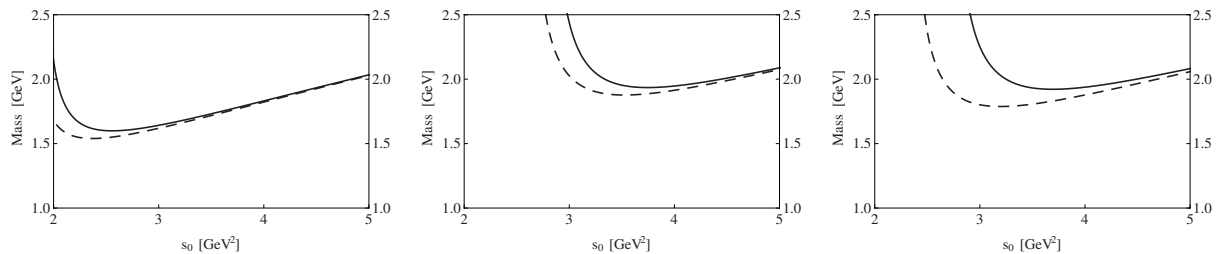}}
\caption{The mass calculated using the method of finite energy sum rules. The left, middle and right panels are for the masses obtained using the
currents $\eta^{--}_{S,i}(qq\bar q \bar q)$, $\eta^{--}_{S,i}(qs\bar q \bar s)$ and $\eta^{--}_{A,i}(qs\bar q \bar s)$, respectively. The solid and dashed curves are for $i=1$ and $i=2$, respectively.} \label{fig:fesrNN}
\end{center}
\end{figure}

The masses obtained using tetraquark currents of $I^G J^{PC} = 1^+1^{--}$ are shown in Fig.~\ref{fig:fesrNN} as functions of the threshold value $s_0$, where $n$ is chosen to be 1. We find that there is a mass minimum for all curves where the stability is the best. This minimum is around 1.6 GeV for the tetraquark currents $\eta^{--}_{S,i}(qq\bar q \bar q)$, whose quark contents are $qq \bar q \bar q$. Consequently, we fix the working region to be 2.6 GeV$^2 <s_0<$ 3.0 GeV$^2$ where the stability is good, and obtain the similar masses 1.60-1.64 GeV and 1.56-1.62 GeV. This minimum is around 2.0 GeV for the currents $\eta^{--}_{S,i}(qs\bar q \bar s)$ and $\eta^{--}_{A,i}(qs\bar q \bar s)$, whose quark contents are $q s \bar q \bar s$. Consequently, we fix the working region to be 4.0 GeV$^2 <s_0<$ 4.5 GeV$^2$ where the stability is good, and obtain the similar masses 1.95-2.01 GeV, 1.91-1.99 GeV, 1.94-2.00 GeV and 1.88-1.97 GeV. We note that the working regions are set to be the same as those used in SVZ sum rules in the previous section, and the results are also quite similar.

%
%

\begin{figure}[h!t]
\begin{center}
\scalebox{1.2}{\includegraphics{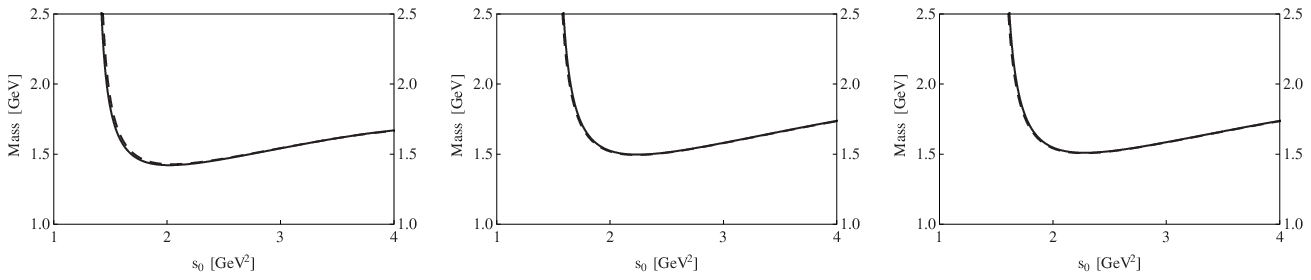}}
\caption{The mass calculated using the method of finite energy sum rules. The left, middle and right panels are for the masses of the
currents $\eta^{++}_{S,i}(qq\bar q \bar q)$, $\eta^{++}_{S,i}(qs\bar q \bar s)$ and $\eta^{++}_{A,i}(qs\bar q \bar s)$, respectively. The solid and dashed curves are for $i=1$ and $i=2$, respectively.} \label{fig:fesrPP}
\end{center}
\end{figure}

The masses obtained using tetraquark currents of $I^G J^{PC} = 1^-1^{++}$ are shown in Fig.~\ref{fig:fesrPP} as functions of the threshold value $s_0$. We find that there is a mass minimum around 1.5 GeV for all curves where the stability is the best. Consequently, we fix the working region to be 2.6 GeV$^2 <s_0<$ 3.0 GeV$^2$ where the stability is good. Using these currents, we obtain the similar masses 1.48-1.54 GeV, 1.48-1.54 GeV, 1.52-1.58 GeV, 1.52-1.58 GeV, 1.53-1.59 GeV and 1.53-1.58 GeV. Again, we have set the same working regions as those used in SVZ sum rules, and the results are also similar.
%
%

\begin{figure}[h!t]
\begin{center}
\scalebox{1.2}{\includegraphics{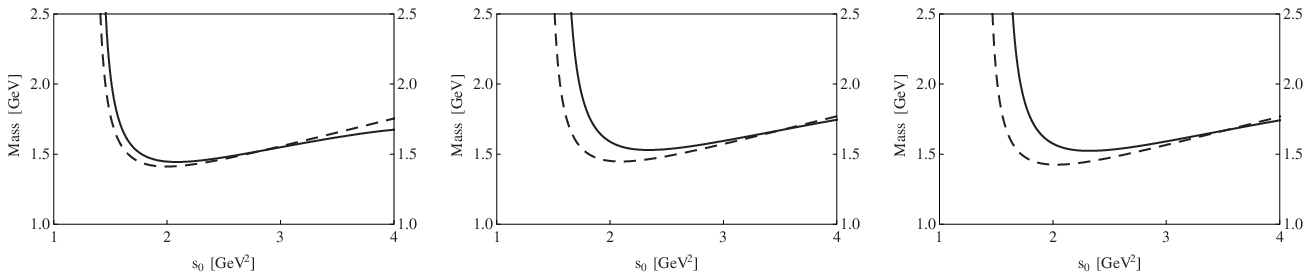}}
\caption{The mass calculated using the method of finite energy sum rules. The left, middle and right panels are for the masses of the
currents $\eta^{+-}_{S,i}(qq\bar q \bar q)$, $\eta^{+-}_{S,i}(qs\bar q \bar s)$ and $\eta^{+-}_{A,i}(qs\bar q \bar s)$, respectively. The solid and dashed curves are for $i=1$ and $i=2$, respectively.} \label{fig:fesrPN}
\end{center}
\end{figure}

The masses obtained using tetraquark currents of $I^G J^{PC} = 1^+1^{+-}$ are shown in Fig.~\ref{fig:fesrPN} as functions of the threshold value $s_0$. We find that there is a mass minimum around 1.5 GeV for all curves where the stability is the best. Consequently, we fix the working region to be 2.6 GeV$^2 <s_0<$ 3.0 GeV$^2$ where the stability is good. Using these currents, we obtain the similar masses 1.49-1.55 GeV, 1.48-1.56 GeV, 1.54-1.59 GeV, 1.50-1.57 GeV, 1.54-1.59 GeV and 1.49-1.57 GeV. Again, we have set the same working regions as those used in SVZ sum rules, and the results are also similar.
%
%

In a short summary, we arrive at the results similar to those obtained using SVZ sum rules in the previous section.

\section{Summary}
\label{sec:summary}

We have systematically investigated the chiral structure of local vector and axial-vector tetraquark currents, and studied their chiral transformation properties. We have classified all the possible local tetraquark currents constructed using diquarks and antidiquarks, and found that there is a one to one correspondence between these vector and axial-vector tetraquark currents. Then we used the left-handed quark field $L_A^a$ and the right-handed quark field $R_A^a$ to rewrite these currents, from which we can clearly see their chiral structure. After making proper combinations, we have obtained their chiral representations and chirality. We have also studied their chiral transformation properties, from which we found that vector and axial-vector tetraquark currents are closely related.

We have considered the charge-conjugation parity and classified all the local isovector vector and axial-vector tetraquark currents of quantum numbers $I^GJ^{PC} = 1^-1^{-+}$, $I^GJ^{PC} = 1^+1^{--}$, $I^GJ^{PC} = 1^-1^{++}$ and $I^GJ^{PC} = 1^+1^{+-}$. Again, we found that there is a one to one correspondence among them. However, experiments in the energy region around 1.6 GeV only observed mesons of quantum numbers $I^GJ^{PC}=1^-1^{-+}$, $I^GJ^{PC}=1^+1^{--}$ and $I^GJ^{PC}=1^-1^{++}$, but did not observe mesons of quantum numbers $I^GJ^{PC}=1^+1^{+-}$. Accordingly, we proposed that there might be a missing state having $I^GJ^{PC}=1^+1^{+-}$ in this energy region.

To verify this, we have performed SVZ sum rule analyses and finite energy sum rule analyses. The tetraquark currents $\eta^{-+}_{S,i,1} \equiv \eta^{-+}_{S,i}(qq\bar q \bar q)$ have been studied in Ref.~\cite{Chen:2008qw}. They lead to masses around 1.6 GeV, and so they can couple to the exotic meson $\pi_1(1600)$. The tetraquark currents $\eta^{--}_{S,i,1} \equiv \eta^{--}_{S,i}(qq\bar q \bar q)$ lead to masses about 1.56-1.73 GeV, and so they can couple to $\rho(1570)$ or $\rho(1700)$. The tetraquark currents $\eta^{++}_{A/S,i,j}$  all lead to masses about 1.48-1.63 GeV, and so they can couple to $a_1(1640)$. The tetraquark currents $\eta^{+-}_{A/S,i,j}$ all lead to masses around 1.47-1.66 GeV, which suggests that there is a missing $b_1$ state having $I^GJ^{PC}=1^+1^{+-}$ and a mass around 1.47-1.66 GeV.


Finally we would like to note that although we have only used tetraquark currents, we are not suggesting this missing $b_1$ state is a tetraquark state. Our results only suggest that it contains some tetraquark (multi-quark) components, because the tetraquark currents $\eta^{+-}_{A/S,i,j}$ can couple to it. We would also like to note that other currents representing the $\bar q q$ structure, the hybrid structure and the meson-meson structure can all contribute here (see Res.~\cite{Sugiyama:2007sg,Nakamura:2008zzc,Matheus:2009vq,Wang:2009wk,Nielsen:2010ij,Chen:2013pya} where their mixing is studied for the cases of light scalar mesons, $\Lambda(1405)$, $X(3872)$ and $X(4350)$).


\section*{Acknowledgments}

This work is supported by the National Natural Science Foundation of China under Grant No. 11205011, and the Fundamental Research Funds for the Central Universities.

\appendix

\section{Other Vector and Axial-Vector Tetraquark Currents}
\label{app:othercurrents}

\subsection{Tetraquark currents of flavor singlet and $J^P = 1^+$}
\label{subsec:singletaxialvector}

In this subsection we study flavor singlet tetraquark currents of $J^P = 1^+$. There are altogether eight independent axial-vector currents as listed in the following:
\begin{eqnarray}
\nonumber \eta_1^{\rm AV, \mathbb{S}} &=& q_A^{aT} \mathbb{C} q_B^b (\bar{q}_A^a \gamma_\mu \gamma_5 \mathbb{C} \bar{q}_B^{bT} - \bar{q}_A^b \gamma_\mu \gamma_5 \mathbb{C} \bar{q}_B^{aT}) \, ,
\\ \nonumber \eta_2^{\rm AV, \mathbb{S}} &=& q_A^{aT} \mathbb{C} \gamma_\mu \gamma_5 q_B^b (\bar{q}_A^a \mathbb{C} \bar{q}_B^{bT} - \bar{q}_A^b \mathbb{C} \bar{q}_B^{aT}) \, ,
\\ \nonumber \eta_3^{\rm AV, \mathbb{S}} &=& q_A^{aT} \mathbb{C} \gamma^\nu q_B^b (\bar{q}_A^a \sigma_{\mu\nu} \gamma_5 \mathbb{C} \bar{q}_B^{bT} + \bar{q}_A^b \sigma_{\mu\nu} \gamma_5 \mathbb{C}
\bar{q}_B^{aT}) \, ,
\\ \eta_4^{\rm AV, \mathbb{S}} &=& q_A^{aT} \mathbb{C} \sigma_{\mu\nu} \gamma_5 q_B^b (\bar{q}_A^a \gamma^\nu \mathbb{C} \bar{q}_B^{bT} + \bar{q}_A^b \gamma^\nu \mathbb{C} \bar{q}_B^{aT}) \, ,
\label{eq:singletaxialvector}
\\ \nonumber \eta_5^{\rm AV, \mathbb{S}} &=& q_A^{aT} \mathbb{C} q_B^b (\bar{q}_A^a \gamma_\mu \gamma_5 \mathbb{C} \bar{q}_B^{bT} + \bar{q}_A^b \gamma_\mu \gamma_5 \mathbb{C} \bar{q}_B^{aT}) \, ,
\\ \nonumber \eta_6^{\rm AV, \mathbb{S}} &=& q_A^{aT} \mathbb{C} \gamma_\mu \gamma_5 q_B^b (\bar{q}_A^a \mathbb{C} \bar{q}_B^{bT} + \bar{q}_A^b \mathbb{C} \bar{q}_B^{aT}) \, ,
\\ \nonumber \eta_7^{\rm AV, \mathbb{S}} &=& q_A^{aT} \mathbb{C} \gamma^\nu q_B^b (\bar{q}_A^a \sigma_{\mu\nu} \gamma_5 \mathbb{C} \bar{q}_B^{bT} - \bar{q}_A^b \sigma_{\mu\nu} \gamma_5 \mathbb{C}
\bar{q}_B^{aT}) \, ,
\\ \nonumber \eta_8^{\rm AV, \mathbb{S}} &=& q_A^{aT} \mathbb{C} \sigma_{\mu\nu} \gamma_5 q_B^b (\bar{q}_A^a \gamma^\nu \mathbb{C} \bar{q}_B^{bT} - \bar{q}_A^b \gamma^\nu \mathbb{C} \bar{q}_B^{aT}) \, .
\end{eqnarray}
The former four currents contain diquarks and antidiquarks having the antisymmetric flavor structure $\mathbf{\bar 3} \otimes \mathbf{3}$ and the latter four currents contain diquarks and antidiquarks having the symmetric flavor structure $\mathbf{6} \otimes \mathbf{\bar6}$. From the following combinations we can clearly see their chiral structure:
\begin{eqnarray}
\nonumber \eta_1^{\rm AV, \mathbb{S}} &=& - 2 L_A^{aT} \mathbb{C} L_B^b (\bar{L}_A^a \gamma_\mu \mathbb{C} \bar{R}_B^{bT} - \bar{L}_A^b \gamma_\mu \mathbb{C} \bar{R}_B^{aT})
+ 2 R_A^{aT} \mathbb{C} R_B^b (\bar{R}_A^a \gamma_\mu \mathbb{C} \bar{L}_B^{bT} - \bar{R}_A^b \gamma_\mu \mathbb{C} \bar{L}_B^{aT}) \, ,
\\ \nonumber \eta_2^{\rm AV, \mathbb{S}} &=& 2 L_A^{aT} \mathbb{C} \gamma_\mu R_B^b (\bar{L}_A^a \mathbb{C} \bar{L}_B^{bT} - \bar{L}_A^b \mathbb{C} \bar{L}_B^{aT})
- 2 R_A^{aT} \mathbb{C} \gamma_\mu L_B^b (\bar{R}_A^a \mathbb{C} \bar{R}_B^{bT} - \bar{R}_A^b \mathbb{C} \bar{R}_B^{aT}) \, ,
\\ \nonumber \eta_3^{\rm AV, \mathbb{S}} &=& 2 L_A^{aT} \mathbb{C} \gamma^\nu R_B^b (\bar{L}_A^a \sigma_{\mu\nu} \mathbb{C} \bar{L}_B^{bT} + \bar{L}_A^b \sigma_{\mu\nu} \mathbb{C} \bar{L}_B^{aT})
- 2 R_A^{aT} \mathbb{C} \gamma^\nu L_B^b (\bar{R}_A^a \sigma_{\mu\nu} \mathbb{C} \bar{R}_B^{bT} + \bar{R}_A^b \sigma_{\mu\nu} \mathbb{C} \bar{R}_B^{aT}) \, ,
\\ \eta_4^{\rm AV, \mathbb{S}} &=& - 2 L_A^{aT} \mathbb{C} \sigma_{\mu\nu} L_B^b (\bar{L}_A^a \gamma^\nu \mathbb{C} \bar{R}_B^{bT} + \bar{L}_A^b \gamma^\nu \mathbb{C} \bar{R}_B^{aT})
+ 2 R_A^{aT} \mathbb{C} \sigma_{\mu\nu} R_B^b (\bar{R}_A^a \gamma^\nu \mathbb{C} \bar{L}_B^{bT} + \bar{R}_A^b \gamma^\nu \mathbb{C} \bar{L}_B^{aT}) \, ,
\\ \nonumber \eta_5^{\rm AV, \mathbb{S}} &=& - 2 L_A^{aT} \mathbb{C} L_B^b (\bar{L}_A^a \gamma_\mu \mathbb{C} \bar{R}_B^{bT} + \bar{L}_A^b \gamma_\mu \mathbb{C} \bar{R}_B^{aT})
\bar{R}_B^{aT})
+ 2 R_A^{aT} \mathbb{C} R_B^b (\bar{R}_A^a \gamma_\mu \mathbb{C} \bar{L}_B^{bT} + \bar{R}_A^b \gamma_\mu \mathbb{C} \bar{L}_B^{aT}) \, ,
\\ \nonumber \eta_6^{\rm AV, \mathbb{S}} &=& 2 L_A^{aT} \mathbb{C} \gamma_\mu R_B^b (\bar{L}_A^a \mathbb{C} \bar{L}_B^{bT} + \bar{L}_A^b \mathbb{C} \bar{L}_B^{aT})
- 2 R_A^{aT} \mathbb{C} \gamma_\mu L_B^b (\bar{R}_A^a \mathbb{C} \bar{R}_B^{bT} + \bar{R}_A^b \mathbb{C} \bar{R}_B^{aT}) \, ,
\\ \nonumber \eta_7^{\rm AV, \mathbb{S}} &=& 2 L_A^{aT} \mathbb{C} \gamma^\nu R_B^b (\bar{L}_A^a \sigma_{\mu\nu} \mathbb{C} \bar{L}_B^{bT} - \bar{L}_A^b \sigma_{\mu\nu} \mathbb{C} \bar{L}_B^{aT})
- 2 R_A^{aT} \mathbb{C} \gamma^\nu L_B^b (\bar{R}_A^a \sigma_{\mu\nu} \mathbb{C} \bar{R}_B^{bT} - \bar{R}_A^b \sigma_{\mu\nu} \mathbb{C} \bar{R}_B^{aT}) \, ,
\\ \nonumber \eta_8^{\rm AV, \mathbb{S}} &=& - 2 L_A^{aT} \mathbb{C} \sigma_{\mu\nu} L_B^b (\bar{L}_A^a \gamma^\nu \mathbb{C} \bar{R}_B^{bT} - \bar{L}_A^b \gamma^\nu \mathbb{C} \bar{R}_B^{aT})
+ 2 R_A^{aT} \mathbb{C} \sigma_{\mu\nu} R_B^b (\bar{R}_A^a \gamma^\nu \mathbb{C} \bar{L}_B^{bT} - \bar{R}_A^b \gamma^\nu \mathbb{C} \bar{L}_B^{aT}) \, .
\end{eqnarray}
Among these currents, $\eta_{1,4,5,8}^{\rm AV, \mathbb{S}}$ belong to the chiral representation $[({\mathbf 3}, \bar {\mathbf 3}) + (\bar {\mathbf 3}, {\mathbf 3})]$ and their chirality is $L L \bar L \bar R + R R \bar R \bar L$; $\eta_{2,3,6,7}^{\rm AV, \mathbb{S}}$ belong to the mirror one $[(\bar {\mathbf 3}, {\mathbf 3}) + ({\mathbf 3}, \bar {\mathbf 3})]$ and their chirality $L R \bar L \bar L + R L \bar R \bar R$. Again in this case we do not find any ``exotic'' chiral structure.

\subsection{Tetraquark currents of flavor octet and $J^P = 1^-$}
\label{subsec:8vector}

In this subsection we study flavor octet tetraquark currents of $J^P = 1^-$. There are altogether sixteen independent vector currents as listed in the following:
\begin{eqnarray}
\nonumber \eta_{1,N}^{\rm V, \mathbb{O}} &=& \lambda^N_{BD} ( q_A^{aT} \mathbb{C} \gamma_5 q_B^b ) (\bar{q}_A^a \gamma_\mu \gamma_5 \mathbb{C} \bar{q}_D^{bT} - \bar{q}_A^b \gamma_\mu \gamma_5 \mathbb{C} \bar{q}_D^{aT}) \, ,
\\ \nonumber \eta_{2,N}^{\rm V, \mathbb{O}} &=& \lambda^N_{BD} ( q_A^{aT} \mathbb{C} \gamma_\mu \gamma_5 q_B^b ) (\bar{q}_A^a \gamma_5 \mathbb{C} \bar{q}_D^{bT} - \bar{q}_A^b \gamma_5 \mathbb{C} \bar{q}_D^{aT}) \, ,
\\ \nonumber \eta_{3,N}^{\rm V, \mathbb{O}} &=& \lambda^N_{BD} ( q_A^{aT} \mathbb{C} \gamma^\nu q_B^b ) (\bar{q}_A^a \sigma_{\mu\nu} \mathbb{C} \bar{q}_D^{bT} + \bar{q}_A^b \sigma_{\mu\nu} \mathbb{C} \bar{q}_D^{aT}) \, ,
\\ \nonumber \eta_{4,N}^{\rm V, \mathbb{O}} &=& \lambda^N_{BD} ( q_A^{aT} \mathbb{C} \sigma_{\mu\nu} q_B^b ) (\bar{q}_A^a \gamma^\nu \mathbb{C} \bar{q}_D^{bT} + \bar{q}_A^b \gamma^\nu \mathbb{C} \bar{q}_D^{aT}) \, ,
\\ \nonumber \eta_{5,N}^{\rm V, \mathbb{O}} &=& \lambda^N_{BD} ( q_A^{aT} \mathbb{C} \gamma_5 q_B^b ) (\bar{q}_A^a \gamma_\mu \gamma_5 \mathbb{C} \bar{q}_D^{bT} + \bar{q}_A^b \gamma_\mu \gamma_5 \mathbb{C} \bar{q}_D^{aT}) \, ,
\\ \nonumber \eta_{6,N}^{\rm V, \mathbb{O}} &=& \lambda^N_{BD} ( q_A^{aT} \mathbb{C} \gamma_\mu \gamma_5 q_B^b ) (\bar{q}_A^a \gamma_5 \mathbb{C} \bar{q}_D^{bT} + \bar{q}_A^b \gamma_5 \mathbb{C} \bar{q}_D^{aT}) \, ,
\\ \nonumber \eta_{7,N}^{\rm V, \mathbb{O}} &=& \lambda^N_{BD} ( q_A^{aT} \mathbb{C} \gamma^\nu q_B^b ) (\bar{q}_A^a \sigma_{\mu\nu} \mathbb{C} \bar{q}_D^{bT} - \bar{q}_A^b \sigma_{\mu\nu} \mathbb{C} \bar{q}_D^{aT}) \, ,
\\ \eta_{8,N}^{\rm V, \mathbb{O}} &=& \lambda^N_{BD} ( q_A^{aT} \mathbb{C} \sigma_{\mu\nu} q_B^b ) (\bar{q}_A^a \gamma^\nu \mathbb{C} \bar{q}_D^{bT} - \bar{q}_A^b \gamma^\nu \mathbb{C} \bar{q}_D^{aT}) \, ,
\label{eq:8vector}
\\ \nonumber \eta_{9,N}^{\rm V, \mathbb{O}} &=& \lambda^N_{BD} ( q_A^{aT} \mathbb{C} q_B^b ) (\bar{q}_A^a \gamma_\mu \mathbb{C} \bar{q}_D^{bT} - \bar{q}_A^b \gamma_\mu \mathbb{C} \bar{q}_D^{aT} ) \, ,
\\ \nonumber \eta_{10,N}^{\rm V, \mathbb{O}} &=& \lambda^N_{BD} ( q_A^{aT} \mathbb{C} \gamma_\mu q_B^b ) (\bar{q}_A^a \mathbb{C} \bar{q}_D^{bT} + \bar{q}_A^b \mathbb{C} \bar{q}_D^{aT} ) \, ,
\\ \nonumber \eta_{11,N}^{\rm V, \mathbb{O}} &=& \lambda^N_{BD} ( q_A^{aT} \mathbb{C} \gamma_\nu \gamma_5 q_B^b ) (\bar{q}_A^a \sigma_{\mu\nu} \gamma_5 \mathbb{C} \bar{q}_D^{bT} - \bar{q}_A^b \sigma_{\mu\nu} \gamma_5 \mathbb{C} \bar{q}_D^{aT} ) \, ,
\\ \nonumber \eta_{12,N}^{\rm V, \mathbb{O}} &=& \lambda^N_{BD} ( q_A^{aT} \mathbb{C} \sigma_{\mu\nu} \gamma_5 q_B^b ) (\bar{q}_A^a \gamma_\nu \gamma_5 \mathbb{C} \bar{q}_D^{bT} + \bar{q}_A^b \gamma_\nu \gamma_5 \mathbb{C} \bar{q}_D^{aT} ) \, ,
\\ \nonumber \eta_{13,N}^{\rm V, \mathbb{O}} &=& \lambda^N_{BD} ( q_A^{aT} \mathbb{C} q_B^b ) (\bar{q}_A^a \gamma_\mu \mathbb{C} \bar{q}_D^{bT} + \bar{q}_A^b \gamma_\mu \mathbb{C} \bar{q}_D^{aT} ) \, ,
\\ \nonumber \eta_{14,N}^{\rm V, \mathbb{O}} &=& \lambda^N_{BD} ( q_A^{aT} \mathbb{C} \gamma_\mu q_B^b ) (\bar{q}_A^a \mathbb{C} \bar{q}_D^{bT} - \bar{q}_A^b \mathbb{C} \bar{q}_D^{aT} ) \, ,
\\ \nonumber \eta_{15,N}^{\rm V, \mathbb{O}} &=& \lambda^N_{BD} ( q_A^{aT} \mathbb{C} \gamma_\nu \gamma_5 q_B^b ) (\bar{q}_A^a \sigma_{\mu\nu} \gamma_5 \mathbb{C} \bar{q}_D^{bT} + \bar{q}_A^b \sigma_{\mu\nu} \gamma_5 \mathbb{C} \bar{q}_D^{aT} ) \, ,
\\ \nonumber \eta_{16,N}^{\rm V, \mathbb{O}} &=& \lambda^N_{BD} ( q_A^{aT} \mathbb{C} \sigma_{\mu\nu} \gamma_5 q_B^b ) (\bar{q}_A^a \gamma_\nu \gamma_5 \mathbb{C} \bar{q}_D^{bT} - \bar{q}_A^b \gamma_\nu \gamma_5 \mathbb{C} \bar{q}_D^{aT} ) \, .
\end{eqnarray}
Among these currents, $\eta_{1,2,3,4,N}^{\rm V, \mathbb{O}}$ contain diquarks and antidiquarks having both the antisymmetric flavor structure $\mathbf{\bar 3} \otimes \mathbf{3}$; $\eta_{5,6,7,8,N}^{\rm V, \mathbb{O}}$ contain diquarks and antidiquarks having both the symmetric flavor structure $\mathbf{6} \otimes \mathbf{\bar6}$; $\eta_{9,10,11,12,N}^{\rm V, \mathbb{O}}$ contain diquarks having the antisymmetric flavor structure and antidiquarks the symmetric flavor structure $\mathbf{\bar3} \otimes \mathbf{\bar6}$; $\eta_{13,14,15,16,N}^{\rm V, \mathbb{O}}$ contain diquarks having the symmetric flavor structure and antidiquarks the antisymmetric flavor structure $\mathbf{6} \otimes \mathbf{3}$. From the following combinations we can clearly see their chiral structure:
\begin{eqnarray}
\nonumber \eta_{1,N}^{\rm V, \mathbb{O}} - \eta_{9,N}^{\rm V, \mathbb{O}} &=& - 2 \lambda_N^{DB} L_A^{aT} \mathbb{C} L_B^b (\bar{R}_A^a \gamma_\mu \mathbb{C} \bar{L}_D^{bT} - \bar{R}_A^b \gamma_\mu \mathbb{C} \bar{L}_D^{aT})
- 2 \lambda_N^{DB} R_A^{aT} \mathbb{C} R_B^b (\bar{L}_A^a \gamma_\mu \mathbb{C} \bar{R}_D^{bT} - \bar{L}_A^b \gamma_\mu \mathbb{C} \bar{R}_D^{aT}) \, ,
\\ \nonumber \eta_{1,N}^{\rm V, \mathbb{O}} + \eta_{9,N}^{\rm V, \mathbb{O}} &=& 2 \lambda_N^{DB} L_A^{aT} \mathbb{C} L_B^b (\bar{L}_A^a \gamma_\mu \mathbb{C} \bar{R}_D^{bT} - \bar{L}_A^b \gamma_\mu \mathbb{C} \bar{R}_D^{aT})
+ 2 \lambda_N^{DB} R_A^{aT} \mathbb{C} R_B^b (\bar{R}_A^a \gamma_\mu \mathbb{C} \bar{L}_D^{bT} - \bar{R}_A^b \gamma_\mu \mathbb{C} \bar{L}_D^{aT}) \, ,
\\ \nonumber \eta_{2,N}^{\rm V, \mathbb{O}} - \eta_{14,N}^{\rm V, \mathbb{O}} &=& - 2 \lambda_N^{DB} L_A^{aT} \mathbb{C} \gamma_\mu R_B^b (\bar{R}_A^a \mathbb{C} \bar{R}_D^{bT} - \bar{R}_A^b \mathbb{C} \bar{R}_D^{aT})
- 2 \lambda_N^{DB} R_A^{aT} \mathbb{C} \gamma_\mu L_B^b (\bar{L}_A^a \mathbb{C} \bar{L}_D^{bT} - \bar{L}_A^b \mathbb{C} \bar{L}_D^{aT})\, ,
\\ \nonumber \eta_{2,N}^{\rm V, \mathbb{O}} + \eta_{14,N}^{\rm V, \mathbb{O}} &=& 2 \lambda_N^{DB} L_A^{aT} \mathbb{C} \gamma_\mu R_B^b (\bar{L}_A^a \mathbb{C} \bar{L}_D^{bT} - \bar{L}_A^b \mathbb{C} \bar{L}_D^{aT})
+ 2 \lambda_N^{DB} R_A^{aT} \mathbb{C} \gamma_\mu L_B^b (\bar{R}_A^a \mathbb{C} \bar{R}_D^{bT} - \bar{R}_A^b \mathbb{C} \bar{R}_D^{aT}) \, ,
\\ \nonumber \eta_{3,N}^{\rm V, \mathbb{O}} - \eta_{15,N}^{\rm V, \mathbb{O}} &=& 2 \lambda_N^{DB} L_A^{aT} \mathbb{C} \gamma^\nu R_B^b (\bar{R}_A^a \sigma_{\mu\nu} \mathbb{C} \bar{R}_D^{bT} + \bar{R}_A^b \sigma_{\mu\nu} \mathbb{C} \bar{R}_D^{aT})
+ 2 \lambda_N^{DB} R_A^{aT} \mathbb{C} \gamma^\nu L_B^b (\bar{L}_A^a \sigma_{\mu\nu} \mathbb{C} \bar{L}_D^{bT} + \bar{L}_A^b \sigma_{\mu\nu} \mathbb{C} \bar{L}_D^{aT}) \, ,
\\ \nonumber \eta_{3,N}^{\rm V, \mathbb{O}} + \eta_{15,N}^{\rm V, \mathbb{O}} &=& 2 \lambda_N^{DB} L_A^{aT} \mathbb{C} \gamma^\nu R_B^b (\bar{L}_A^a \sigma_{\mu\nu} \mathbb{C} \bar{L}_D^{bT} + \bar{L}_A^b \sigma_{\mu\nu} \mathbb{C} \bar{L}_D^{aT})
+ 2 \lambda_N^{DB} R_A^{aT} \mathbb{C} \gamma^\nu L_B^b (\bar{R}_A^a \sigma_{\mu\nu} \mathbb{C} \bar{R}_D^{bT} + \bar{R}_A^b \sigma_{\mu\nu} \mathbb{C} \bar{R}_D^{aT}) \, ,
\\ \nonumber \eta_{4,N}^{\rm V, \mathbb{O}} - \eta_{12,N}^{\rm V, \mathbb{O}} &=& 2 \lambda_N^{DB} L_A^{aT} \mathbb{C} \sigma_{\mu\nu} L_B^b (\bar{R}_A^a \gamma^\nu \mathbb{C} \bar{L}_D^{bT} + \bar{R}_A^b \gamma^\nu \mathbb{C} \bar{L}_D^{aT})
+ 2 \lambda_N^{DB} R_A^{aT} \mathbb{C} \sigma_{\mu\nu} R_B^b (\bar{L}_A^a \gamma^\nu \mathbb{C} \bar{R}_D^{bT} + \bar{L}_A^b \gamma^\nu \mathbb{C} \bar{R}_D^{aT}) \, ,
\\ \nonumber \eta_{4,N}^{\rm V, \mathbb{O}} + \eta_{12,N}^{\rm V, \mathbb{O}} &=& 2 \lambda_N^{DB} L_A^{aT} \mathbb{C} \sigma_{\mu\nu} L_B^b (\bar{L}_A^a \gamma^\nu \mathbb{C} \bar{R}_D^{bT} + \bar{L}_A^b \gamma^\nu \mathbb{C} \bar{R}_D^{aT})
+ 2 \lambda_N^{DB} R_A^{aT} \mathbb{C} \sigma_{\mu\nu} R_B^b (\bar{R}_A^a \gamma^\nu \mathbb{C} \bar{L}_D^{bT} + \bar{R}_A^b \gamma^\nu \mathbb{C} \bar{L}_D^{aT}) \, ,
\\ \eta_{5,N}^{\rm V, \mathbb{O}} - \eta_{13,N}^{\rm V, \mathbb{O}} &=& - 2 \lambda_N^{DB} L_A^{aT} \mathbb{C} L_B^b (\bar{R}_A^a \gamma_\mu \mathbb{C} \bar{L}_D^{bT} + \bar{R}_A^b \gamma_\mu \mathbb{C} \bar{L}_D^{aT})
- 2 \lambda_N^{DB} R_A^{aT} \mathbb{C} R_B^b (\bar{L}_A^a \gamma_\mu \mathbb{C} \bar{R}_D^{bT} + \bar{L}_A^b \gamma_\mu \mathbb{C} \bar{R}_D^{aT}) \, ,
\\ \nonumber \eta_{5,N}^{\rm V, \mathbb{O}} + \eta_{13,N}^{\rm V, \mathbb{O}} &=& 2 \lambda_N^{DB} L_A^{aT} \mathbb{C} L_B^b (\bar{L}_A^a \gamma_\mu \mathbb{C} \bar{R}_D^{bT} + \bar{L}_A^b \gamma_\mu \mathbb{C} \bar{R}_D^{aT})
+ 2 \lambda_N^{DB} R_A^{aT} \mathbb{C} R_B^b (\bar{R}_A^a \gamma_\mu \mathbb{C} \bar{L}_D^{bT} + \bar{R}_A^b \gamma_\mu \mathbb{C} \bar{L}_D^{aT}) \, ,
\\ \nonumber \eta_{6,N}^{\rm V, \mathbb{O}} - \eta_{10,N}^{\rm V, \mathbb{O}} &=& - 2 \lambda_N^{DB} L_A^{aT} \mathbb{C} \gamma_\mu R_B^b (\bar{R}_A^a \mathbb{C} \bar{R}_D^{bT} + \bar{R}_A^b \mathbb{C} \bar{R}_D^{aT})
- 2 \lambda_N^{DB} R_A^{aT} \mathbb{C} \gamma_\mu L_B^b (\bar{L}_A^a \mathbb{C} \bar{L}_D^{bT} + \bar{L}_A^b \mathbb{C} \bar{L}_D^{aT}) \, ,
\\ \nonumber \eta_{6,N}^{\rm V, \mathbb{O}} + \eta_{10,N}^{\rm V, \mathbb{O}} &=& 2 \lambda_N^{DB} L_A^{aT} \mathbb{C} \gamma_\mu R_B^b (\bar{L}_A^a \mathbb{C} \bar{L}_D^{bT} + \bar{L}_A^b \mathbb{C} \bar{L}_D^{aT})
+ 2 \lambda_N^{DB} R_A^{aT} \mathbb{C} \gamma_\mu L_B^b (\bar{R}_A^a \mathbb{C} \bar{R}_D^{bT} + \bar{R}_A^b \mathbb{C} \bar{R}_D^{aT}) \, ,
\\ \nonumber \eta_{7,N}^{\rm V, \mathbb{O}} - \eta_{11,N}^{\rm V, \mathbb{O}} &=& 2 \lambda_N^{DB} L_A^{aT} \mathbb{C} \gamma^\nu R_B^b (\bar{R}_A^a \sigma_{\mu\nu} \mathbb{C} \bar{R}_D^{bT} - \bar{R}_A^b \sigma_{\mu\nu} \mathbb{C} \bar{R}_D^{aT})
+ 2 \lambda_N^{DB} R_A^{aT} \mathbb{C} \gamma^\nu L_B^b (\bar{L}_A^a \sigma_{\mu\nu} \mathbb{C} \bar{L}_D^{bT} - \bar{L}_A^b \sigma_{\mu\nu} \mathbb{C} \bar{L}_D^{aT}) \, ,
\\ \nonumber \eta_{7,N}^{\rm V, \mathbb{O}} + \eta_{11,N}^{\rm V, \mathbb{O}} &=& 2 \lambda_N^{DB} L_A^{aT} \mathbb{C} \gamma^\nu R_B^b (\bar{L}_A^a \sigma_{\mu\nu} \mathbb{C} \bar{L}_D^{bT} - \bar{L}_A^b \sigma_{\mu\nu} \mathbb{C} \bar{L}_D^{aT})
+ 2 \lambda_N^{DB} R_A^{aT} \mathbb{C} \gamma^\nu L_B^b (\bar{R}_A^a \sigma_{\mu\nu} \mathbb{C} \bar{R}_D^{bT} - \bar{R}_A^b \sigma_{\mu\nu} \mathbb{C} \bar{R}_D^{aT}) \, ,
\\ \nonumber \eta_{8,N}^{\rm V, \mathbb{O}} - \eta_{16,N}^{\rm V, \mathbb{O}} &=& 2 \lambda_N^{DB} L_A^{aT} \mathbb{C} \sigma_{\mu\nu} L_B^b (\bar{R}_A^a \gamma^\nu \mathbb{C} \bar{L}_D^{bT} - \bar{R}_A^b \gamma^\nu \mathbb{C} \bar{L}_D^{aT})
+ 2 \lambda_N^{DB} R_A^{aT} \mathbb{C} \sigma_{\mu\nu} R_B^b (\bar{L}_A^a \gamma^\nu \mathbb{C} \bar{R}_D^{bT} - \bar{L}_A^b \gamma^\nu \mathbb{C} \bar{R}_D^{aT}) \, ,
\\ \nonumber \eta_{8,N}^{\rm V, \mathbb{O}} + \eta_{16,N}^{\rm V, \mathbb{O}} &=& 2 \lambda_N^{DB} L_A^{aT} \mathbb{C} \sigma_{\mu\nu} L_B^b (\bar{L}_A^a \gamma^\nu \mathbb{C} \bar{R}_D^{bT} - \bar{L}_A^b \gamma^\nu \mathbb{C} \bar{R}_D^{aT})
+ 2 \lambda_N^{DB} R_A^{aT} \mathbb{C} \sigma_{\mu\nu} R_B^b (\bar{R}_A^a \gamma^\nu \mathbb{C} \bar{L}_D^{bT} - \bar{R}_A^b \gamma^\nu \mathbb{C} \bar{L}_D^{aT}) \, .
\end{eqnarray}
We list their chirality and chiral representations in Table~\ref{tab:8vector}.
\begin{table}[!hbt]
\renewcommand{\arraystretch}{1.3}
\begin{center}
\caption{Flavor octet tetraquark currents of $J^P = 1^-$ classified in subsection~\ref{subsec:8vector}.}
\begin{tabular}{c c c}
\hline\hline
Tetraquark Currents of $(\mathbf 8_F, J^P = 1^-)$ & Chiral Representations & Chirality
\\ \hline \hline
$\eta_{1,N}^{\rm V, \mathbb{O}} - \eta_{9,N}^{\rm V, \mathbb{O}}$, $\eta_{4,N}^{\rm V, \mathbb{O}} - \eta_{12,N}^{\rm V, \mathbb{O}}$ & $[(\bar {\mathbf 6}, \bar {\mathbf 3}) + (\bar {\mathbf 3}, \bar {\mathbf 6})]\oplus[({\mathbf 3}, \bar {\mathbf 3}) + (\bar {\mathbf 3}, {\mathbf 3})]$ & $L L \bar R \bar L + R R \bar L \bar R$
\\ \hline
$\eta_{1,N}^{\rm V, \mathbb{O}} + \eta_{9,N}^{\rm V, \mathbb{O}}$, $\eta_{4,N}^{\rm V, \mathbb{O}} + \eta_{12,N}^{\rm V, \mathbb{O}}$ & $[({\mathbf 3}, \bar {\mathbf 3}) + (\bar {\mathbf 3}, {\mathbf 3})]$ & $L L \bar L \bar R + R R \bar R \bar L$
\\ \hline
$\eta_{2,N}^{\rm V, \mathbb{O}} - \eta_{14,N}^{\rm V, \mathbb{O}}$, $\eta_{3,N}^{\rm V, \mathbb{O}} - \eta_{15,N}^{\rm V, \mathbb{O}}$ & $[({\mathbf 3}, {\mathbf 6}) + ({\mathbf 6}, {\mathbf 3})]\oplus[({\mathbf 3}, \bar {\mathbf 3}) + (\bar {\mathbf 3}, {\mathbf 3})]$ & $L R \bar R \bar R + R L \bar L \bar L$
\\ \hline
$\eta_{2,N}^{\rm V, \mathbb{O}} + \eta_{14,N}^{\rm V, \mathbb{O}}$, $\eta_{3,N}^{\rm V, \mathbb{O}} + \eta_{15,N}^{\rm V, \mathbb{O}}$ & $[(\bar {\mathbf 3}, {\mathbf 3}) + ({\mathbf 3}, \bar {\mathbf 3})]$ & $L R \bar L \bar L + R L \bar R \bar R$
\\ \hline
$\eta_{5,N}^{\rm V, \mathbb{O}} - \eta_{13,N}^{\rm V, \mathbb{O}}$, $\eta_{8,N}^{\rm V, \mathbb{O}} - \eta_{16,N}^{\rm V, \mathbb{O}}$ & $[({\mathbf {15}}, \bar {\mathbf 3}) + (\bar {\mathbf 3}, {\mathbf {15}})]\oplus[({\mathbf 3}, \bar {\mathbf 3}) + (\bar {\mathbf 3}, {\mathbf 3})]$ & $L L \bar R \bar L + R R \bar L \bar R$
\\ \hline
$\eta_{5,N}^{\rm V, \mathbb{O}} + \eta_{13,N}^{\rm V, \mathbb{O}}$, $\eta_{8,N}^{\rm V, \mathbb{O}} + \eta_{16,N}^{\rm V, \mathbb{O}}$ & $[({\mathbf 3}, \bar {\mathbf 3}) + (\bar {\mathbf 3}, {\mathbf 3})]$ & $L L \bar L \bar R + R R \bar R \bar L$
\\ \hline
$\eta_{6,N}^{\rm V, \mathbb{O}} - \eta_{10,N}^{\rm V, \mathbb{O}}$, $\eta_{7,N}^{\rm V, \mathbb{O}} - \eta_{11,N}^{\rm V, \mathbb{O}}$ & $[({\mathbf 3}, \overline{\mathbf {15}}) + (\overline{\mathbf {15}}, {\mathbf 3})]\oplus[({\mathbf 3}, \bar {\mathbf 3}) + (\bar {\mathbf 3}, {\mathbf 3})]$ & $L R \bar R \bar R + R L \bar L \bar L$
\\ \hline
$\eta_{6,N}^{\rm V, \mathbb{O}} + \eta_{10,N}^{\rm V, \mathbb{O}}$, $\eta_{7,N}^{\rm V, \mathbb{O}} + \eta_{11,N}^{\rm V, \mathbb{O}}$ & $[(\bar {\mathbf 3}, {\mathbf 3}) + ({\mathbf 3}, \bar {\mathbf 3})]$ & $L R \bar L \bar L + R L \bar R \bar R$
\\ \hline\hline
\end{tabular}
\label{tab:8vector}
\end{center}
\renewcommand{\arraystretch}{1}
\end{table}

\subsection{Tetraquark currents of flavor octet and $J^P = 1^+$}
\label{subsec:8axialvector}

In this subsection we study flavor octet tetraquark currents of $J^P = 1^+$. There are altogether sixteen independent axial-vector currents as listed in the following:
\begin{eqnarray}
\nonumber \eta_{1,N}^{\rm AV, \mathbb{O}} &=& \lambda^N_{BD} ( q_A^{aT} \mathbb{C} q_B^b ) (\bar{q}_A^a \gamma_\mu \gamma_5 \mathbb{C} \bar{q}_D^{bT} - \bar{q}_A^b \gamma_\mu \gamma_5 \mathbb{C} \bar{q}_D^{aT}) \, ,
\\ \nonumber \eta_{2,N}^{\rm AV, \mathbb{O}} &=& \lambda^N_{BD} ( q_A^{aT} \mathbb{C} \gamma_\mu \gamma_5 q_B^b ) (\bar{q}_A^a \mathbb{C} \bar{q}_D^{bT} - \bar{q}_A^b \mathbb{C} \bar{q}_D^{aT}) \, ,
\\ \nonumber \eta_{3,N}^{\rm AV, \mathbb{O}} &=& \lambda^N_{BD} ( q_A^{aT} \mathbb{C} \gamma^\nu q_B^b ) (\bar{q}_A^a \sigma_{\mu\nu} \gamma_5 \mathbb{C} \bar{q}_D^{bT} + \bar{q}_A^b \sigma_{\mu\nu} \gamma_5 \mathbb{C} \bar{q}_D^{aT}) \, ,
\\ \nonumber \eta_{4,N}^{\rm AV, \mathbb{O}} &=& \lambda^N_{BD} ( q_A^{aT} \mathbb{C} \sigma_{\mu\nu} \gamma_5 q_B^b ) (\bar{q}_A^a \gamma^\nu \mathbb{C} \bar{q}_D^{bT} + \bar{q}_A^b \gamma^\nu \mathbb{C} \bar{q}_D^{aT}) \, ,
\\ \nonumber \eta_{5,N}^{\rm AV, \mathbb{O}} &=& \lambda^N_{BD} ( q_A^{aT} \mathbb{C} q_B^b ) (\bar{q}_A^a \gamma_\mu \gamma_5 \mathbb{C} \bar{q}_D^{bT} + \bar{q}_A^b \gamma_\mu \gamma_5 \mathbb{C} \bar{q}_D^{aT}) \, ,
\\ \nonumber \eta_{6,N}^{\rm AV, \mathbb{O}} &=& \lambda^N_{BD} ( q_A^{aT} \mathbb{C} \gamma_\mu \gamma_5 q_B^b ) (\bar{q}_A^a \mathbb{C} \bar{q}_D^{bT} + \bar{q}_A^b \mathbb{C} \bar{q}_D^{aT}) \, ,
\\ \nonumber \eta_{7,N}^{\rm AV, \mathbb{O}} &=& \lambda^N_{BD} ( q_A^{aT} \mathbb{C} \gamma^\nu q_B^b ) (\bar{q}_A^a \sigma_{\mu\nu} \gamma_5 \mathbb{C} \bar{q}_D^{bT} - \bar{q}_A^b \sigma_{\mu\nu} \gamma_5 \mathbb{C} \bar{q}_D^{aT}) \, ,
\\ \eta_{8,N}^{\rm AV, \mathbb{O}} &=& \lambda^N_{BD} ( q_A^{aT} \mathbb{C} \sigma_{\mu\nu} \gamma_5 q_B^b ) (\bar{q}_A^a \gamma^\nu \mathbb{C} \bar{q}_D^{bT} - \bar{q}_A^b \gamma^\nu \mathbb{C} \bar{q}_D^{aT}) \, ,
\label{eq:8axialvector}
\\ \nonumber \eta_{9,N}^{\rm AV, \mathbb{O}} &=& \lambda^N_{BD} ( q_A^{aT} \mathbb{C} \gamma_5 q_B^b ) (\bar{q}_A^a \gamma_\mu \mathbb{C} \bar{q}_D^{bT} - \bar{q}_A^b \gamma_\mu \mathbb{C} \bar{q}_D^{aT} ) \, ,
\\ \nonumber \eta_{10,N}^{\rm AV, \mathbb{O}} &=& \lambda^N_{BD} ( q_A^{aT} \mathbb{C} \gamma_\mu q_B^b ) (\bar{q}_A^a \gamma_5 \mathbb{C} \bar{q}_D^{bT} + \bar{q}_A^b \gamma_5 \mathbb{C} \bar{q}_D^{aT} ) \, ,
\\ \nonumber \eta_{11,N}^{\rm AV, \mathbb{O}} &=& \lambda^N_{BD} ( q_A^{aT} \mathbb{C} \gamma_\nu \gamma_5 q_B^b ) (\bar{q}_A^a \sigma_{\mu\nu} \mathbb{C} \bar{q}_D^{bT} - \bar{q}_A^b \sigma_{\mu\nu} \mathbb{C} \bar{q}_D^{aT} ) \, ,
\\ \nonumber \eta_{12,N}^{\rm AV, \mathbb{O}} &=& \lambda^N_{BD} ( q_A^{aT} \mathbb{C} \sigma_{\mu\nu} q_B^b ) (\bar{q}_A^a \gamma_\nu \gamma_5 \mathbb{C} \bar{q}_D^{bT} + \bar{q}_A^b \gamma_\nu \gamma_5 \mathbb{C} \bar{q}_D^{aT}) \, ,
\\ \nonumber \eta_{13,N}^{\rm AV, \mathbb{O}} &=& \lambda^N_{BD} ( q_A^{aT} \mathbb{C} \gamma_5 q_B^b ) (\bar{q}_A^a \gamma_\mu \mathbb{C} \bar{q}_D^{bT} + \bar{q}_A^b \gamma_\mu \mathbb{C} \bar{q}_D^{aT} ) \, ,
\\ \nonumber \eta_{14,N}^{\rm AV, \mathbb{O}} &=& \lambda^N_{BD} ( q_A^{aT} \mathbb{C} \gamma_\mu q_B^b ) (\bar{q}_A^a \gamma_5 \mathbb{C} \bar{q}_D^{bT} - \bar{q}_A^b \gamma_5 \mathbb{C} \bar{q}_D^{aT} ) \, ,
\\ \nonumber \eta_{15,N}^{\rm AV, \mathbb{O}} &=& \lambda^N_{BD} ( q_A^{aT} \mathbb{C} \gamma_\nu \gamma_5 q_B^b ) (\bar{q}_A^a \sigma_{\mu\nu} \mathbb{C} \bar{q}_D^{bT} + \bar{q}_A^b \sigma_{\mu\nu} \mathbb{C} \bar{q}_D^{aT} ) \, ,
\\ \nonumber \eta_{16,N}^{\rm AV, \mathbb{O}} &=& \lambda^N_{BD} ( q_A^{aT} \mathbb{C} \sigma_{\mu\nu} q_B^b ) (\bar{q}_A^a \gamma_\nu \gamma_5 \mathbb{C} \bar{q}_D^{bT} - \bar{q}_A^b \gamma_\nu \gamma_5 \mathbb{C} \bar{q}_D^{aT}) \, .
\end{eqnarray}
Among these currents, $\eta_{1,2,3,4,N}^{\rm AV, \mathbb{O}}$ contain diquarks and antidiquarks having both the antisymmetric flavor structure $\mathbf{\bar 3} \otimes \mathbf{3}$; $\eta_{5,6,7,8,N}^{\rm AV, \mathbb{O}}$ contain diquarks and antidiquarks having both the symmetric flavor structure $\mathbf{6} \otimes \mathbf{\bar6}$; $\eta_{9,10,11,12,N}^{\rm AV, \mathbb{O}}$ contain diquarks having the antisymmetric flavor structure and antidiquarks the symmetric flavor structure $\mathbf{\bar3} \otimes \mathbf{\bar6}$; $\eta_{13,14,15,16,N}^{\rm AV, \mathbb{O}}$ contain diquarks having the symmetric flavor structure and antidiquarks the antisymmetric flavor structure $\mathbf{6} \otimes \mathbf{3}$. From the following combinations we can clearly see their chiral structure:
\begin{eqnarray}
\nonumber \eta_{1,N}^{\rm AV, \mathbb{O}} - \eta_{9,N}^{\rm AV, \mathbb{O}} &=& 2 \lambda_N^{DB} L_A^{aT} \mathbb{C} L_B^b (\bar{R}_A^a \gamma_\mu \mathbb{C} \bar{L}_D^{bT} - \bar{R}_A^b \gamma_\mu \mathbb{C} \bar{L}_D^{aT})
- 2 \lambda_N^{DB} R_A^{aT} \mathbb{C} R_B^b (\bar{L}_A^a \gamma_\mu \mathbb{C} \bar{R}_D^{bT} - \bar{L}_A^b \gamma_\mu \mathbb{C} \bar{R}_D^{aT}) \, ,
\\ \nonumber \eta_{1,N}^{\rm AV, \mathbb{O}} + \eta_{9,N}^{\rm AV, \mathbb{O}} &=& - 2 \lambda_N^{DB} L_A^{aT} \mathbb{C} L_B^b (\bar{L}_A^a \gamma_\mu \mathbb{C} \bar{R}_D^{bT} - \bar{L}_A^b \gamma_\mu \mathbb{C} \bar{R}_D^{aT})
+ 2 \lambda_N^{DB} R_A^{aT} \mathbb{C} R_B^b (\bar{R}_A^a \gamma_\mu \mathbb{C} \bar{L}_D^{bT} - \bar{R}_A^b \gamma_\mu \mathbb{C} \bar{L}_D^{aT}) \, ,
\\ \nonumber \eta_{2,N}^{\rm AV, \mathbb{O}} - \eta_{14,N}^{\rm AV, \mathbb{O}} &=& 2 \lambda_N^{DB} L_A^{aT} \mathbb{C} \gamma_\mu R_B^b (\bar{R}_A^a \mathbb{C} \bar{R}_D^{bT} - \bar{R}_A^b \mathbb{C} \bar{R}_D^{aT})
- 2 \lambda_N^{DB} R_A^{aT} \mathbb{C} \gamma_\mu L_B^b (\bar{L}_A^a \mathbb{C} \bar{L}_D^{bT} - \bar{L}_A^b \mathbb{C} \bar{L}_D^{aT})\, ,
\\ \nonumber \eta_{2,N}^{\rm AV, \mathbb{O}} + \eta_{14,N}^{\rm AV, \mathbb{O}} &=& 2 \lambda_N^{DB} L_A^{aT} \mathbb{C} \gamma_\mu R_B^b (\bar{L}_A^a \mathbb{C} \bar{L}_D^{bT} - \bar{L}_A^b \mathbb{C} \bar{L}_D^{aT})
- 2 \lambda_N^{DB} R_A^{aT} \mathbb{C} \gamma_\mu L_B^b (\bar{R}_A^a \mathbb{C} \bar{R}_D^{bT} - \bar{R}_A^b \mathbb{C} \bar{R}_D^{aT}) \, ,
\\ \nonumber \eta_{3,N}^{\rm AV, \mathbb{O}} - \eta_{15,N}^{\rm AV, \mathbb{O}} &=& - 2 \lambda_N^{DB} L_A^{aT} \mathbb{C} \gamma^\nu R_B^b (\bar{R}_A^a \sigma_{\mu\nu} \mathbb{C} \bar{R}_D^{bT} + \bar{R}_A^b \sigma_{\mu\nu} \mathbb{C} \bar{R}_D^{aT})
+ 2 \lambda_N^{DB} R_A^{aT} \mathbb{C} \gamma^\nu L_B^b (\bar{L}_A^a \sigma_{\mu\nu} \mathbb{C} \bar{L}_D^{bT} + \bar{L}_A^b \sigma_{\mu\nu} \mathbb{C} \bar{L}_D^{aT}) \, ,
\\ \nonumber \eta_{3,N}^{\rm AV, \mathbb{O}} + \eta_{15,N}^{\rm AV, \mathbb{O}} &=& 2 \lambda_N^{DB} L_A^{aT} \mathbb{C} \gamma^\nu R_B^b (\bar{L}_A^a \sigma_{\mu\nu} \mathbb{C} \bar{L}_D^{bT} + \bar{L}_A^b \sigma_{\mu\nu} \mathbb{C} \bar{L}_D^{aT})
- 2 \lambda_N^{DB} R_A^{aT} \mathbb{C} \gamma^\nu L_B^b (\bar{R}_A^a \sigma_{\mu\nu} \mathbb{C} \bar{R}_D^{bT} + \bar{R}_A^b \sigma_{\mu\nu} \mathbb{C} \bar{R}_D^{aT}) \, ,
\\ \nonumber \eta_{4,N}^{\rm AV, \mathbb{O}} - \eta_{12,N}^{\rm AV, \mathbb{O}} &=& - 2 \lambda_N^{DB} L_A^{aT} \mathbb{C} \sigma_{\mu\nu} L_B^b (\bar{R}_A^a \gamma^\nu \mathbb{C} \bar{L}_D^{bT} + \bar{R}_A^b \gamma^\nu \mathbb{C} \bar{L}_D^{aT})
+ 2 \lambda_N^{DB} R_A^{aT} \mathbb{C} \sigma_{\mu\nu} R_B^b (\bar{L}_A^a \gamma^\nu \mathbb{C} \bar{R}_D^{bT} + \bar{L}_A^b \gamma^\nu \mathbb{C} \bar{R}_D^{aT}) \, ,
\\ \nonumber \eta_{4,N}^{\rm AV, \mathbb{O}} + \eta_{12,N}^{\rm AV, \mathbb{O}} &=& - 2 \lambda_N^{DB} L_A^{aT} \mathbb{C} \sigma_{\mu\nu} L_B^b (\bar{L}_A^a \gamma^\nu \mathbb{C} \bar{R}_D^{bT} + \bar{L}_A^b \gamma^\nu \mathbb{C} \bar{R}_D^{aT})
+ 2 \lambda_N^{DB} R_A^{aT} \mathbb{C} \sigma_{\mu\nu} R_B^b (\bar{R}_A^a \gamma^\nu \mathbb{C} \bar{L}_D^{bT} + \bar{R}_A^b \gamma^\nu \mathbb{C} \bar{L}_D^{aT}) \, ,
\\ \eta_{5,N}^{\rm AV, \mathbb{O}} - \eta_{13,N}^{\rm AV, \mathbb{O}} &=& 2 \lambda_N^{DB} L_A^{aT} \mathbb{C} L_B^b (\bar{R}_A^a \gamma_\mu \mathbb{C} \bar{L}_D^{bT} + \bar{R}_A^b \gamma_\mu \mathbb{C} \bar{L}_D^{aT})
- 2 \lambda_N^{DB} R_A^{aT} \mathbb{C} R_B^b (\bar{L}_A^a \gamma_\mu \mathbb{C} \bar{R}_D^{bT} + \bar{L}_A^b \gamma_\mu \mathbb{C} \bar{R}_D^{aT}) \, ,
\\ \nonumber \eta_{5,N}^{\rm AV, \mathbb{O}} + \eta_{13,N}^{\rm AV, \mathbb{O}} &=& - 2 \lambda_N^{DB} L_A^{aT} \mathbb{C} L_B^b (\bar{L}_A^a \gamma_\mu \mathbb{C} \bar{R}_D^{bT} + \bar{L}_A^b \gamma_\mu \mathbb{C} \bar{R}_D^{aT})
+ 2 \lambda_N^{DB} R_A^{aT} \mathbb{C} R_B^b (\bar{R}_A^a \gamma_\mu \mathbb{C} \bar{L}_D^{bT} + \bar{R}_A^b \gamma_\mu \mathbb{C} \bar{L}_D^{aT}) \, ,
\\ \nonumber \eta_{6,N}^{\rm AV, \mathbb{O}} - \eta_{10,N}^{\rm AV, \mathbb{O}} &=& 2 \lambda_N^{DB} L_A^{aT} \mathbb{C} \gamma_\mu R_B^b (\bar{R}_A^a \mathbb{C} \bar{R}_D^{bT} + \bar{R}_A^b \mathbb{C} \bar{R}_D^{aT})
- 2 \lambda_N^{DB} R_A^{aT} \mathbb{C} \gamma_\mu L_B^b (\bar{L}_A^a \mathbb{C} \bar{L}_D^{bT} + \bar{L}_A^b \mathbb{C} \bar{L}_D^{aT})\, ,
\\ \nonumber \eta_{6,N}^{\rm AV, \mathbb{O}} + \eta_{10,N}^{\rm AV, \mathbb{O}} &=& 2 \lambda_N^{DB} L_A^{aT} \mathbb{C} \gamma_\mu R_B^b (\bar{L}_A^a \mathbb{C} \bar{L}_D^{bT} + \bar{L}_A^b \mathbb{C} \bar{L}_D^{aT})
- 2 \lambda_N^{DB} R_A^{aT} \mathbb{C} \gamma_\mu L_B^b (\bar{R}_A^a \mathbb{C} \bar{R}_D^{bT} + \bar{R}_A^b \mathbb{C} \bar{R}_D^{aT}) \, ,
\\ \nonumber \eta_{7,N}^{\rm AV, \mathbb{O}} - \eta_{11,N}^{\rm AV, \mathbb{O}} &=& - 2 \lambda_N^{DB} L_A^{aT} \mathbb{C} \gamma^\nu R_B^b (\bar{R}_A^a \sigma_{\mu\nu} \mathbb{C} \bar{R}_D^{bT} - \bar{R}_A^b \sigma_{\mu\nu} \mathbb{C} \bar{R}_D^{aT})
+ 2 \lambda_N^{DB} R_A^{aT} \mathbb{C} \gamma^\nu L_B^b (\bar{L}_A^a \sigma_{\mu\nu} \mathbb{C} \bar{L}_D^{bT} - \bar{L}_A^b \sigma_{\mu\nu} \mathbb{C} \bar{L}_D^{aT}) \, ,
\\ \nonumber \eta_{7,N}^{\rm AV, \mathbb{O}} + \eta_{11,N}^{\rm AV, \mathbb{O}} &=& 2 \lambda_N^{DB} L_A^{aT} \mathbb{C} \gamma^\nu R_B^b (\bar{L}_A^a \sigma_{\mu\nu} \mathbb{C} \bar{L}_D^{bT} - \bar{L}_A^b \sigma_{\mu\nu} \mathbb{C} \bar{L}_D^{aT})
- 2 \lambda_N^{DB} R_A^{aT} \mathbb{C} \gamma^\nu L_B^b (\bar{R}_A^a \sigma_{\mu\nu} \mathbb{C} \bar{R}_D^{bT} - \bar{R}_A^b \sigma_{\mu\nu} \mathbb{C} \bar{R}_D^{aT}) \, ,
\\ \nonumber \eta_{8,N}^{\rm AV, \mathbb{O}} - \eta_{16,N}^{\rm AV, \mathbb{O}} &=& - 2 \lambda_N^{DB} L_A^{aT} \mathbb{C} \sigma_{\mu\nu} L_B^b (\bar{R}_A^a \gamma^\nu \mathbb{C} \bar{L}_D^{bT} - \bar{R}_A^b \gamma^\nu \mathbb{C} \bar{L}_D^{aT})
+ 2 \lambda_N^{DB} R_A^{aT} \mathbb{C} \sigma_{\mu\nu} R_B^b (\bar{L}_A^a \gamma^\nu \mathbb{C} \bar{R}_D^{bT} - \bar{L}_A^b \gamma^\nu \mathbb{C} \bar{R}_D^{aT}) \, ,
\\ \nonumber \eta_{8,N}^{\rm AV, \mathbb{O}} + \eta_{16,N}^{\rm AV, \mathbb{O}} &=& - 2 \lambda_N^{DB} L_A^{aT} \mathbb{C} \sigma_{\mu\nu} L_B^b (\bar{L}_A^a \gamma^\nu \mathbb{C} \bar{R}_D^{bT} - \bar{L}_A^b \gamma^\nu \mathbb{C} \bar{R}_D^{aT})
+ 2 \lambda_N^{DB} R_A^{aT} \mathbb{C} \sigma_{\mu\nu} R_B^b (\bar{R}_A^a \gamma^\nu \mathbb{C} \bar{L}_D^{bT} - \bar{R}_A^b \gamma^\nu \mathbb{C} \bar{L}_D^{aT}) \, .
\end{eqnarray}
We list their chirality and chiral representations in Table~\ref{tab:8axialvector}.
\begin{table}[!hbt]
\renewcommand{\arraystretch}{1.3}
\begin{center}
\caption{Flavor octet tetraquark currents of $J^P = 1^+$ classified in subsection~\ref{subsec:8axialvector}.}
\begin{tabular}{c c c}
\hline\hline
Tetraquark Currents of $(\mathbf 8_F, J^P = 1^+)$ & Chiral Representations & Chirality
\\ \hline \hline
$\eta_{1,N}^{\rm AV, \mathbb{O}} - \eta_{9,N}^{\rm AV, \mathbb{O}}$, $\eta_{4,N}^{\rm AV, \mathbb{O}} - \eta_{12,N}^{\rm AV, \mathbb{O}}$ & $[(\bar {\mathbf 6}, \bar {\mathbf 3}) + (\bar {\mathbf 3}, \bar {\mathbf 6})]\oplus[({\mathbf 3}, \bar {\mathbf 3}) + (\bar {\mathbf 3}, {\mathbf 3})]$ & $L L \bar R \bar L + R R \bar L \bar R$
\\ \hline
$\eta_{1,N}^{\rm AV, \mathbb{O}} + \eta_{9,N}^{\rm AV, \mathbb{O}}$, $\eta_{4,N}^{\rm AV, \mathbb{O}} + \eta_{12,N}^{\rm AV, \mathbb{O}}$ & $[({\mathbf 3}, \bar {\mathbf 3}) + (\bar {\mathbf 3}, {\mathbf 3})]$ & $L L \bar L \bar R + R R \bar R \bar L$
\\ \hline
$\eta_{2,N}^{\rm AV, \mathbb{O}} - \eta_{14,N}^{\rm AV, \mathbb{O}}$, $\eta_{3,N}^{\rm AV, \mathbb{O}} - \eta_{15,N}^{\rm AV, \mathbb{O}}$ & $[({\mathbf 3}, {\mathbf 6}) + ({\mathbf 6}, {\mathbf 3})]\oplus[({\mathbf 3}, \bar {\mathbf 3}) + (\bar {\mathbf 3}, {\mathbf 3})]$ & $L R \bar R \bar R + R L \bar L \bar L$
\\ \hline
$\eta_{2,N}^{\rm AV, \mathbb{O}} + \eta_{14,N}^{\rm AV, \mathbb{O}}$, $\eta_{3,N}^{\rm AV, \mathbb{O}} + \eta_{15,N}^{\rm AV, \mathbb{O}}$ & $[(\bar {\mathbf 3}, {\mathbf 3}) + ({\mathbf 3}, \bar {\mathbf 3})]$ & $L R \bar L \bar L + R L \bar R \bar R$
\\ \hline
$\eta_{5,N}^{\rm AV, \mathbb{O}} - \eta_{13,N}^{\rm AV, \mathbb{O}}$, $\eta_{8,N}^{\rm AV, \mathbb{O}} - \eta_{16,N}^{\rm AV, \mathbb{O}}$ & $[({\mathbf {15}}, \bar {\mathbf 3}) + (\bar {\mathbf 3}, {\mathbf {15}})]\oplus[({\mathbf 3}, \bar {\mathbf 3}) + (\bar {\mathbf 3}, {\mathbf 3})]$ & $L L \bar R \bar L + R R \bar L \bar R$
\\ \hline
$\eta_{5,N}^{\rm AV, \mathbb{O}} + \eta_{13,N}^{\rm AV, \mathbb{O}}$, $\eta_{8,N}^{\rm AV, \mathbb{O}} + \eta_{16,N}^{\rm AV, \mathbb{O}}$ & $[({\mathbf 3}, \bar {\mathbf 3}) + (\bar {\mathbf 3}, {\mathbf 3})]$ & $L L \bar L \bar R + R R \bar R \bar L$
\\ \hline
$\eta_{6,N}^{\rm AV, \mathbb{O}} - \eta_{10,N}^{\rm AV, \mathbb{O}}$, $\eta_{7,N}^{\rm AV, \mathbb{O}} - \eta_{11,N}^{\rm AV, \mathbb{O}}$ & $[({\mathbf 3}, \overline{\mathbf {15}}) + (\overline{\mathbf {15}}, {\mathbf 3})]\oplus[({\mathbf 3}, \bar {\mathbf 3}) + (\bar {\mathbf 3}, {\mathbf 3})]$ & $L R \bar R \bar R + R L \bar L \bar L$
\\ \hline
$\eta_{6,N}^{\rm AV, \mathbb{O}} + \eta_{10,N}^{\rm AV, \mathbb{O}}$, $\eta_{7,N}^{\rm AV, \mathbb{O}} + \eta_{11,N}^{\rm AV, \mathbb{O}}$ & $[(\bar {\mathbf 3}, {\mathbf 3}) + ({\mathbf 3}, \bar {\mathbf 3})]$ & $L R \bar L \bar L + R L \bar R \bar R$
\\ \hline\hline
\end{tabular}
\label{tab:8axialvector}
\end{center}
\renewcommand{\arraystretch}{1}
\end{table}

\subsection{Tetraquark currents of flavor $\mathbf{27}_F$ and $J^P = 1^-$}
\label{subsec:27vector}

In this subsection we study flavor $\mathbf{27}_F$ tetraquark currents of $J^P = 1^-$. There are altogether four independent vector currents as listed in the following:
\begin{eqnarray}
\nonumber \eta_1^{\rm V, \mathbb{TS}} &=& S_P^{ABCD} ( q_A^{aT} \mathbb{C} \gamma_5 q_B^b ) (\bar{q}_C^a \gamma_\mu \gamma_5 \mathbb{C} \bar{q}_D^{bT} ) \, ,
\\ \eta_2^{\rm V, \mathbb{TS}} &=& S_P^{ABCD} ( q_A^{aT} \mathbb{C} \gamma_\mu \gamma_5 q_B^b ) (\bar{q}_C^a \gamma_5 \mathbb{C} \bar{q}_D^{bT} ) \, ,
\label{eq:27vector}
\\ \nonumber \eta_3^{\rm V, \mathbb{TS}} &=& S_P^{ABCD} ( q_A^{aT} \mathbb{C} \gamma^\nu q_B^b ) (\bar{q}_C^a \sigma_{\mu\nu} \mathbb{C} \bar{q}_D^{bT} ) \, ,
\\ \nonumber \eta_4^{\rm V, \mathbb{TS}} &=& S_P^{ABCD} ( q_A^{aT} \mathbb{C} \sigma_{\mu\nu} q_B^b ) (\bar{q}_C^a \gamma^\nu \mathbb{C} \bar{q}_D^{bT} ) \, .
\end{eqnarray}
All these four currents contain diquarks and antidiquarks having the symmetric flavor structure $\mathbf{6} \otimes \mathbf{\bar6}$. From the following combinations we can clearly see their chiral structure:
\begin{eqnarray}
\nonumber \eta_1^{\rm V, \mathbb{TS}} &=& 2 S_P^{ABCD} L_A^{aT} \mathbb{C} L_B^b \bar{L}_C^a \gamma_\mu \mathbb{C} \bar{R}_D^{bT} + 2 S_P^{ABCD} R_A^{aT} \mathbb{C} R_B^b \bar{R}_C^a \gamma_\mu \mathbb{C} \bar{L}_D^{bT} \, ,
\\ \eta_2^{\rm V, \mathbb{TS}} &=& 2 S_P^{ABCD} L_A^{aT} \mathbb{C} \gamma_\mu R_B^b \bar{L}_C^a \mathbb{C} \bar{L}_D^{bT} + 2 S_P^{ABCD} R_A^{aT} \mathbb{C} \gamma_\mu L_B^b \bar{R}_C^a \mathbb{C} \bar{R}_D^{bT} \, ,
\\ \nonumber \eta_3^{\rm V, \mathbb{TS}} &=& 2 S_P^{ABCD} L_A^{aT} \mathbb{C} \gamma^\nu R_B^b \bar{L}_C^a \sigma_{\mu\nu} \mathbb{C} \bar{L}_D^{bT} + 2 S_P^{ABCD} R_A^{aT} \mathbb{C} \gamma^\nu L_B^b \bar{R}_C^a \sigma_{\mu\nu} \mathbb{C} \bar{R}_D^{bT} \, ,
\\ \nonumber \eta_4^{\rm V, \mathbb{TS}} &=& 2 S_P^{ABCD} L_A^{aT} \mathbb{C} \sigma_{\mu\nu} L_B^b \bar{L}_C^a \gamma^\nu \mathbb{C} \bar{R}_D^{bT} + 2 S_P^{ABCD} R_A^{aT} \mathbb{C} \sigma_{\mu\nu} R_B^b \bar{R}_C^a \gamma^\nu \mathbb{C} \bar{L}_D^{bT} \, .
\end{eqnarray}
Among these currents, $\eta_1^{\rm V, \mathbb{TS}}$ and $\eta_4^{\rm V, \mathbb{TS}}$ belong to the chiral representation $[({\mathbf {15}}, \bar {\mathbf 3}) + (\bar {\mathbf 3}, {\mathbf {15}})]$ and their chirality is $L L \bar L \bar R + R R \bar R \bar L$; $\eta_2^{\rm V, \mathbb{TS}}$ and $\eta_3^{\rm V, \mathbb{TS}}$ belong to the chiral representation $[(\overline{\mathbf {15}}, {\mathbf 3}) + ({\mathbf 3}, \overline{\mathbf {15}})]$ and their chirality is $L R \bar L \bar L + R L \bar R \bar R$.

\subsection{Tetraquark currents of flavor $\mathbf{27}_F$ and $J^P = 1^+$}
\label{subsec:27axialvector}

In this subsection we study flavor $\mathbf{27}_F$ tetraquark currents of $J^P = 1^+$. There are altogether four independent axial-vector currents as listed in the following:
\begin{eqnarray}
\nonumber \eta_{1,U}^{\rm AV, \mathbb{TS}} &=& S_P^{ABCD} ( q_A^{aT} \mathbb{C} q_B^b ) (\bar{q}_C^a \gamma_\mu \gamma_5 \mathbb{C} \bar{q}_D^{bT} ) \, ,
\\ \eta_{2,U}^{\rm AV, \mathbb{TS}} &=& S_P^{ABCD} ( q_A^{aT} \mathbb{C} \gamma_\mu \gamma_5 q_B^b ) (\bar{q}_C^a \mathbb{C} \bar{q}_D^{bT} ) \, ,
\label{eq:27axialvector}
\\ \nonumber \eta_{3,U}^{\rm AV, \mathbb{TS}} &=& S_P^{ABCD} ( q_A^{aT} \mathbb{C} \gamma^\nu q_B^b ) (\bar{q}_C^a \sigma_{\mu\nu} \gamma_5 \mathbb{C} \bar{q}_D^{bT} ) \, ,
\\ \nonumber \eta_{4,U}^{\rm AV, \mathbb{TS}} &=& S_P^{ABCD} ( q_A^{aT} \mathbb{C} \sigma_{\mu\nu} \gamma_5 q_B^b ) (\bar{q}_C^a \gamma^\nu \mathbb{C} \bar{q}_D^{bT} ) \, .
\end{eqnarray}
All these four currents contain diquarks and antidiquarks having the symmetric flavor structure $\mathbf{6} \otimes \mathbf{\bar6}$. From the following combinations we can clearly see their chiral structure:
\begin{eqnarray}
\nonumber \eta_{1,U}^{\rm AV, \mathbb{TS}} &=& - 2 S_P^{ABCD} L_A^{aT} \mathbb{C} L_B^b \bar{L}_C^a \gamma_\mu \mathbb{C} \bar{R}_D^{bT} + 2 S_P^{ABCD} R_A^{aT} \mathbb{C} R_B^b \bar{R}_C^a \gamma_\mu \mathbb{C} \bar{L}_D^{bT} \, ,
\\ \eta_{2,U}^{\rm AV, \mathbb{TS}} &=& 2 S_P^{ABCD} L_A^{aT} \mathbb{C} \gamma_\mu R_B^b \bar{L}_C^a \mathbb{C} \bar{L}_D^{bT} - 2 S_P^{ABCD} R_A^{aT} \mathbb{C} \gamma_\mu L_B^b \bar{R}_C^a \mathbb{C} \bar{R}_D^{bT} \, ,
\\ \nonumber \eta_{3,U}^{\rm AV, \mathbb{TS}} &=& 2 S_P^{ABCD} L_A^{aT} \mathbb{C} \gamma^\nu R_B^b \bar{L}_C^a \sigma_{\mu\nu} \mathbb{C} \bar{L}_D^{bT} - 2 S_P^{ABCD} R_A^{aT} \mathbb{C} \gamma^\nu L_B^b \bar{R}_C^a \sigma_{\mu\nu} \mathbb{C} \bar{R}_D^{bT} \, ,
\\ \nonumber \eta_{4,U}^{\rm AV, \mathbb{TS}} &=& - 2 S_P^{ABCD} L_A^{aT} \mathbb{C} \sigma_{\mu\nu} L_B^b \bar{L}_C^a \gamma^\nu \mathbb{C} \bar{R}_D^{bT} + 2 S_P^{ABCD} R_A^{aT} \mathbb{C} \sigma_{\mu\nu} R_B^b \bar{R}_C^a \gamma^\nu \mathbb{C} \bar{L}_D^{bT} \, .
\end{eqnarray}
Among these currents, $\eta_{1,U}^{\rm AV, \mathbb{TS}}$ and $\eta_{4,U}^{\rm AV, \mathbb{TS}}$ belong to the chiral representation $[({\mathbf {15}}, \bar {\mathbf 3}) + (\bar {\mathbf 3}, {\mathbf {15}})]$ and their chirality is $L L \bar L \bar R + R R \bar R \bar L$; $\eta_{2,U}^{\rm AV, \mathbb{TS}}$ and $\eta_{3,U}^{\rm AV, \mathbb{TS}}$ belong to the chiral representation $[(\overline{\mathbf {15}}, {\mathbf 3}) + ({\mathbf 3}, \overline{\mathbf {15}})]$ and their chirality is $L R \bar L \bar L + R L \bar R \bar R$.

\subsection{Tetraquark currents of flavor $\overline{\mathbf{10}}_F$ and $J^P = 1^-$}
\label{subsec:10vector}

In this subsection and the following subsection we study flavor $\overline{\mathbf{10}}_F$ tetraquark currents. Those of flavor ${\mathbf{10}}_F$ can be similarly obtained by simply replacing the flavor coefficient from $\epsilon^{ABE} S_P^{CDE}$ to $S_P^{ABE} \epsilon^{CDE}$, and we use $\eta_{i,P}^{\rm V, \mathbb{D}}$ and $\eta_{i,P}^{\rm AV, \mathbb{D}}$ to denote them.

In this subsection we study flavor $\overline{\mathbf{10}}_F$ tetraquark currents of $J^P = 1^-$. There are altogether four independent vector currents as listed in the following:
\begin{eqnarray}
\nonumber \eta_{1,P}^{\rm V, \bar\mathbb{D}} &=& \epsilon^{ABE} S_P^{CDE} ( q_A^{aT} \mathbb{C} q_B^b ) (\bar{q}_C^a \gamma_\mu \mathbb{C} \bar{q}_D^{bT} ) \, ,
\\ \eta_{2,P}^{\rm V, \bar\mathbb{D}} &=& \epsilon^{ABE} S_P^{CDE} ( q_A^{aT} \mathbb{C} \gamma_\mu q_B^b ) (\bar{q}_C^a \mathbb{C} \bar{q}_D^{bT} ) \, ,
\label{eq:10vector}
\\ \nonumber \eta_{3,P}^{\rm V, \bar\mathbb{D}} &=& \epsilon^{ABE} S_P^{CDE} ( q_A^{aT} \mathbb{C} \gamma_\nu \gamma_5 q_B^b ) (\bar{q}_C^a \sigma_{\mu\nu} \gamma_5 \mathbb{C} \bar{q}_D^{bT} ) \, ,
\\ \nonumber \eta_{4,P}^{\rm V, \bar\mathbb{D}} &=& \epsilon^{ABE} S_P^{CDE} ( q_A^{aT} \mathbb{C} \sigma_{\mu\nu} \gamma_5 q_B^b ) (\bar{q}_C^a \gamma_\nu \gamma_5 \mathbb{C} \bar{q}_D^{bT} ) \, .
\end{eqnarray}
All these four currents contain diquarks having the antisymmetric flavor structure and antidiquarks the symmetric flavor structure $\mathbf{\bar3} \otimes \mathbf{\bar6}$. From the following combinations we can clearly see their chiral structure:
\begin{eqnarray}
\nonumber \eta_{1,P}^{\rm V, \bar\mathbb{D}} &=& 2 \epsilon^{ABE} S_P^{CDE} L_A^{aT} \mathbb{C} L_B^b \bar{L}_C^a \gamma_\mu \mathbb{C} \bar{R}_D^{bT} + 2 \epsilon^{ABE} S_P^{CDE} R_A^{aT} \mathbb{C} R_B^b \bar{R}_C^a \gamma_\mu \mathbb{C} \bar{L}_D^{bT} \, ,
\\ \eta_{2,P}^{\rm V, \bar\mathbb{D}} &=& 2 \epsilon^{ABE} S_P^{CDE} L_A^{aT} \mathbb{C} \gamma_\mu R_B^b \bar{L}_C^a \mathbb{C} \bar{L}_D^{bT} + 2 \epsilon^{ABE} S_P^{CDE} R_A^{aT} \mathbb{C} \gamma_\mu L_B^b \bar{R}_C^a \mathbb{C} \bar{R}_D^{bT} \, ,
\\ \nonumber \eta_{3,P}^{\rm V, \bar\mathbb{D}} &=& 2 \epsilon^{ABE} S_P^{CDE} L_A^{aT} \mathbb{C} \gamma^\nu R_B^b \bar{L}_C^a \sigma_{\mu\nu} \mathbb{C} \bar{L}_D^{bT} + 2 \epsilon^{ABE} S_P^{CDE} R_A^{aT} \mathbb{C} \gamma^\nu L_B^b \bar{R}_C^a \sigma_{\mu\nu} \mathbb{C} \bar{R}_D^{bT} \, ,
\\ \nonumber \eta_{4,P}^{\rm V, \bar\mathbb{D}} &=& 2 \epsilon^{ABE} S_P^{CDE} L_A^{aT} \mathbb{C} \sigma_{\mu\nu} L_B^b \bar{L}_C^a \gamma^\nu \mathbb{C} \bar{R}_D^{bT} + 2 \epsilon^{ABE} S_P^{CDE} R_A^{aT} \mathbb{C} \sigma_{\mu\nu} R_B^b \bar{R}_C^a \gamma^\nu \mathbb{C} \bar{L}_D^{bT} \, .
\end{eqnarray}
The currents $\eta_{1,P}^{\rm V, \bar\mathbb{D}}$ and $\eta_{4,P}^{\rm V, \bar\mathbb{D}}$ belong to the chiral representation $[(\bar {\mathbf 6}, \bar {\mathbf 3}) + (\bar {\mathbf 3}, \bar {\mathbf 6})]$ and their chirality is $L L \bar L \bar R + R R \bar R \bar L$, and the currents $\eta_{2,P}^{\rm V, \bar\mathbb{D}}$ and $\eta_{3,P}^{\rm V, \bar\mathbb{D}}$ belong to the chiral representation $[(\overline{\mathbf {15}}, {\mathbf 3}) + ({\mathbf 3}, \overline{\mathbf {15}})]$ and their chirality is $L R \bar L \bar L + R L \bar R \bar R$.

\subsection{Tetraquark currents of flavor $\overline{\mathbf{10}}_F$ and $J^P = 1^+$}
\label{subsec:10axialvector}

In this subsection we study flavor $\overline{\mathbf{10}}_F$ tetraquark currents of $J^P = 1^+$. There are altogether four independent axial-vector currents as listed in the following:
\begin{eqnarray}
\nonumber \eta_{1,P}^{\rm AV, \bar\mathbb{D}} &=& \epsilon^{ABE} S_P^{CDE} ( q_A^{aT} \mathbb{C} \gamma_5 q_B^b ) (\bar{q}_C^a \gamma_\mu \mathbb{C} \bar{q}_D^{bT} ) \, ,
\\ \eta_{2,P}^{\rm AV, \bar\mathbb{D}} &=& \epsilon^{ABE} S_P^{CDE} ( q_A^{aT} \mathbb{C} \gamma_\mu q_B^b ) (\bar{q}_C^a \gamma_5 \mathbb{C} \bar{q}_D^{bT} ) \, ,
\label{eq:10axialvector}
\\ \nonumber \eta_{3,P}^{\rm AV, \bar\mathbb{D}} &=& \epsilon^{ABE} S_P^{CDE} ( q_A^{aT} \mathbb{C} \gamma_\nu \gamma_5 q_B^b ) (\bar{q}_C^a \sigma_{\mu\nu} \mathbb{C} \bar{q}_D^{bT} ) \, ,
\\ \nonumber \eta_{4,P}^{\rm AV, \bar\mathbb{D}} &=& \epsilon^{ABE} S_P^{CDE} ( q_A^{aT} \mathbb{C} \sigma_{\mu\nu} q_B^b ) (\bar{q}_C^a \gamma_\nu \gamma_5 \mathbb{C} \bar{q}_D^{bT} ) \, .
\end{eqnarray}
All these four currents contain diquarks having the antisymmetric flavor structure and antidiquarks the symmetric flavor structure $\mathbf{\bar3} \otimes \mathbf{\bar6}$. From the following combinations we can clearly see their chiral structure:
\begin{eqnarray}
\nonumber \eta_{1,P}^{\rm AV, \bar\mathbb{D}} &=& - 2 \epsilon^{ABE} S_P^{CDE} L_A^{aT} \mathbb{C} L_B^b \bar{L}_C^a \gamma_\mu \mathbb{C} \bar{R}_D^{bT} + 2 \epsilon^{ABE} S_P^{CDE} R_A^{aT} \mathbb{C} R_B^b \bar{R}_C^a \gamma_\mu \mathbb{C} \bar{L}_D^{bT} \, ,
\\ \eta_{2,P}^{\rm AV, \bar\mathbb{D}} &=& 2 \epsilon^{ABE} S_P^{CDE} L_A^{aT} \mathbb{C} \gamma_\mu R_B^b \bar{L}_C^a \mathbb{C} \bar{L}_D^{bT} - 2 \epsilon^{ABE} S_P^{CDE} R_A^{aT} \mathbb{C} \gamma_\mu L_B^b \bar{R}_C^a \mathbb{C} \bar{R}_D^{bT} \, ,
\\ \nonumber \eta_{3,P}^{\rm AV, \bar\mathbb{D}} &=& 2 \epsilon^{ABE} S_P^{CDE} L_A^{aT} \mathbb{C} \gamma^\nu R_B^b \bar{L}_C^a \sigma_{\mu\nu} \mathbb{C} \bar{L}_D^{bT} - 2 \epsilon^{ABE} S_P^{CDE} R_A^{aT} \mathbb{C} \gamma^\nu L_B^b \bar{R}_C^a \sigma_{\mu\nu} \mathbb{C} \bar{R}_D^{bT} \, ,
\\ \nonumber \eta_{4,P}^{\rm AV, \bar\mathbb{D}} &=& - 2 \epsilon^{ABE} S_P^{CDE} L_A^{aT} \mathbb{C} \sigma_{\mu\nu} L_B^b \bar{L}_C^a \gamma^\nu \mathbb{C} \bar{R}_D^{bT} + 2 \epsilon^{ABE} S_P^{CDE} R_A^{aT} \mathbb{C} \sigma_{\mu\nu} R_B^b \bar{R}_C^a \gamma^\nu \mathbb{C} \bar{L}_D^{bT} \, .
\end{eqnarray}
The currents $\eta_{1,P}^{\rm AV, \bar\mathbb{D}}$ and $\eta_{4,P}^{\rm AV, \bar\mathbb{D}}$ belong to the chiral representation $[(\bar {\mathbf 6}, \bar {\mathbf 3}) + (\bar {\mathbf 3}, \bar {\mathbf 6})]$ and their chirality is $L L \bar L \bar R + R R \bar R \bar L$, and the currents $\eta_{2,P}^{\rm AV, \bar\mathbb{D}}$ and $\eta_{3,P}^{\rm AV, \bar\mathbb{D}}$ belong to the chiral representation $[(\overline{\mathbf {15}}, {\mathbf 3}) + ({\mathbf 3}, \overline{\mathbf {15}})]$ and their chirality is $L R \bar L \bar L + R L \bar R \bar R$.

\section{Chiral Transformations}
\label{app:chiraltransformation}

There are two $[(\bar{\mathbf 3},\bar{\mathbf 6}) \oplus (\bar{\mathbf 6},\bar{\mathbf 3})]$ chiral multiplets, $\big ( 3 \eta_{1,N}^{\rm V, \mathbb{O}} - \eta_{9,N}^{\rm V, \mathbb{O}}, 3 \eta_{1,N}^{\rm AV, \mathbb{O}} - \eta_{9,N}^{\rm AV, \mathbb{O}}, \eta_{1,P}^{\rm V, \bar\mathbb{D}}, \eta_{1,P}^{\rm AV, \bar\mathbb{D}} \big )$, $\big ( 3 \eta_{4,N}^{\rm V, \mathbb{O}} - \eta_{12,N}^{\rm V, \mathbb{O}}, 3 \eta_{4,N}^{\rm AV, \mathbb{O}} - \eta_{12,N}^{\rm AV, \mathbb{O}}, \eta_{4,P}^{\rm V, \bar\mathbb{D}}, \eta_{4,P}^{\rm AV, \bar\mathbb{D}} \big )$. We use $\big ( \eta_{(\bar{\mathbf 3},\bar{\mathbf 6}),N}^{\rm V, \mathbb{O}}, \eta_{(\bar{\mathbf 3},\bar{\mathbf 6}),N}^{\rm AV, \mathbb{O}}, \eta_{(\bar{\mathbf 3},\bar{\mathbf 6}),P}^{\rm V, \bar\mathbb{D}}, \eta_{(\bar{\mathbf 3},\bar{\mathbf 6}),P}^{\rm AV, \bar\mathbb{D}} \big )$ to denote them, and their chiral transformation properties are
\begin{eqnarray}
\nonumber \delta_5 \eta_{(\bar{\mathbf 3},\bar{\mathbf 6}),N}^{\rm V, \mathbb{O}} &=& 2 i b \eta_{(\bar{\mathbf 3},\bar{\mathbf 6}),N}^{\rm AV, \mathbb{O}} \, ,
\\ \nonumber \delta^{\vec a} \eta_{(\bar{\mathbf 3},\bar{\mathbf 6}),N}^{\rm V, \mathbb{O}} &=& 2 a^N f_{NMO} \eta_{(\bar{\mathbf 3},\bar{\mathbf 6}),O}^{\rm V, \mathbb{O}} \, ,
\\ \nonumber \delta_5^{\vec b} \eta_{(\bar{\mathbf 3},\bar{\mathbf 6}),N}^{\rm V, \mathbb{O}} &=& i b^N \big ( - 2 d_{NMO} - {4 i\over3}f_{NMO} \big ) \eta_{(\bar{\mathbf 3},\bar{\mathbf 6}),O}^{\rm AV, \mathbb{O}} - 8 i b^N \big ( {\bf T}^*_{8\times10} \big )^N_{MP} \eta_{(\bar{\mathbf 3},\bar{\mathbf 6}),P}^{\rm AV, \bar\mathbb{D}} \, ,
\\ \nonumber \delta_5 \eta_{(\bar{\mathbf 3},\bar{\mathbf 6}),N}^{\rm AV, \mathbb{O}} &=& 2 i b \eta_{(\bar{\mathbf 3},\bar{\mathbf 6}),N}^{\rm V, \mathbb{O}} \, ,
\\ \nonumber \delta^{\vec a} \eta_{(\bar{\mathbf 3},\bar{\mathbf 6}),N}^{\rm AV, \mathbb{O}} &=& 2 a^N f_{NMO} \eta_{(\bar{\mathbf 3},\bar{\mathbf 6}),O}^{\rm AV, \mathbb{O}} \, ,
\\ \nonumber \delta_5^{\vec b} \eta_{(\bar{\mathbf 3},\bar{\mathbf 6}),N}^{\rm AV, \mathbb{O}} &=& i b^N \big ( - 2 d_{NMO} - {4 i\over3}f_{NMO} \big ) \eta_{(\bar{\mathbf 3},\bar{\mathbf 6}),O}^{\rm V, \mathbb{O}} - 8 i b^N \big ( {\bf T}^*_{8\times10} \big )^N_{MP} \eta_{(\bar{\mathbf 3},\bar{\mathbf 6}),P}^{\rm V, \bar\mathbb{D}} \, ,
\\ \nonumber \delta_5 \eta_{(\bar{\mathbf 3},\bar{\mathbf 6}),P}^{\rm V, \bar\mathbb{D}} &=& 2 i b \eta_{(\bar{\mathbf 3},\bar{\mathbf 6}),P}^{\rm AV, \bar\mathbb{D}} \, ,
\\ \nonumber \delta^{\vec a} \eta_{(\bar{\mathbf 3},\bar{\mathbf 6}),P}^{\rm V, \bar\mathbb{D}} &=& -3 i a^N \big ( {\bf T}^*_{10\times10} \big )^N_{PQ} \eta_{(\bar{\mathbf 3},\bar{\mathbf 6}),Q}^{\rm V, \bar\mathbb{D}} \, ,
\\ \nonumber \delta_5^{\vec b} \eta_{(\bar{\mathbf 3},\bar{\mathbf 6}),P}^{\rm V, \bar\mathbb{D}} &=& - {2 i \over3} b^N \big ( {\bf T}^{\dagger*}_{8\times10} \big )^N_{PO} \eta_{(\bar{\mathbf 3},\bar{\mathbf 6}),O}^{\rm AV, \mathbb{O}} - i b^N \big ( {\bf T}^*_{10\times10} \big )^N_{PQ} \eta_{(\bar{\mathbf 3},\bar{\mathbf 6}),Q}^{\rm AV, \bar\mathbb{D}} \, ,
\\ \nonumber \delta_5 \eta_{(\bar{\mathbf 3},\bar{\mathbf 6}),P}^{\rm AV, \bar\mathbb{D}} &=& 2 i b \eta_{(\bar{\mathbf 3},\bar{\mathbf 6}),P}^{\rm V, \bar\mathbb{D}} \, ,
\\ \nonumber \delta^{\vec a} \eta_{(\bar{\mathbf 3},\bar{\mathbf 6}),P}^{\rm AV, \bar\mathbb{D}} &=& -3 i a^N \big ( {\bf T}^*_{10\times10} \big )^N_{PQ} \eta_{(\bar{\mathbf 3},\bar{\mathbf 6}),Q}^{\rm AV, \bar\mathbb{D}} \, ,
\\ \nonumber \delta_5^{\vec b} \eta_{(\bar{\mathbf 3},\bar{\mathbf 6}),P}^{\rm AV, \bar\mathbb{D}} &=& - {2 i \over3} b^N \big ( {\bf T}^{\dagger*}_{8\times10} \big )^N_{PO} \eta_{(\bar{\mathbf 3},\bar{\mathbf 6}),O}^{\rm V, \mathbb{O}} - i b^N \big ( {\bf T}^*_{10\times10} \big )^N_{PQ} \eta_{(\bar{\mathbf 3},\bar{\mathbf 6}),Q}^{\rm V, \bar\mathbb{D}} \, ,
\end{eqnarray}
There are two $[({\mathbf 6},{\mathbf 3}) \oplus ({\mathbf 3},{\mathbf 6})]$ chiral multiplets, $\big ( 3 \eta_{2,N}^{\rm V, \mathbb{O}} - \eta_{14,N}^{\rm V, \mathbb{O}}, 3 \eta_{2,N}^{\rm AV, \mathbb{O}} - \eta_{14,N}^{\rm AV, \mathbb{O}}, \eta_{2,P}^{\rm V, \mathbb{D}}, \eta_{2,P}^{\rm AV, \mathbb{D}} \big )$, $\big ( 3 \eta_{3,N}^{\rm V, \mathbb{O}} - \eta_{15,N}^{\rm V, \mathbb{O}}, 3 \eta_{3,N}^{\rm AV, \mathbb{O}} - \eta_{15,N}^{\rm AV, \mathbb{O}}, \eta_{3,P}^{\rm V, \mathbb{D}}, \eta_{3,P}^{\rm AV, \mathbb{D}} \big )$. We use $\big ( \eta_{({\mathbf 6},{\mathbf 3}),N}^{\rm V, \mathbb{O}}, \eta_{({\mathbf 6},{\mathbf 3}),N}^{\rm AV, \mathbb{O}}, \eta_{({\mathbf 6},{\mathbf 3}),P}^{\rm V, \mathbb{D}}, \eta_{({\mathbf 6},{\mathbf 3}),P}^{\rm AV, \mathbb{D}} \big )$ to denote them, and their chiral transformation properties are
\begin{eqnarray}
\nonumber \delta_5 \eta_{({\mathbf 6},{\mathbf 3}),N}^{\rm V, \mathbb{O}} &=& 2 i b \eta_{({\mathbf 6},{\mathbf 3}),N}^{\rm AV, \mathbb{O}} \, ,
\\ \nonumber \delta^{\vec a} \eta_{({\mathbf 6},{\mathbf 3}),N}^{\rm V, \mathbb{O}} &=& 2 a^N f_{NMO} \eta_{({\mathbf 6},{\mathbf 3}),O}^{\rm V, \mathbb{O}} \, ,
\\ \nonumber \delta_5^{\vec b} \eta_{({\mathbf 6},{\mathbf 3}),N}^{\rm V, \mathbb{O}} &=& i b^N \big ( - 2 d_{NMO} + {4 i\over3}f_{NMO} \big ) \eta_{({\mathbf 6},{\mathbf 3}),O}^{\rm AV, \mathbb{O}} - 8 i b^N \big ( {\bf T}_{8\times10} \big )^N_{MP} \eta_{({\mathbf 6},{\mathbf 3}),P}^{\rm AV, \mathbb{D}} \, ,
\\ \nonumber \delta_5 \eta_{({\mathbf 6},{\mathbf 3}),N}^{\rm AV, \mathbb{O}} &=& 2 i b \eta_{({\mathbf 6},{\mathbf 3}),N}^{\rm V, \mathbb{O}} \, ,
\\ \nonumber \delta^{\vec a} \eta_{({\mathbf 6},{\mathbf 3}),N}^{\rm AV, \mathbb{O}} &=& 2 a^N f_{NMO} \eta_{({\mathbf 6},{\mathbf 3}),O}^{\rm AV, \mathbb{O}} \, ,
\\ \nonumber \delta_5^{\vec b} \eta_{({\mathbf 6},{\mathbf 3}),N}^{\rm AV, \mathbb{O}} &=& i b^N \big ( - 2 d_{NMO} + {4 i\over3}f_{NMO} \big ) \eta_{({\mathbf 6},{\mathbf 3}),O}^{\rm V, \mathbb{O}} - 8 i b^N \big ( {\bf T}_{8\times10} \big )^N_{MP} \eta_{({\mathbf 6},{\mathbf 3}),P}^{\rm V, \mathbb{D}} \, ,
\\ \nonumber \delta_5 \eta_{({\mathbf 6},{\mathbf 3}),P}^{\rm V, \mathbb{D}} &=& 2 i b \eta_{({\mathbf 6},{\mathbf 3}),P}^{\rm AV, \mathbb{D}} \, ,
\\ \nonumber \delta^{\vec a} \eta_{({\mathbf 6},{\mathbf 3}),P}^{\rm V, \mathbb{D}} &=& 3 i a^N \big ( {\bf T}_{10\times10} \big )^N_{PQ} \eta_{({\mathbf 6},{\mathbf 3}),Q}^{\rm V, \mathbb{D}} \, ,
\\ \nonumber \delta_5^{\vec b} \eta_{({\mathbf 6},{\mathbf 3}),P}^{\rm V, \mathbb{D}} &=& - {2 i \over3} b^N \big ( {\bf T}^{\dagger}_{8\times10} \big )^N_{PO} \eta_{({\mathbf 6},{\mathbf 3}),O}^{\rm AV, \mathbb{O}} - i b^N \big ( {\bf T}_{10\times10} \big )^N_{PQ} \eta_{({\mathbf 6},{\mathbf 3}),Q}^{\rm AV, \mathbb{D}} \, ,
\\ \nonumber \delta_5 \eta_{({\mathbf 6},{\mathbf 3}),P}^{\rm AV, \mathbb{D}} &=& 2 i b \eta_{({\mathbf 6},{\mathbf 3}),P}^{\rm V, \mathbb{D}} \, ,
\\ \nonumber \delta^{\vec a} \eta_{({\mathbf 6},{\mathbf 3}),P}^{\rm AV, \mathbb{D}} &=& 3 i a^N \big ( {\bf T}_{10\times10} \big )^N_{PQ} \eta_{({\mathbf 6},{\mathbf 3}),Q}^{\rm AV, \mathbb{D}} \, ,
\\ \nonumber \delta_5^{\vec b} \eta_{({\mathbf 6},{\mathbf 3}),P}^{\rm AV, \mathbb{D}} &=& - {2 i \over3} b^N \big ( {\bf T}^{\dagger}_{8\times10} \big )^N_{PO} \eta_{({\mathbf 6},{\mathbf 3}),O}^{\rm V, \mathbb{O}} - i b^N \big ( {\bf T}_{10\times10} \big )^N_{PQ} \eta_{({\mathbf 6},{\mathbf 3}),Q}^{\rm V, \mathbb{D}} \, ,
\end{eqnarray}
There are two $[({\mathbf{15}},\bar{\mathbf 3}) \oplus (\bar{\mathbf 3},{\mathbf{15}})]$ chiral multiplets, $\big ( 3 \eta_{5,N}^{\rm V, \mathbb{O}} - 5 \eta_{13,N}^{\rm V, \mathbb{O}}, 3 \eta_{5,N}^{\rm AV, \mathbb{O}} - 5 \eta_{13,N}^{\rm AV, \mathbb{O}}, \eta_{1,P}^{\rm V, \mathbb{D}}, \eta_{1,P}^{\rm AV, \mathbb{D}}, \eta_{1,U}^{\rm V, \mathbb{TS}}, \eta_{1,U}^{\rm AV, \mathbb{TS}} \big )$, $\big ( 3 \eta_{8,N}^{\rm V, \mathbb{O}} - 5 \eta_{16,N}^{\rm V, \mathbb{O}}, 3 \eta_{8,N}^{\rm AV, \mathbb{O}} - 5 \eta_{16,N}^{\rm AV, \mathbb{O}}, \eta_{4,P}^{\rm V, \mathbb{D}}, \eta_{4,P}^{\rm AV, \mathbb{D}}, \eta_{4,U}^{\rm V, \mathbb{TS}}, \eta_{4,U}^{\rm AV, \mathbb{TS}} \big )$. We use $\big ( \eta_{({\mathbf{15}},\bar{\mathbf 3}),N}^{\rm V, \mathbb{O}}, \eta_{({\mathbf{15}},\bar{\mathbf 3}),N}^{\rm AV, \mathbb{O}}, \eta_{({\mathbf{15}},\bar{\mathbf 3}),P}^{\rm V, \mathbb{D}}, \eta_{({\mathbf{15}},\bar{\mathbf 3}),P}^{\rm AV, \mathbb{D}}, \eta_{({\mathbf{15}},\bar{\mathbf 3}),U}^{\rm V, \mathbb{TS}}, \eta_{({\mathbf{15}},\bar{\mathbf 3}),U}^{\rm AV, \mathbb{TS}} \big )$ to denote them, and their chiral transformation properties are
\begin{eqnarray}
\nonumber \delta_5 \eta_{({\mathbf{15}},\bar{\mathbf 3}),N}^{\rm V, \mathbb{O}} &=& 2 i b \eta_{({\mathbf{15}},\bar{\mathbf 3}),N}^{\rm AV, \mathbb{O}} \, ,
\\ \nonumber \delta^{\vec a} \eta_{({\mathbf{15}},\bar{\mathbf 3}),N}^{\rm V, \mathbb{O}} &=& 2 a^N f_{NMO} \eta_{({\mathbf{15}},\bar{\mathbf 3}),O}^{\rm V, \mathbb{O}} \, ,
\\ \nonumber \delta_5^{\vec b} \eta_{({\mathbf{15}},\bar{\mathbf 3}),N}^{\rm V, \mathbb{O}} &=& i b^N \big ( {2\over5} d_{NMO} - {8 i\over3}f_{NMO} \big ) \eta_{({\mathbf{15}},\bar{\mathbf 3}),O}^{\rm AV, \mathbb{O}} - 16 i b^N \big ( {\bf T}_{8\times10} \big )^N_{MP} \eta_{({\mathbf{15}},\bar{\mathbf 3}),P}^{\rm AV, \mathbb{D}} + 16 i b^N \big ( {\bf T}_{8\times27} \big )^N_{MU} \eta_{({\mathbf{15}},\bar{\mathbf 3}),U}^{\rm AV, \mathbb{TS}} \, ,
\\ \nonumber \delta_5 \eta_{({\mathbf{15}},\bar{\mathbf 3}),N}^{\rm AV, \mathbb{O}} &=& 2 i b \eta_{({\mathbf{15}},\bar{\mathbf 3}),N}^{\rm V, \mathbb{O}} \, ,
\\ \nonumber \delta^{\vec a} \eta_{({\mathbf{15}},\bar{\mathbf 3}),N}^{\rm AV, \mathbb{O}} &=& 2 a^N f_{NMO} \eta_{({\mathbf{15}},\bar{\mathbf 3}),O}^{\rm AV, \mathbb{O}} \, ,
\\ \nonumber \delta_5^{\vec b} \eta_{({\mathbf{15}},\bar{\mathbf 3}),N}^{\rm AV, \mathbb{O}} &=& i b^N \big ( {2\over5} d_{NMO} - {8 i\over3}f_{NMO} \big ) \eta_{({\mathbf{15}},\bar{\mathbf 3}),O}^{\rm V, \mathbb{O}} - 16 i b^N \big ( {\bf T}_{8\times10} \big )^N_{MP} \eta_{({\mathbf{15}},\bar{\mathbf 3}),P}^{\rm V, \mathbb{D}} + 16 i b^N \big ( {\bf T}_{8\times27} \big )^N_{MU} \eta_{({\mathbf{15}},\bar{\mathbf 3}),U}^{\rm V, \mathbb{TS}} \, ,
\\ \nonumber \delta_5 \eta_{({\mathbf{15}},\bar{\mathbf 3}),P}^{\rm V, \mathbb{D}} &=& 2 i b \eta_{({\mathbf{15}},\bar{\mathbf 3}),P}^{\rm AV, \mathbb{D}} \, ,
\\ \nonumber \delta^{\vec a} \eta_{({\mathbf{15}},\bar{\mathbf 3}),P}^{\rm V, \mathbb{D}} &=& 3 i a^N \big ( {\bf T}_{10\times10} \big )^N_{PQ} \eta_{({\mathbf{15}},\bar{\mathbf 3}),Q}^{\rm V, \mathbb{D}} \, ,
\\ \nonumber \delta_5^{\vec b} \eta_{({\mathbf{15}},\bar{\mathbf 3}),P}^{\rm V, \mathbb{D}} &=& - {2 i \over15} b^N \big ( {\bf T}^{\dagger}_{8\times10} \big )^N_{PO} \eta_{({\mathbf{15}},\bar{\mathbf 3}),O}^{\rm AV, \mathbb{O}} + 2 i b^N \big ( {\bf T}_{10\times10} \big )^N_{PQ} \eta_{({\mathbf{15}},\bar{\mathbf 3}),Q}^{\rm AV, \mathbb{D}} + 2 i b^N \big ( {\bf T}^{\bf A}_{10\times27} \big )^N_{PU} \eta_{({\mathbf{15}},\bar{\mathbf 3}),U}^{\rm AV, \mathbb{TS}} \, ,
\\ \nonumber \delta_5 \eta_{({\mathbf{15}},\bar{\mathbf 3}),P}^{\rm AV, \mathbb{D}} &=& 2 i b \eta_{({\mathbf{15}},\bar{\mathbf 3}),P}^{\rm V, \mathbb{D}} \, ,
\\ \nonumber \delta^{\vec a} \eta_{({\mathbf{15}},\bar{\mathbf 3}),P}^{\rm AV, \mathbb{D}} &=& 3 i a^N \big ( {\bf T}_{10\times10} \big )^N_{PQ} \eta_{({\mathbf{15}},\bar{\mathbf 3}),Q}^{\rm AV, \mathbb{D}} \, ,
\\ \nonumber \delta_5^{\vec b} \eta_{({\mathbf{15}},\bar{\mathbf 3}),P}^{\rm AV, \mathbb{D}} &=& - {2 i \over15} b^N \big ( {\bf T}^{\dagger}_{8\times10} \big )^N_{PO} \eta_{({\mathbf{15}},\bar{\mathbf 3}),O}^{\rm V, \mathbb{O}} + 2 i b^N \big ( {\bf T}_{10\times10} \big )^N_{PQ} \eta_{({\mathbf{15}},\bar{\mathbf 3}),Q}^{\rm V, \mathbb{D}} + 2 i b^N \big ( {\bf T}^{\bf A}_{10\times27} \big )^N_{PU} \eta_{({\mathbf{15}},\bar{\mathbf 3}),U}^{\rm V, \mathbb{TS}} \, ,
\\ \nonumber \delta_5 \eta_{({\mathbf{15}},\bar{\mathbf 3}),U}^{\rm V, \mathbb{TS}} &=& 2 i b \eta_{({\mathbf{15}},\bar{\mathbf 3}),U}^{\rm AV, \mathbb{TS}} \, ,
\\ \nonumber \delta^{\vec a} \eta_{({\mathbf{15}},\bar{\mathbf 3}),U}^{\rm V, \mathbb{TS}} &=& 2 i a^N \big ( {\bf T}^{\bf A}_{27\times27} - {\bf T}^{\bf B}_{27\times27} \big )^N_{UV} \eta_{({\mathbf{15}},\bar{\mathbf 3}),V}^{\rm V, \mathbb{TS}} \, ,
\\ \nonumber \delta_5^{\vec b} \eta_{({\mathbf{15}},\bar{\mathbf 3}),U}^{\rm V, \mathbb{TS}} &=& {i \over 5} b^N \big ( {\bf T}^{\dagger}_{8\times27} \big )^N_{UO} \eta_{({\mathbf{15}},\bar{\mathbf 3}),O}^{\rm AV, \mathbb{O}} + 3 i b^N \big ( {\bf T}^{\bf A\dagger}_{10\times27} \big )^N_{UP} \eta_{({\mathbf{15}},\bar{\mathbf 3}),P}^{\rm AV, \mathbb{D}} + 2 i b^N \big ( {\bf T}^{\bf A}_{27\times27} \big )^N_{UV} \eta_{({\mathbf{15}},\bar{\mathbf 3}),V}^{\rm AV, \mathbb{TS}} \, ,
\\ \nonumber \delta_5 \eta_{({\mathbf{15}},\bar{\mathbf 3}),U}^{\rm AV, \mathbb{TS}} &=& 2 i b \eta_{({\mathbf{15}},\bar{\mathbf 3}),U}^{\rm V, \mathbb{TS}} \, ,
\\ \nonumber \delta^{\vec a} \eta_{({\mathbf{15}},\bar{\mathbf 3}),U}^{\rm AV, \mathbb{TS}} &=& 2 i a^N \big ( {\bf T}^{\bf A}_{27\times27} - {\bf T}^{\bf B}_{27\times27} \big )^N_{UV} \eta_{({\mathbf{15}},\bar{\mathbf 3}),V}^{\rm AV, \mathbb{TS}} \, ,
\\ \nonumber \delta_5^{\vec b} \eta_{({\mathbf{15}},\bar{\mathbf 3}),U}^{\rm AV, \mathbb{TS}} &=& {i \over 5} b^N \big ( {\bf T}^{\dagger}_{8\times27} \big )^N_{UO} \eta_{({\mathbf{15}},\bar{\mathbf 3}),O}^{\rm V, \mathbb{O}} + 3 i b^N \big ( {\bf T}^{\bf A\dagger}_{10\times27} \big )^N_{UP} \eta_{({\mathbf{15}},\bar{\mathbf 3}),P}^{\rm V, \mathbb{D}} + 2 i b^N \big ( {\bf T}^{\bf A}_{27\times27} \big )^N_{UV} \eta_{({\mathbf{15}},\bar{\mathbf 3}),V}^{\rm V, \mathbb{TS}} \, ,
\end{eqnarray}
There are two $[(\overline{\mathbf{15}},{\mathbf 3}) \oplus ({\mathbf 3},\overline{\mathbf{15}})]$ chiral multiplets, $\big ( 3 \eta_{6,N}^{\rm V, \mathbb{O}} - 5 \eta_{10,N}^{\rm V, \mathbb{O}}, 3 \eta_{6,N}^{\rm AV, \mathbb{O}} - 5 \eta_{10,N}^{\rm AV, \mathbb{O}}, \eta_{2,P}^{\rm V, \bar\mathbb{D}}, \eta_{2,P}^{\rm AV, \bar\mathbb{D}}, \eta_{2,U}^{\rm V, \mathbb{TS}}, \eta_{2,U}^{\rm AV, \mathbb{TS}} \big )$, $\big ( 3 \eta_{7,N}^{\rm V, \mathbb{O}} - 5 \eta_{11,N}^{\rm V, \mathbb{O}}, 3 \eta_{7,N}^{\rm AV, \mathbb{O}} - 5 \eta_{11,N}^{\rm AV, \mathbb{O}}, \eta_{3,P}^{\rm V, \bar\mathbb{D}}, \eta_{3,P}^{\rm AV, \bar\mathbb{D}}, \eta_{3,U}^{\rm V, \mathbb{TS}}, \eta_{3,U}^{\rm AV, \mathbb{TS}} \big )$. We use $\big ( \eta_{(\overline{\mathbf{15}},{\mathbf 3}),N}^{\rm V, \mathbb{O}}, \eta_{(\overline{\mathbf{15}},{\mathbf 3}),N}^{\rm AV, \mathbb{O}}, \eta_{(\overline{\mathbf{15}},{\mathbf 3}),P}^{\rm V, \bar\mathbb{D}}, \eta_{(\overline{\mathbf{15}},{\mathbf 3}),P}^{\rm AV, \bar\mathbb{D}}, \eta_{(\overline{\mathbf{15}},{\mathbf 3}),U}^{\rm V, \mathbb{TS}}, \eta_{(\overline{\mathbf{15}},{\mathbf 3}),U}^{\rm AV, \mathbb{TS}} \big )$ to denote them, and their chiral transformation properties are
\begin{eqnarray}
\nonumber \delta_5 \eta_{(\overline{\mathbf{15}},{\mathbf 3}),N}^{\rm V, \mathbb{O}} &=& 2 i b \eta_{(\overline{\mathbf{15}},{\mathbf 3}),N}^{\rm AV, \mathbb{O}} \, ,
\\ \nonumber \delta^{\vec a} \eta_{(\overline{\mathbf{15}},{\mathbf 3}),N}^{\rm V, \mathbb{O}} &=& 2 a^N f_{NMO} \eta_{(\overline{\mathbf{15}},{\mathbf 3}),O}^{\rm V, \mathbb{O}} \, ,
\\ \nonumber \delta_5^{\vec b} \eta_{(\overline{\mathbf{15}},{\mathbf 3}),N}^{\rm V, \mathbb{O}} &=& i b^N \big ( {2\over5} d_{NMO} + {8 i\over3}f_{NMO} \big ) \eta_{(\overline{\mathbf{15}},{\mathbf 3}),O}^{\rm AV, \mathbb{O}} - 16 i b^N \big ( {\bf T}_{8\times10}^* \big )^N_{MP} \eta_{(\overline{\mathbf{15}},{\mathbf 3}),P}^{\rm AV, \bar\mathbb{D}} + 16 i b^N \big ( {\bf T}_{8\times27} \big )^N_{MU} \eta_{(\overline{\mathbf{15}},{\mathbf 3}),U}^{\rm AV, \mathbb{TS}} \, ,
\\ \nonumber \delta_5 \eta_{(\overline{\mathbf{15}},{\mathbf 3}),N}^{\rm AV, \mathbb{O}} &=& 2 i b \eta_{(\overline{\mathbf{15}},{\mathbf 3}),N}^{\rm V, \mathbb{O}} \, ,
\\ \nonumber \delta^{\vec a} \eta_{(\overline{\mathbf{15}},{\mathbf 3}),N}^{\rm AV, \mathbb{O}} &=& 2 a^N f_{NMO} \eta_{(\overline{\mathbf{15}},{\mathbf 3}),O}^{\rm AV, \mathbb{O}} \, ,
\\ \nonumber \delta_5^{\vec b} \eta_{(\overline{\mathbf{15}},{\mathbf 3}),N}^{\rm AV, \mathbb{O}} &=& i b^N \big ( {2\over5} d_{NMO} + {8 i\over3}f_{NMO} \big ) \eta_{(\overline{\mathbf{15}},{\mathbf 3}),O}^{\rm V, \mathbb{O}} - 16 i b^N \big ( {\bf T}_{8\times10}^* \big )^N_{MP} \eta_{(\overline{\mathbf{15}},{\mathbf 3}),P}^{\rm V, \bar\mathbb{D}} + 16 i b^N \big ( {\bf T}_{8\times27} \big )^N_{MU} \eta_{(\overline{\mathbf{15}},{\mathbf 3}),U}^{\rm V, \mathbb{TS}} \, ,
\\ \nonumber \delta_5 \eta_{(\overline{\mathbf{15}},{\mathbf 3}),P}^{\rm V, \bar\mathbb{D}} &=& 2 i b \eta_{(\overline{\mathbf{15}},{\mathbf 3}),P}^{\rm AV, \bar\mathbb{D}} \, ,
\\ \nonumber \delta^{\vec a} \eta_{(\overline{\mathbf{15}},{\mathbf 3}),P}^{\rm V, \bar\mathbb{D}} &=& - 3 i a^N \big ( {\bf T}^*_{10\times10} \big )^N_{PQ} \eta_{(\overline{\mathbf{15}},{\mathbf 3}),Q}^{\rm V, \bar\mathbb{D}} \, ,
\\ \nonumber \delta_5^{\vec b} \eta_{(\overline{\mathbf{15}},{\mathbf 3}),P}^{\rm V, \bar\mathbb{D}} &=& - {2 i \over15} b^N \big ( {\bf T}^{\dagger*}_{8\times10} \big )^N_{PO} \eta_{(\overline{\mathbf{15}},{\mathbf 3}),O}^{\rm AV, \mathbb{O}} + 2 i b^N \big ( {\bf T}^*_{10\times10} \big )^N_{PQ} \eta_{(\overline{\mathbf{15}},{\mathbf 3}),Q}^{\rm AV, \bar\mathbb{D}} + 2 i b^N \big ( {\bf T}^{\bf B}_{10\times27} \big )^N_{PU} \eta_{(\overline{\mathbf{15}},{\mathbf 3}),U}^{\rm AV, \mathbb{TS}} \, ,
\\ \nonumber \delta_5 \eta_{(\overline{\mathbf{15}},{\mathbf 3}),P}^{\rm AV, \bar\mathbb{D}} &=& 2 i b \eta_{(\overline{\mathbf{15}},{\mathbf 3}),P}^{\rm V, \bar\mathbb{D}} \, ,
\\ \nonumber \delta^{\vec a} \eta_{(\overline{\mathbf{15}},{\mathbf 3}),P}^{\rm AV, \bar\mathbb{D}} &=& - 3 i a^N \big ( {\bf T}^*_{10\times10} \big )^N_{PQ} \eta_{(\overline{\mathbf{15}},{\mathbf 3}),Q}^{\rm AV, \bar\mathbb{D}} \, ,
\\ \nonumber \delta_5^{\vec b} \eta_{(\overline{\mathbf{15}},{\mathbf 3}),P}^{\rm AV, \bar\mathbb{D}} &=& - {2 i \over15} b^N \big ( {\bf T}^{\dagger*}_{8\times10} \big )^N_{PO} \eta_{(\overline{\mathbf{15}},{\mathbf 3}),O}^{\rm V, \mathbb{O}} + 2 i b^N \big ( {\bf T}^*_{10\times10} \big )^N_{PQ} \eta_{(\overline{\mathbf{15}},{\mathbf 3}),Q}^{\rm V, \bar\mathbb{D}} + 2 i b^N \big ( {\bf T}^{\bf B}_{10\times27} \big )^N_{PU} \eta_{(\overline{\mathbf{15}},{\mathbf 3}),U}^{\rm V, \mathbb{TS}} \, ,
\\ \nonumber \delta_5 \eta_{(\overline{\mathbf{15}},{\mathbf 3}),U}^{\rm V, \mathbb{TS}} &=& 2 i b \eta_{(\overline{\mathbf{15}},{\mathbf 3}),U}^{\rm AV, \mathbb{TS}} \, ,
\\ \nonumber \delta^{\vec a} \eta_{(\overline{\mathbf{15}},{\mathbf 3}),U}^{\rm V, \mathbb{TS}} &=& 2 i a^N \big ( {\bf T}^{\bf A}_{27\times27} - {\bf T}^{\bf B}_{27\times27} \big )^N_{UV} \eta_{(\overline{\mathbf{15}},{\mathbf 3}),V}^{\rm V, \mathbb{TS}} \, ,
\\ \nonumber \delta_5^{\vec b} \eta_{(\overline{\mathbf{15}},{\mathbf 3}),U}^{\rm V, \mathbb{TS}} &=& {i \over 5} b^N \big ( {\bf T}^{\dagger}_{8\times27} \big )^N_{UO} \eta_{({\mathbf{15}},\bar{\mathbf 3}),O}^{\rm AV, \mathbb{O}} + 3 i b^N \big ( {\bf T}^{\bf B\dagger}_{10\times27} \big )^N_{UP} \eta_{(\overline{\mathbf{15}},{\mathbf 3}),P}^{\rm AV, \bar\mathbb{D}} + 2 i b^N \big ( {\bf T}^{\bf B}_{27\times27} \big )^N_{UV} \eta_{(\overline{\mathbf{15}},{\mathbf 3}),V}^{\rm AV, \mathbb{TS}} \, ,
\\ \nonumber \delta_5 \eta_{(\overline{\mathbf{15}},{\mathbf 3}),U}^{\rm AV, \mathbb{TS}} &=& 2 i b \eta_{(\overline{\mathbf{15}},{\mathbf 3}),U}^{\rm V, \mathbb{TS}} \, ,
\\ \nonumber \delta^{\vec a} \eta_{(\overline{\mathbf{15}},{\mathbf 3}),U}^{\rm AV, \mathbb{TS}} &=& 2 i a^N \big ( {\bf T}^{\bf A}_{27\times27} - {\bf T}^{\bf B}_{27\times27} \big )^N_{UV} \eta_{(\overline{\mathbf{15}},{\mathbf 3}),V}^{\rm AV, \mathbb{TS}} \, ,
\\ \nonumber \delta_5^{\vec b} \eta_{(\overline{\mathbf{15}},{\mathbf 3}),U}^{\rm AV, \mathbb{TS}} &=& {i \over 5} b^N \big ( {\bf T}^{\dagger}_{8\times27} \big )^N_{UO} \eta_{({\mathbf{15}},\bar{\mathbf 3}),O}^{\rm V, \mathbb{O}} + 3 i b^N \big ( {\bf T}^{\bf B\dagger}_{10\times27} \big )^N_{UP} \eta_{(\overline{\mathbf{15}},{\mathbf 3}),P}^{\rm V, \bar\mathbb{D}} + 2 i b^N \big ( {\bf T}^{\bf B}_{27\times27} \big )^N_{UV} \eta_{(\overline{\mathbf{15}},{\mathbf 3}),V}^{\rm V, \mathbb{TS}} \, ,
\end{eqnarray}

\section{Two-point Correlation Functions}
\label{app:ope}

In this appendix we show the results for the Borel transformed correlation functions as defined in Eq.~(\ref{eq:borel}). Results for tetraquark currents having quantum numbers $I^G J^{PC} = 1^+1^{--}$, $1^+1^{+-}$ and $1^-1^{++}$ are separately listed in the following subsections.

\subsection{Tetraquark currents of $I^G J^{PC} = 1^+1^{--}$}

%
\begin{eqnarray}
\nonumber \Pi^{--}_{A,2,1}(M_B^2) &=& \int^{s_0}_{4 m_s^2} \Bigg [ {1 \over 6144
\pi^6} s^4 - { 3 m_s^2 \over 512 \pi^6 } s^3 + \Big ( { 11 \langle
g_s^2 G G \rangle \over 18432 \pi^6 } + {m_s \langle \bar s s
\rangle \over 32 \pi^4} \Big ) s^2 + \Big ( { \langle \bar q q \rangle^2 \over 12
\pi^2 } + { \langle \bar s s \rangle^2 \over 12 \pi^2 }
\\ \nonumber && - { m_s
\langle g_s \bar s \sigma G s \rangle \over 32 \pi^4 }
+ { 7 m_s^2
\langle g_s^2 G G \rangle \over 6144 \pi^6 } \Big ) s + { \langle
\bar q q \rangle \langle g_s \bar q \sigma G q \rangle \over 8
\pi^2 }
+ { \langle \bar s s \rangle \langle g_s \bar s
\sigma G s \rangle \over 8 \pi^2 }
\\ && - { m_s \langle g_s^2 G G
\rangle \langle \bar q q \rangle \over 64 \pi^4} - { 5 m_s \langle g_s^2 G G
\rangle \langle \bar s s \rangle \over 256 \pi^4} - { 3 m_s^2 \langle
\bar q q \rangle^2 \over 2 \pi^2 } + { m_s^2 \langle
\bar s s \rangle^2 \over 8 \pi^2 } \Bigg ] e^{-s/M_B^2} ds
\\ \nonumber && + \Big ( { \langle g_s \bar q \sigma G q \rangle^2 \over 32 \pi^2
} + { \langle g_s \bar s \sigma G
s \rangle^2 \over 32 \pi^2 } + { 25 \langle g_s^2 GG \rangle \langle
\bar q q \rangle^2 \over 1728 \pi^2 } + { 5 \langle g_s^2 GG \rangle \langle
\bar q q \rangle \langle \bar s s \rangle \over 216 \pi^2 } + { 25 \langle g_s^2 GG \rangle \langle
\bar s s \rangle^2 \over 1728 \pi^2 }
\\ \nonumber && + { 10 m_s
\langle \bar q q \rangle^2 \langle \bar s s \rangle \over 3 }
- { 5 m_s \langle g_s^2 GG \rangle \langle g_s \bar q
\sigma G q \rangle \over 1152 \pi^4 }
- { 25 m_s \langle g_s^2 GG \rangle \langle g_s \bar s
\sigma G s \rangle \over 4608 \pi^4 }
- { m_s^2 \langle \bar q q
\rangle \langle g_s \bar q \sigma G q \rangle \over \pi^2 }\Big)
\\ \nonumber && + {1 \over M_B^2} \Big( - {32 g_s^2 \langle
\bar q q \rangle ^2 \langle \bar s s \rangle^2 \over 27 } - { 5 \langle g_s^2 GG \rangle \langle \bar q q \rangle
\langle g_s \bar q \sigma G q \rangle \over 1152 \pi^2 }
- { \langle
g_s^2 GG \rangle \langle \bar s s \rangle \langle g_s \bar q \sigma
G q \rangle \over 288 \pi^2 }
\\ \nonumber && - { \langle
g_s^2 GG \rangle \langle \bar q q \rangle \langle g_s \bar s \sigma
G s \rangle \over 288 \pi^2 } - { 5 \langle
g_s^2 GG \rangle \langle \bar s s \rangle \langle g_s \bar s \sigma
G s \rangle \over 1152 \pi^2 } - { 2 m_s \langle \bar q q \rangle^2
\langle g_s \bar s \sigma G s \rangle \over 3}
\\ \nonumber && - { m_s \langle
\bar q q \rangle \langle \bar s s \rangle \langle g_s \bar q \sigma
G q \rangle} - { 5 m_s^2 \langle g_s^2 GG \rangle \langle \bar s s \rangle^2 \over 1152 \pi^2 } \Big) \, .
\end{eqnarray}
%
%
\begin{eqnarray}
\nonumber \Pi^{--}_{S,1,1}(M_B^2) &=& \int^{s_0}_{4 m_s^2} \Bigg [ {1 \over 18432
\pi^6} s^4 - { m_s^2 \over 512 \pi^6 } s^3 + \Big ( - { \langle
g_s^2 G G \rangle \over 18432 \pi^6 } + {m_s \langle \bar s s
\rangle \over 96 \pi^4} \Big ) s^2 + \Big ( { \langle \bar q q \rangle^2 \over 36
\pi^2 } + { \langle \bar s s \rangle^2 \over 36 \pi^2 }
\\ \nonumber && - { m_s
\langle g_s \bar s \sigma G s \rangle \over 96 \pi^4 }
+ { m_s^2
\langle g_s^2 G G \rangle \over 4608 \pi^6 } \Big ) s + { \langle
\bar q q \rangle \langle g_s \bar q \sigma G q \rangle \over 24
\pi^2 }
+ { \langle \bar s s \rangle \langle g_s \bar s
\sigma G s \rangle \over 24 \pi^2 }
\\ && - { m_s \langle g_s^2 G G
\rangle \langle \bar q q \rangle \over 256 \pi^4} - { m_s^2 \langle
\bar q q \rangle^2 \over 2 \pi^2 } + { m_s^2 \langle
\bar s s \rangle^2 \over 24 \pi^2 } \Bigg ] e^{-s/M_B^2} ds
\\ \nonumber && + \Big ( { \langle g_s \bar q \sigma G q \rangle^2 \over 96 \pi^2
} + { \langle g_s \bar s \sigma G
s \rangle^2 \over 96 \pi^2 } + { 5 \langle g_s^2 GG \rangle \langle
\bar q q \rangle \langle \bar s s \rangle \over 864 \pi^2 } + { 10 m_s
\langle \bar q q \rangle^2 \langle \bar s s \rangle \over 9 }
- { 5 m_s \langle g_s^2 GG \rangle \langle g_s \bar q
\sigma G q \rangle \over 4608 \pi^4 }
\\ \nonumber &&  - { m_s^2 \langle \bar q q
\rangle \langle g_s \bar q \sigma G q \rangle \over 3 \pi^2 }\Big) + {1 \over M_B^2} \Big( - {32 g_s^2 \langle
\bar q q \rangle ^2 \langle \bar s s \rangle^2 \over 81 } - { \langle g_s^2 GG \rangle \langle \bar q q \rangle
\langle g_s \bar s \sigma G s \rangle \over 1152 \pi^2 }
\\ \nonumber && - { \langle
g_s^2 GG \rangle \langle \bar s s \rangle \langle g_s \bar q \sigma
G q \rangle \over 1152 \pi^2 } - { 2 m_s \langle \bar q q \rangle^2
\langle g_s \bar s \sigma G s \rangle \over 9} - { m_s \langle
\bar q q \rangle \langle \bar s s \rangle \langle g_s \bar q \sigma
G q \rangle \over 3} \Big)\, .
\end{eqnarray}
%
%
\begin{eqnarray}
\nonumber \Pi^{--}_{S,2,1}(M_B^2) &=& \int^{s_0}_{4 m_s^2} \Bigg [ {1 \over 12288
\pi^6} s^4 - { 3 m_s^2 \over 1024 \pi^6 } s^3 + \Big ( { \langle
g_s^2 G G \rangle \over 18432 \pi^6 } + {m_s \langle \bar s s
\rangle \over 64 \pi^4} \Big ) s^2 + \Big ( { \langle \bar q q \rangle^2 \over 24
\pi^2 } + { \langle \bar s s \rangle^2 \over 24 \pi^2 }
\\ \nonumber && - { m_s
\langle g_s \bar s \sigma G s \rangle \over 64 \pi^4 }
+ { m_s^2
\langle g_s^2 G G \rangle \over 2048 \pi^6 } \Big ) s + { \langle
\bar q q \rangle \langle g_s \bar q \sigma G q \rangle \over 16
\pi^2 }
+ { \langle \bar s s \rangle \langle g_s \bar s
\sigma G s \rangle \over 16 \pi^2 }
\\ && - { m_s \langle g_s^2 G G
\rangle \langle \bar q q \rangle \over 128 \pi^4} - { m_s \langle g_s^2 G G
\rangle \langle \bar s s \rangle \over 256 \pi^4} - { 3 m_s^2 \langle
\bar q q \rangle^2 \over 4 \pi^2 } + { m_s^2 \langle
\bar s s \rangle^2 \over 16 \pi^2 } \Bigg ] e^{-s/M_B^2} ds
\\ \nonumber && + \Big ( { \langle g_s \bar q \sigma G q \rangle^2 \over 64 \pi^2
} + { \langle g_s \bar s \sigma G
s \rangle^2 \over 64 \pi^2 } + { 5 \langle g_s^2 GG \rangle \langle
\bar q q \rangle^2 \over 1728 \pi^2 } + { 5 \langle g_s^2 GG \rangle \langle
\bar q q \rangle \langle \bar s s \rangle \over 432 \pi^2 } + { 5 \langle g_s^2 GG \rangle \langle
\bar s s \rangle^2 \over 1728 \pi^2 }
\\ \nonumber && + { 5 m_s
\langle \bar q q \rangle^2 \langle \bar s s \rangle \over 3 }
- { 5 m_s \langle g_s^2 GG \rangle \langle g_s \bar q
\sigma G q \rangle \over 2304 \pi^4 }
- { 5 m_s \langle g_s^2 GG \rangle \langle g_s \bar s
\sigma G s \rangle \over 4608 \pi^4 }
- { m_s^2 \langle \bar q q
\rangle \langle g_s \bar q \sigma G q \rangle \over 2 \pi^2 }\Big)
\\ \nonumber && + {1 \over M_B^2} \Big( - {16 g_s^2 \langle
\bar q q \rangle ^2 \langle \bar s s \rangle^2 \over 27 } - { \langle g_s^2 GG \rangle \langle \bar q q \rangle
\langle g_s \bar q \sigma G q \rangle \over 1152 \pi^2 }
- { \langle
g_s^2 GG \rangle \langle \bar s s \rangle \langle g_s \bar q \sigma
G q \rangle \over 576 \pi^2 }
\\ \nonumber && - { \langle
g_s^2 GG \rangle \langle \bar q q \rangle \langle g_s \bar s \sigma
G s \rangle \over 576 \pi^2 } - { \langle
g_s^2 GG \rangle \langle \bar s s \rangle \langle g_s \bar s \sigma
G s \rangle \over 1152 \pi^2 } - { m_s \langle \bar q q \rangle^2
\langle g_s \bar s \sigma G s \rangle \over 3}
\\ \nonumber && - { m_s \langle
\bar q q \rangle \langle \bar s s \rangle \langle g_s \bar q \sigma
G q \rangle \over 2} - { m_s^2 \langle g_s^2 GG \rangle \langle \bar s s \rangle^2 \over 1152 \pi^2 } \Big) \, .
\end{eqnarray}
%
%
\begin{eqnarray}
\Pi^{--}_{S,1,2}(M_B^2) &=& \int^{s_0}_{0} \Bigg [ {1 \over 18432
\pi^6} s^4 - { \langle
g_s^2 G G \rangle \over 18432 \pi^6 } s^2 + { \langle \bar q q \rangle^2 \over 18
\pi^2 } s + { \langle
\bar q q \rangle \langle g_s \bar q \sigma G q \rangle \over 12
\pi^2 } \Bigg ] e^{-s/M_B^2} ds
\\ \nonumber && + \Big ( { \langle g_s \bar q \sigma G q \rangle^2 \over 48 \pi^2
} + { 5 \langle g_s^2 GG \rangle \langle
\bar q q \rangle^2 \over 864 \pi^2 } \Big) + {1 \over M_B^2} \Big( - {32 g_s^2 \langle
\bar q q \rangle ^4 \over 81 } - { \langle g_s^2 GG \rangle \langle \bar q q \rangle
\langle g_s \bar q \sigma G q \rangle \over 576 \pi^2 } \Big)\, .
\end{eqnarray}
%
%
\begin{eqnarray}
\Pi^{--}_{S,2,2}(M_B^2) &=& \int^{s_0}_{0} \Bigg [ {1 \over 12288
\pi^6} s^4 + { \langle
g_s^2 G G \rangle \over 18432 \pi^6 } s^2 + { \langle \bar q q \rangle^2 \over 12
\pi^2 } s + { \langle
\bar q q \rangle \langle g_s \bar q \sigma G q \rangle \over 8
\pi^2 } \Bigg ] e^{-s/M_B^2} ds
\\ \nonumber && + \Big ( { \langle g_s \bar q \sigma G q \rangle^2 \over 32 \pi^2
} + { 5 \langle g_s^2 GG \rangle \langle
\bar q q \rangle^2 \over 288 \pi^2 } \Big)
+ {1 \over M_B^2} \Big( - {16 g_s^2 \langle
\bar q q \rangle ^4 \over 27 } - { \langle g_s^2 GG \rangle \langle \bar q q \rangle
\langle g_s \bar q \sigma G q \rangle \over 192 \pi^2 } \Big) \, .
\end{eqnarray}
%

\subsection{Tetraquark currents of $I^G J^{PC} = 1^+1^{+-}$}

%
\begin{eqnarray}
\nonumber \Pi^{+-}_{A,1,1}(M_B^2) &=& \int^{s_0}_{4 m_s^2} \Bigg [ {1 \over 36864
\pi^6} s^4 - { m_s^2 \over 960 \pi^6 } s^3 + \Big ( { \langle
g_s^2 G G \rangle \over 18432 \pi^6 } + {7 m_s \langle \bar q q
\rangle \over 384 \pi^4} + {m_s \langle \bar s s
\rangle \over 128 \pi^4} \Big ) s^2 + \Big ( - { 5 \langle \bar q q \rangle \langle \bar s s \rangle \over 36 \pi^2 }
\\ \nonumber && + { 5 m_s
\langle g_s \bar q \sigma G q \rangle \over 192 \pi^4 }
- { m_s^2
\langle g_s^2 G G \rangle \over 12288 \pi^6 } \Big ) s - { \langle
\bar q q \rangle \langle g_s \bar s \sigma G s \rangle \over 16
\pi^2 }
- { \langle \bar s s \rangle \langle g_s \bar q
\sigma G q \rangle \over 16 \pi^2 }
\\ && + { m_s \langle g_s^2 G G
\rangle \langle \bar q q \rangle \over 1536 \pi^4} - { m_s \langle g_s^2 G G
\rangle \langle \bar s s \rangle \over 1536 \pi^4} + { m_s^2 \langle
\bar q q \rangle^2 \over 6 \pi^2 } + { 3 m_s^2 \langle
\bar q q \rangle \langle
\bar s s \rangle \over 8 \pi^2 }+ { m_s^2 \langle
\bar s s \rangle^2 \over 48 \pi^2 }  \Bigg ] e^{-s/M_B^2} ds
\\ \nonumber && + \Big ( - { \langle g_s \bar q \sigma G q \rangle \langle g_s \bar s \sigma G s \rangle \over 96 \pi^2
} +  { \langle g_s^2 GG \rangle \langle
\bar q q \rangle^2 \over 3456 \pi^2 } - { \langle g_s^2 GG \rangle \langle
\bar q q \rangle \langle \bar s s \rangle \over 1728 \pi^2 } +  { \langle g_s^2 GG \rangle \langle
\bar s s \rangle^2 \over 3456 \pi^2 } - { 4 m_s
\langle \bar q q \rangle^2 \langle \bar s s \rangle \over 9 }
\\ \nonumber &&  - { m_s
\langle \bar q q \rangle \langle \bar s s \rangle^2 \over 9 }
+ { m_s \langle g_s^2 GG \rangle \langle g_s \bar q
\sigma G q \rangle \over 9216 \pi^4 } - { m_s \langle g_s^2 GG \rangle \langle g_s \bar s
\sigma G s \rangle \over 9216 \pi^4 }
+ { m_s^2 \langle \bar q q
\rangle \langle g_s \bar q \sigma G q \rangle \over 12 \pi^2 }
\\ \nonumber && + { m_s^2 \langle \bar s s
\rangle \langle g_s \bar q \sigma G q \rangle \over 16 \pi^2 } + { m_s^2 \langle \bar q q
\rangle \langle g_s \bar s \sigma G s \rangle \over 24 \pi^2 } \Big) + {1 \over M_B^2} \Big( {16 g_s^2 \langle
\bar q q \rangle ^2 \langle \bar s s \rangle^2 \over 81 } - { \langle g_s^2 GG \rangle \langle \bar q q \rangle
\langle g_s \bar q \sigma G q \rangle \over 2304 \pi^2 }
\\ \nonumber && + { \langle g_s^2 GG \rangle \langle \bar q q \rangle
\langle g_s \bar s \sigma G s \rangle \over 2304 \pi^2 } + { \langle g_s^2 GG \rangle \langle \bar s s \rangle
\langle g_s \bar q \sigma G q \rangle \over 2304 \pi^2 } - { \langle g_s^2 GG \rangle \langle \bar s s \rangle
\langle g_s \bar s \sigma G s \rangle \over 2304 \pi^2 }
\\ \nonumber && + { m_s \langle \bar q q \rangle^2
\langle g_s \bar s \sigma G s \rangle \over 9} + { 2 m_s \langle \bar q q \rangle \langle \bar s s \rangle
\langle g_s \bar q \sigma G q \rangle \over 9} + { m_s \langle \bar q q \rangle \langle \bar s s \rangle
\langle g_s \bar s \sigma G s \rangle \over 12}
\\ \nonumber && + { m_s \langle \bar s s \rangle^2
\langle g_s \bar q \sigma G q \rangle \over 12} - { m_s^2 \langle g_s \bar q \sigma G q \rangle^2 \over 96 \pi^2}
- { m_s^2 \langle g_s \bar q \sigma G q \rangle \langle g_s \bar s \sigma G s \rangle \over 32 \pi^2}  \Big)\, .
\end{eqnarray}
%
%
\begin{eqnarray}
\nonumber \Pi^{+-}_{A,2,1}(M_B^2) &=& \int^{s_0}_{4 m_s^2} \Bigg [ {1 \over 6144
\pi^6} s^4 - { m_s^2 \over 256 \pi^6 } s^3 + \Big ( { 11 \langle
g_s^2 G G \rangle \over 18432 \pi^6 } - {7 m_s \langle \bar q q
\rangle \over 64 \pi^4} - {5 m_s \langle \bar s s
\rangle \over 192 \pi^4} \Big ) s^2 + \Big ( { 5 \langle \bar q q \rangle^2 \over 18 \pi^2 }
+ { 5 \langle \bar q q \rangle \langle \bar s s \rangle \over 6 \pi^2 }
\\ \nonumber && + { 5 \langle \bar s s \rangle^2 \over 18 \pi^2 }
- { 5 m_s
\langle g_s \bar q \sigma G q \rangle \over 32 \pi^4 } - { 5 m_s
\langle g_s \bar s \sigma G s \rangle \over 48 \pi^4 }
- { 13 m_s^2
\langle g_s^2 G G \rangle \over 36864 \pi^6 } \Big ) s + { \langle
\bar q q \rangle \langle g_s \bar q \sigma G q \rangle \over 4
\pi^2 } + { 3 \langle
\bar q q \rangle \langle g_s \bar s \sigma G s \rangle \over 8
\pi^2 }
\\ \nonumber && + {3 \langle \bar s s \rangle \langle g_s \bar q
\sigma G q \rangle \over 8 \pi^2 } + { \langle
\bar s s \rangle \langle g_s \bar s \sigma G s \rangle \over 4
\pi^2 }
- { 7 m_s \langle g_s^2 G G
\rangle \langle \bar q q \rangle \over 512 \pi^4} - { 5 m_s \langle g_s^2 G G
\rangle \langle \bar s s \rangle \over 512 \pi^4}
- { 9 m_s^2 \langle
\bar q q \rangle \langle
\bar s s \rangle \over 4 \pi^2 }
\\ && + { m_s^2 \langle
\bar s s \rangle^2 \over 8 \pi^2 }\Bigg ] e^{-s/M_B^2} ds
+ \Big ( { \langle g_s \bar q \sigma G q \rangle^2 \over 48 \pi^2
} + { \langle g_s \bar q \sigma G q \rangle \langle g_s \bar s \sigma G s \rangle \over 16 \pi^2
} + { \langle g_s \bar s \sigma G s \rangle^2 \over 48 \pi^2
}
\\ \nonumber && +  { 5 \langle g_s^2 GG \rangle \langle
\bar q q \rangle^2 \over 1152 \pi^2 } + { 7 \langle g_s^2 GG \rangle \langle
\bar q q \rangle \langle \bar s s \rangle \over 576 \pi^2 } +  { 5 \langle g_s^2 GG \rangle \langle
\bar s s \rangle^2 \over 1152 \pi^2 } - { 20 m_s
\langle \bar q q \rangle^2 \langle \bar s s \rangle \over 9 }
\\ \nonumber &&  + { 2 m_s
\langle \bar q q \rangle \langle \bar s s \rangle^2 \over 3 }
- { 7 m_s \langle g_s^2 GG \rangle \langle g_s \bar q
\sigma G q \rangle \over 3072 \pi^4 } - { 5 m_s \langle g_s^2 GG \rangle \langle g_s \bar s
\sigma G s \rangle \over 3072 \pi^4 }
+ { m_s^2 \langle \bar q q
\rangle \langle g_s \bar q \sigma G q \rangle \over 6 \pi^2 }
\\ \nonumber && - { 3 m_s^2 \langle \bar s s
\rangle \langle g_s \bar q \sigma G q \rangle \over 8 \pi^2 } - { m_s^2 \langle \bar q q
\rangle \langle g_s \bar s \sigma G s \rangle \over 4 \pi^2 } \Big) + {1 \over M_B^2} \Big( {32 g_s^2 \langle
\bar q q \rangle ^2 \langle \bar s s \rangle^2 \over 27 } - { 5 \langle g_s^2 GG \rangle \langle \bar q q \rangle
\langle g_s \bar q \sigma G q \rangle \over 768 \pi^2 }
\\ \nonumber && - { 7 \langle g_s^2 GG \rangle \langle \bar q q \rangle
\langle g_s \bar s \sigma G s \rangle \over 768 \pi^2 } - { 7 \langle g_s^2 GG \rangle \langle \bar s s \rangle
\langle g_s \bar q \sigma G q \rangle \over 768 \pi^2 } - { 5 \langle g_s^2 GG \rangle \langle \bar s s \rangle
\langle g_s \bar s \sigma G s \rangle \over 768 \pi^2 }
\\ \nonumber && + { 2 m_s \langle \bar q q \rangle^2
\langle g_s \bar s \sigma G s \rangle \over 3} + { 2 m_s \langle \bar q q \rangle \langle \bar s s \rangle
\langle g_s \bar q \sigma G q \rangle \over 3} - { m_s \langle \bar q q \rangle \langle \bar s s \rangle
\langle g_s \bar s \sigma G s \rangle \over 2}
\\ \nonumber && - { m_s \langle \bar s s \rangle^2
\langle g_s \bar q \sigma G q \rangle \over 2} + { m_s^2 \langle g_s \bar q \sigma G q \rangle^2 \over 16 \pi^2}
+ {3 m_s^2 \langle g_s \bar q \sigma G q \rangle \langle g_s \bar s \sigma G s \rangle \over 16 \pi^2} - { 5 m_s^2 \langle g_s^2 GG \rangle \langle \bar s s \rangle^2 \over 1152 \pi^2 }  \Big)\, .
\end{eqnarray}
%
%
\begin{eqnarray}
\nonumber \Pi^{+-}_{S,1,1}(M_B^2) &=& \int^{s_0}_{4 m_s^2} \Bigg [ {1 \over 18432
\pi^6} s^4 - { m_s^2 \over 480 \pi^6 } s^3 + \Big ( - { \langle
g_s^2 G G \rangle \over 18432 \pi^6 } + {7 m_s \langle \bar q q
\rangle \over 192 \pi^4} + {m_s \langle \bar s s
\rangle \over 64 \pi^4} \Big ) s^2 + \Big ( - { 5 \langle \bar q q \rangle \langle \bar s s \rangle \over 18 \pi^2 }
\\ \nonumber && + { 5 m_s
\langle g_s \bar q \sigma G q \rangle \over 96 \pi^4 }
+ {11  m_s^2
\langle g_s^2 G G \rangle \over 12288 \pi^6 } \Big ) s - { \langle
\bar q q \rangle \langle g_s \bar s \sigma G s \rangle \over 8
\pi^2 }
- { \langle \bar s s \rangle \langle g_s \bar q
\sigma G q \rangle \over 8 \pi^2 }
\\ && - { 7 m_s \langle g_s^2 G G
\rangle \langle \bar q q \rangle \over 1536 \pi^4} - { 5 m_s \langle g_s^2 G G
\rangle \langle \bar s s \rangle \over 1536 \pi^4} + { m_s^2 \langle
\bar q q \rangle^2 \over 3 \pi^2 } + { 3 m_s^2 \langle
\bar q q \rangle \langle
\bar s s \rangle \over 4 \pi^2 }+ { m_s^2 \langle
\bar s s \rangle^2 \over 24 \pi^2 }  \Bigg ] e^{-s/M_B^2} ds
\\ \nonumber && + \Big ( - { \langle g_s \bar q \sigma G q \rangle \langle g_s \bar s \sigma G s \rangle \over 48 \pi^2
} +  { 5 \langle g_s^2 GG \rangle \langle
\bar q q \rangle^2 \over 3456 \pi^2 } + { 7 \langle g_s^2 GG \rangle \langle
\bar q q \rangle \langle \bar s s \rangle \over 1728 \pi^2 } +  { 5 \langle g_s^2 GG \rangle \langle
\bar s s \rangle^2 \over 3456 \pi^2 } - { 8 m_s
\langle \bar q q \rangle^2 \langle \bar s s \rangle \over 9 }
\\ \nonumber &&  - { 2 m_s
\langle \bar q q \rangle \langle \bar s s \rangle^2 \over 9 }
- { 7 m_s \langle g_s^2 GG \rangle \langle g_s \bar q
\sigma G q \rangle \over 9216 \pi^4 } - { 5 m_s \langle g_s^2 GG \rangle \langle g_s \bar s
\sigma G s \rangle \over 9216 \pi^4 }
+ { m_s^2 \langle \bar q q
\rangle \langle g_s \bar q \sigma G q \rangle \over 6 \pi^2 }
\\ \nonumber && + { m_s^2 \langle \bar s s
\rangle \langle g_s \bar q \sigma G q \rangle \over 8 \pi^2 } + { m_s^2 \langle \bar q q
\rangle \langle g_s \bar s \sigma G s \rangle \over 12 \pi^2 } \Big) + {1 \over M_B^2} \Big( {32 g_s^2 \langle
\bar q q \rangle ^2 \langle \bar s s \rangle^2 \over 81 } - { 5 \langle g_s^2 GG \rangle \langle \bar q q \rangle
\langle g_s \bar q \sigma G q \rangle \over 2304 \pi^2 }
\\ \nonumber && - { 7 \langle g_s^2 GG \rangle \langle \bar q q \rangle
\langle g_s \bar s \sigma G s \rangle \over 2304 \pi^2 } - { 7 \langle g_s^2 GG \rangle \langle \bar s s \rangle
\langle g_s \bar q \sigma G q \rangle \over 2304 \pi^2 } - { 5 \langle g_s^2 GG \rangle \langle \bar s s \rangle
\langle g_s \bar s \sigma G s \rangle \over 2304 \pi^2 }
\\ \nonumber && + { 2 m_s \langle \bar q q \rangle^2
\langle g_s \bar s \sigma G s \rangle \over 9} + { 4 m_s \langle \bar q q \rangle \langle \bar s s \rangle
\langle g_s \bar q \sigma G q \rangle \over 9} + { m_s \langle \bar q q \rangle \langle \bar s s \rangle
\langle g_s \bar s \sigma G s \rangle \over 6}
\\ \nonumber && + { m_s \langle \bar s s \rangle^2
\langle g_s \bar q \sigma G q \rangle \over 6} - { m_s^2 \langle g_s \bar q \sigma G q \rangle^2 \over 48 \pi^2}
- { m_s^2 \langle g_s \bar q \sigma G q \rangle \langle g_s \bar s \sigma G s \rangle \over 16 \pi^2}  \Big)\, .
\end{eqnarray}
%
%
\begin{eqnarray}
\nonumber \Pi^{+-}_{S,2,1}(M_B^2) &=& \int^{s_0}_{4 m_s^2} \Bigg [ {1 \over 12288
\pi^6} s^4 - { m_s^2 \over 512 \pi^6 } s^3 + \Big ( { \langle
g_s^2 G G \rangle \over 18432 \pi^6 } - {7 m_s \langle \bar q q
\rangle \over 128 \pi^4} - {5 m_s \langle \bar s s
\rangle \over 384 \pi^4} \Big ) s^2 + \Big ( { 5 \langle \bar q q \rangle^2 \over 36 \pi^2 }
+ { 5 \langle \bar q q \rangle \langle \bar s s \rangle \over 12 \pi^2 }
\\ \nonumber && + { 5 \langle \bar s s \rangle^2 \over 36 \pi^2 }
- { 5 m_s
\langle g_s \bar q \sigma G q \rangle \over 64 \pi^4 } - { 5 m_s
\langle g_s \bar s \sigma G s \rangle \over 96 \pi^4 }
+ { 7 m_s^2
\langle g_s^2 G G \rangle \over 36864 \pi^6 } \Big ) s + { \langle
\bar q q \rangle \langle g_s \bar q \sigma G q \rangle \over 8
\pi^2 } + { 3 \langle
\bar q q \rangle \langle g_s \bar s \sigma G s \rangle \over 16
\pi^2 }
\\ \nonumber && + {3 \langle \bar s s \rangle \langle g_s \bar q
\sigma G q \rangle \over 16 \pi^2 } + { \langle
\bar s s \rangle \langle g_s \bar s \sigma G s \rangle \over 8
\pi^2 }
+ { m_s \langle g_s^2 G G
\rangle \langle \bar q q \rangle \over 512 \pi^4} - { m_s \langle g_s^2 G G
\rangle \langle \bar s s \rangle \over 512 \pi^4}
- { 9 m_s^2 \langle
\bar q q \rangle \langle
\bar s s \rangle \over 8 \pi^2 }
\\ && + { m_s^2 \langle
\bar s s \rangle^2 \over 16 \pi^2 }\Bigg ] e^{-s/M_B^2} ds
+ \Big ( { \langle g_s \bar q \sigma G q \rangle^2 \over 96 \pi^2
} + { \langle g_s \bar q \sigma G q \rangle \langle g_s \bar s \sigma G s \rangle \over 32 \pi^2
} + { \langle g_s \bar s \sigma G s \rangle^2 \over 96 \pi^2
}
\\ \nonumber && +  { \langle g_s^2 GG \rangle \langle
\bar q q \rangle^2 \over 1152 \pi^2 } - { \langle g_s^2 GG \rangle \langle
\bar q q \rangle \langle \bar s s \rangle \over 576 \pi^2 } +  { \langle g_s^2 GG \rangle \langle
\bar s s \rangle^2 \over 1152 \pi^2 } - { 10 m_s
\langle \bar q q \rangle^2 \langle \bar s s \rangle \over 9 }
\\ \nonumber &&  + { m_s
\langle \bar q q \rangle \langle \bar s s \rangle^2 \over 3 }
+ { m_s \langle g_s^2 GG \rangle \langle g_s \bar q
\sigma G q \rangle \over 3072 \pi^4 } - { m_s \langle g_s^2 GG \rangle \langle g_s \bar s
\sigma G s \rangle \over 3072 \pi^4 }
+ { m_s^2 \langle \bar q q
\rangle \langle g_s \bar q \sigma G q \rangle \over 12 \pi^2 }
\\ \nonumber && - { 3 m_s^2 \langle \bar s s
\rangle \langle g_s \bar q \sigma G q \rangle \over 16 \pi^2 } - { m_s^2 \langle \bar q q
\rangle \langle g_s \bar s \sigma G s \rangle \over 8 \pi^2 } \Big) + {1 \over M_B^2} \Big( {16 g_s^2 \langle
\bar q q \rangle ^2 \langle \bar s s \rangle^2 \over 27 } - { \langle g_s^2 GG \rangle \langle \bar q q \rangle
\langle g_s \bar q \sigma G q \rangle \over 768 \pi^2 }
\\ \nonumber && + { \langle g_s^2 GG \rangle \langle \bar q q \rangle
\langle g_s \bar s \sigma G s \rangle \over 768 \pi^2 } + { \langle g_s^2 GG \rangle \langle \bar s s \rangle
\langle g_s \bar q \sigma G q \rangle \over 768 \pi^2 } - { \langle g_s^2 GG \rangle \langle \bar s s \rangle
\langle g_s \bar s \sigma G s \rangle \over 768 \pi^2 }
\\ \nonumber && + { m_s \langle \bar q q \rangle^2
\langle g_s \bar s \sigma G s \rangle \over 3} + { m_s \langle \bar q q \rangle \langle \bar s s \rangle
\langle g_s \bar q \sigma G q \rangle \over 3} - { m_s \langle \bar q q \rangle \langle \bar s s \rangle
\langle g_s \bar s \sigma G s \rangle \over 4}
\\ \nonumber && - { m_s \langle \bar s s \rangle^2
\langle g_s \bar q \sigma G q \rangle \over 4} + { m_s^2 \langle g_s \bar q \sigma G q \rangle^2 \over 32 \pi^2}
+ {3 m_s^2 \langle g_s \bar q \sigma G q \rangle \langle g_s \bar s \sigma G s \rangle \over 32 \pi^2} - { m_s^2 \langle g_s^2 GG \rangle \langle \bar s s \rangle^2 \over 1152 \pi^2 }  \Big)\, .
\end{eqnarray}
%
%
\begin{eqnarray}
\Pi^{+-}_{S,1,2}(M_B^2) &=& \int^{s_0}_{0} \Bigg [ {1 \over 18432
\pi^6} s^4 - { \langle
g_s^2 G G \rangle \over 18432 \pi^6 } s^2 - { 5 \langle \bar q q \rangle^2 \over 18 \pi^2 } s - { \langle
\bar q q \rangle \langle g_s \bar q \sigma G q \rangle \over 4
\pi^2 } \Bigg ] e^{-s/M_B^2} ds
\\ \nonumber && + \Big ( - { \langle g_s \bar q \sigma G q \rangle^2 \over 48 \pi^2
} +  { \langle g_s^2 GG \rangle \langle
\bar q q \rangle^2 \over 144 \pi^2 }  \Big) + {1 \over M_B^2} \Big( {32 g_s^2 \langle
\bar q q \rangle^4 \over 81 } - { \langle g_s^2 GG \rangle \langle \bar q q \rangle
\langle g_s \bar q \sigma G q \rangle \over 96 \pi^2 } \Big)\, .
\end{eqnarray}
%
%
\begin{eqnarray}
\Pi^{+-}_{S,2,2}(M_B^2) &=& \int^{s_0}_{0} \Bigg [ {1 \over 12288
\pi^6} s^4 + { \langle
g_s^2 G G \rangle \over 18432 \pi^6 } s^2 + { 25 \langle \bar q q \rangle^2 \over 36 \pi^2 }
s + { 5 \langle
\bar q q \rangle \langle g_s \bar q \sigma G q \rangle \over 8
\pi^2 }  \Bigg ] e^{-s/M_B^2} ds
\\ \nonumber &&
+ { 5 \langle g_s \bar q \sigma G q \rangle^2 \over 96 \pi^2
} + {1 \over M_B^2} {16 g_s^2 \langle
\bar q q \rangle^4 \over 27 } \, .
\end{eqnarray}
%

\subsection{Tetraquark currents of $I^G J^{PC} = 1^-1^{++}$}

%
\begin{eqnarray}
\nonumber \Pi^{++}_{A,1,1}(M_B^2) &=& \int^{s_0}_{4 m_s^2} \Bigg [ {1 \over 36864
\pi^6} s^4 - { m_s^2 \over 960 \pi^6 } s^3 + \Big ( { \langle
g_s^2 G G \rangle \over 18432 \pi^6 } + {7 m_s \langle \bar q q
\rangle \over 384 \pi^4} + {m_s \langle \bar s s
\rangle \over 128 \pi^4} \Big ) s^2 + \Big ( - { 5 \langle \bar q q \rangle \langle \bar s s \rangle \over 36 \pi^2 }
\\ \nonumber && + { 5 m_s
\langle g_s \bar q \sigma G q \rangle \over 192 \pi^4 }
- {13 m_s^2
\langle g_s^2 G G \rangle \over 36864 \pi^6 } \Big ) s - { \langle
\bar q q \rangle \langle g_s \bar s \sigma G s \rangle \over 16
\pi^2 }
- { \langle \bar s s \rangle \langle g_s \bar q
\sigma G q \rangle \over 16 \pi^2 }
\\ && + { m_s \langle g_s^2 G G
\rangle \langle \bar q q \rangle \over 512 \pi^4} + { m_s \langle g_s^2 G G
\rangle \langle \bar s s \rangle \over 1536 \pi^4} + { m_s^2 \langle
\bar q q \rangle^2 \over 6 \pi^2 } + { 3 m_s^2 \langle
\bar q q \rangle \langle
\bar s s \rangle \over 8 \pi^2 }+ { m_s^2 \langle
\bar s s \rangle^2 \over 48 \pi^2 }  \Bigg ] e^{-s/M_B^2} ds
\\ \nonumber && + \Big ( - { \langle g_s \bar q \sigma G q \rangle \langle g_s \bar s \sigma G s \rangle \over 96 \pi^2
} - { \langle g_s^2 GG \rangle \langle
\bar q q \rangle^2 \over 3456 \pi^2 } - { \langle g_s^2 GG \rangle \langle
\bar q q \rangle \langle \bar s s \rangle \over 576 \pi^2 } - { \langle g_s^2 GG \rangle \langle
\bar s s \rangle^2 \over 3456 \pi^2 } - { 4 m_s
\langle \bar q q \rangle^2 \langle \bar s s \rangle \over 9 }
\\ \nonumber &&  - { m_s
\langle \bar q q \rangle \langle \bar s s \rangle^2 \over 9 }
+ { m_s \langle g_s^2 GG \rangle \langle g_s \bar q
\sigma G q \rangle \over 3072 \pi^4 } + { m_s \langle g_s^2 GG \rangle \langle g_s \bar s
\sigma G s \rangle \over 9216 \pi^4 }
+ { m_s^2 \langle \bar q q
\rangle \langle g_s \bar q \sigma G q \rangle \over 12 \pi^2 }
\\ \nonumber && + { m_s^2 \langle \bar s s
\rangle \langle g_s \bar q \sigma G q \rangle \over 16 \pi^2 } + { m_s^2 \langle \bar q q
\rangle \langle g_s \bar s \sigma G s \rangle \over 24 \pi^2 } \Big) + {1 \over M_B^2} \Big( {16 g_s^2 \langle
\bar q q \rangle ^2 \langle \bar s s \rangle^2 \over 81 } + { \langle g_s^2 GG \rangle \langle \bar q q \rangle
\langle g_s \bar q \sigma G q \rangle \over 2304 \pi^2 }
\\ \nonumber && + { \langle g_s^2 GG \rangle \langle \bar q q \rangle
\langle g_s \bar s \sigma G s \rangle \over 768 \pi^2 } + { \langle g_s^2 GG \rangle \langle \bar s s \rangle
\langle g_s \bar q \sigma G q \rangle \over 768 \pi^2 } + { \langle g_s^2 GG \rangle \langle \bar s s \rangle
\langle g_s \bar s \sigma G s \rangle \over 2304 \pi^2 }
\\ \nonumber && + { m_s \langle \bar q q \rangle^2
\langle g_s \bar s \sigma G s \rangle \over 9} + { 2 m_s \langle \bar q q \rangle \langle \bar s s \rangle
\langle g_s \bar q \sigma G q \rangle \over 9} + { m_s \langle \bar q q \rangle \langle \bar s s \rangle
\langle g_s \bar s \sigma G s \rangle \over 12}
\\ \nonumber && + { m_s \langle \bar s s \rangle^2
\langle g_s \bar q \sigma G q \rangle \over 12} - { m_s^2 \langle g_s \bar q \sigma G q \rangle^2 \over 96 \pi^2}
- { m_s^2 \langle g_s \bar q \sigma G q \rangle \langle g_s \bar s \sigma G s \rangle \over 32 \pi^2}  \Big)\, .
\end{eqnarray}
%
%
\begin{eqnarray}
\nonumber \Pi^{++}_{A,2,1}(M_B^2) &=& \int^{s_0}_{4 m_s^2} \Bigg [ {1 \over 6144
\pi^6} s^4 - { 11 m_s^2 \over 1280 \pi^6 } s^3 + \Big ( { 11 \langle
g_s^2 G G \rangle \over 18432 \pi^6 } + {7 m_s \langle \bar q q
\rangle \over 192 \pi^4} + {23 m_s \langle \bar s s
\rangle \over 192 \pi^4} \Big ) s^2 + \Big ( - { 5 \langle \bar q q \rangle^2 \over 18 \pi^2 }
- { 5 \langle \bar q q \rangle \langle \bar s s \rangle \over 18 \pi^2 }
\\ \nonumber && - { 5 \langle \bar s s \rangle^2 \over 18 \pi^2 }
+ { 5 m_s
\langle g_s \bar q \sigma G q \rangle \over 96 \pi^4 } + { 5 m_s
\langle g_s \bar s \sigma G s \rangle \over 48 \pi^4 }
- { 163 m_s^2
\langle g_s^2 G G \rangle \over 36864 \pi^6 } \Big ) s - { \langle
\bar q q \rangle \langle g_s \bar q \sigma G q \rangle \over 4
\pi^2 } - { \langle
\bar q q \rangle \langle g_s \bar s \sigma G s \rangle \over 8
\pi^2 }
\\ \nonumber && - {\langle \bar s s \rangle \langle g_s \bar q
\sigma G q \rangle \over 8 \pi^2 } - { \langle
\bar s s \rangle \langle g_s \bar s \sigma G s \rangle \over 4
\pi^2 }
+ { 3 m_s \langle g_s^2 G G
\rangle \langle \bar q q \rangle \over 512 \pi^4} + { 5 m_s \langle g_s^2 G G
\rangle \langle \bar s s \rangle \over 512 \pi^4} + { 2 m_s^2 \langle
\bar q q \rangle^2 \over \pi^2 }
\\ && + { 3 m_s^2 \langle
\bar q q \rangle \langle
\bar s s \rangle \over 4 \pi^2 }
+ { m_s^2 \langle
\bar s s \rangle^2 \over 8 \pi^2 }\Bigg ] e^{-s/M_B^2} ds
+ \Big ( - { \langle g_s \bar q \sigma G q \rangle^2 \over 48 \pi^2
} - { \langle g_s \bar q \sigma G q \rangle \langle g_s \bar s \sigma G s \rangle \over 48 \pi^2
} - { \langle g_s \bar s \sigma G s \rangle^2 \over 48 \pi^2
}
\\ \nonumber && -  { 5 \langle g_s^2 GG \rangle \langle
\bar q q \rangle^2 \over 1152 \pi^2 } - { \langle g_s^2 GG \rangle \langle
\bar q q \rangle \langle \bar s s \rangle \over 192 \pi^2 } - { 5 \langle g_s^2 GG \rangle \langle
\bar s s \rangle^2 \over 1152 \pi^2 } - { 28 m_s
\langle \bar q q \rangle^2 \langle \bar s s \rangle \over 9 }
\\ \nonumber && - { 2 m_s
\langle \bar q q \rangle \langle \bar s s \rangle^2 \over 9 }
+ { m_s \langle g_s^2 GG \rangle \langle g_s \bar q
\sigma G q \rangle \over 1024 \pi^4 } + { 5 m_s \langle g_s^2 GG \rangle \langle g_s \bar s
\sigma G s \rangle \over 3072 \pi^4 }
+ { 5 m_s^2 \langle \bar q q
\rangle \langle g_s \bar q \sigma G q \rangle \over 6 \pi^2 }
\\ \nonumber && + { m_s^2 \langle \bar s s
\rangle \langle g_s \bar q \sigma G q \rangle \over 8 \pi^2 } + { m_s^2 \langle \bar q q
\rangle \langle g_s \bar s \sigma G s \rangle \over 12 \pi^2 } \Big) + {1 \over M_B^2} \Big( {32 g_s^2 \langle
\bar q q \rangle ^2 \langle \bar s s \rangle^2 \over 27 } + { 5 \langle g_s^2 GG \rangle \langle \bar q q \rangle
\langle g_s \bar q \sigma G q \rangle \over 768 \pi^2 }
\\ \nonumber && + { \langle g_s^2 GG \rangle \langle \bar q q \rangle
\langle g_s \bar s \sigma G s \rangle \over 256 \pi^2 } + { \langle g_s^2 GG \rangle \langle \bar s s \rangle
\langle g_s \bar q \sigma G q \rangle \over 256 \pi^2 } + { 5 \langle g_s^2 GG \rangle \langle \bar s s \rangle
\langle g_s \bar s \sigma G s \rangle \over 768 \pi^2 }
\\ \nonumber && + { 2 m_s \langle \bar q q \rangle^2
\langle g_s \bar s \sigma G s \rangle \over 3} + { 2 m_s \langle \bar q q \rangle \langle \bar s s \rangle
\langle g_s \bar q \sigma G q \rangle } + { m_s \langle \bar q q \rangle \langle \bar s s \rangle
\langle g_s \bar s \sigma G s \rangle \over 6}
\\ \nonumber && + { m_s \langle \bar s s \rangle^2
\langle g_s \bar q \sigma G q \rangle \over 6} - { 3 m_s^2 \langle g_s \bar q \sigma G q \rangle^2 \over 16 \pi^2}
- { m_s^2 \langle g_s \bar q \sigma G q \rangle \langle g_s \bar s \sigma G s \rangle \over 16 \pi^2} - { 5 m_s^2 \langle g_s^2 GG \rangle \langle \bar s s \rangle^2 \over 1152 \pi^2 }  \Big)\, .
\end{eqnarray}
%
%
\begin{eqnarray}
\nonumber \Pi^{++}_{S,1,1}(M_B^2) &=& \int^{s_0}_{4 m_s^2} \Bigg [ {1 \over 18432
\pi^6} s^4 - { m_s^2 \over 480 \pi^6 } s^3 + \Big ( - { \langle
g_s^2 G G \rangle \over 18432 \pi^6 } + {7 m_s \langle \bar q q
\rangle \over 192 \pi^4} + {m_s \langle \bar s s
\rangle \over 64 \pi^4} \Big ) s^2 + \Big ( - { 5 \langle \bar q q \rangle \langle \bar s s \rangle \over 18 \pi^2 }
\\ \nonumber && + { 5 m_s
\langle g_s \bar q \sigma G q \rangle \over 96 \pi^4 }
- {17  m_s^2
\langle g_s^2 G G \rangle \over 36864 \pi^6 } \Big ) s - { \langle
\bar q q \rangle \langle g_s \bar s \sigma G s \rangle \over 8
\pi^2 }
- { \langle \bar s s \rangle \langle g_s \bar q
\sigma G q \rangle \over 8 \pi^2 }
\\ && + { m_s \langle g_s^2 G G
\rangle \langle \bar q q \rangle \over 512 \pi^4} + { 5 m_s \langle g_s^2 G G
\rangle \langle \bar s s \rangle \over 1536 \pi^4} + { m_s^2 \langle
\bar q q \rangle^2 \over 3 \pi^2 } + { 3 m_s^2 \langle
\bar q q \rangle \langle
\bar s s \rangle \over 4 \pi^2 }+ { m_s^2 \langle
\bar s s \rangle^2 \over 24 \pi^2 }  \Bigg ] e^{-s/M_B^2} ds
\\ \nonumber && + \Big ( - { \langle g_s \bar q \sigma G q \rangle \langle g_s \bar s \sigma G s \rangle \over 48 \pi^2
} - { 5 \langle g_s^2 GG \rangle \langle
\bar q q \rangle^2 \over 3456 \pi^2 } - { \langle g_s^2 GG \rangle \langle
\bar q q \rangle \langle \bar s s \rangle \over 576 \pi^2 } -  { 5 \langle g_s^2 GG \rangle \langle
\bar s s \rangle^2 \over 3456 \pi^2 } - { 8 m_s
\langle \bar q q \rangle^2 \langle \bar s s \rangle \over 9 }
\\ \nonumber &&  - { 2 m_s
\langle \bar q q \rangle \langle \bar s s \rangle^2 \over 9 }
+ { m_s \langle g_s^2 GG \rangle \langle g_s \bar q
\sigma G q \rangle \over 3072 \pi^4 } + { 5 m_s \langle g_s^2 GG \rangle \langle g_s \bar s
\sigma G s \rangle \over 9216 \pi^4 }
+ { m_s^2 \langle \bar q q
\rangle \langle g_s \bar q \sigma G q \rangle \over 6 \pi^2 }
\\ \nonumber && + { m_s^2 \langle \bar s s
\rangle \langle g_s \bar q \sigma G q \rangle \over 8 \pi^2 } + { m_s^2 \langle \bar q q
\rangle \langle g_s \bar s \sigma G s \rangle \over 12 \pi^2 } \Big) + {1 \over M_B^2} \Big( {32 g_s^2 \langle
\bar q q \rangle ^2 \langle \bar s s \rangle^2 \over 81 } + { 5 \langle g_s^2 GG \rangle \langle \bar q q \rangle
\langle g_s \bar q \sigma G q \rangle \over 2304 \pi^2 }
\\ \nonumber && + { \langle g_s^2 GG \rangle \langle \bar q q \rangle
\langle g_s \bar s \sigma G s \rangle \over 768 \pi^2 } + { \langle g_s^2 GG \rangle \langle \bar s s \rangle
\langle g_s \bar q \sigma G q \rangle \over 768 \pi^2 } + { 5 \langle g_s^2 GG \rangle \langle \bar s s \rangle
\langle g_s \bar s \sigma G s \rangle \over 2304 \pi^2 }
\\ \nonumber && + { 2 m_s \langle \bar q q \rangle^2
\langle g_s \bar s \sigma G s \rangle \over 9} + { 4 m_s \langle \bar q q \rangle \langle \bar s s \rangle
\langle g_s \bar q \sigma G q \rangle \over 9} + { m_s \langle \bar q q \rangle \langle \bar s s \rangle
\langle g_s \bar s \sigma G s \rangle \over 6}
\\ \nonumber && + { m_s \langle \bar s s \rangle^2
\langle g_s \bar q \sigma G q \rangle \over 6} - { m_s^2 \langle g_s \bar q \sigma G q \rangle^2 \over 48 \pi^2}
- { m_s^2 \langle g_s \bar q \sigma G q \rangle \langle g_s \bar s \sigma G s \rangle \over 16 \pi^2}  \Big)\, .
\end{eqnarray}
%
%
\begin{eqnarray}
\nonumber \Pi^{++}_{S,2,1}(M_B^2) &=& \int^{s_0}_{4 m_s^2} \Bigg [ {1 \over 12288
\pi^6} s^4 - { 11 m_s^2 \over 2560 \pi^6 } s^3 + \Big ( { \langle
g_s^2 G G \rangle \over 18432 \pi^6 } + {7 m_s \langle \bar q q
\rangle \over 384 \pi^4} + {23 m_s \langle \bar s s
\rangle \over 384 \pi^4} \Big ) s^2 + \Big ( - { 5 \langle \bar q q \rangle^2 \over 36 \pi^2 }
- { 5 \langle \bar q q \rangle \langle \bar s s \rangle \over 36 \pi^2 }
\\ \nonumber && - { 5 \langle \bar s s \rangle^2 \over 36 \pi^2 }
+ { 5 m_s
\langle g_s \bar q \sigma G q \rangle \over 192 \pi^4 } + { 5 m_s
\langle g_s \bar s \sigma G s \rangle \over 96 \pi^4 }
- { 23 m_s^2
\langle g_s^2 G G \rangle \over 36864 \pi^6 } \Big ) s - { \langle
\bar q q \rangle \langle g_s \bar q \sigma G q \rangle \over 8
\pi^2 } - { \langle
\bar q q \rangle \langle g_s \bar s \sigma G s \rangle \over 16
\pi^2 }
\\ \nonumber && - {\langle \bar s s \rangle \langle g_s \bar q
\sigma G q \rangle \over 16 \pi^2 } - { \langle
\bar s s \rangle \langle g_s \bar s \sigma G s \rangle \over 8
\pi^2 }
+ { 3 m_s \langle g_s^2 G G
\rangle \langle \bar q q \rangle \over 512 \pi^4} + { m_s \langle g_s^2 G G
\rangle \langle \bar s s \rangle \over 512 \pi^4} + { m_s^2 \langle
\bar q q \rangle^2 \over \pi^2 }
\\ && + { 3 m_s^2 \langle
\bar q q \rangle \langle
\bar s s \rangle \over 8 \pi^2 }
+ { m_s^2 \langle
\bar s s \rangle^2 \over 16 \pi^2 }\Bigg ] e^{-s/M_B^2} ds
+ \Big ( - { \langle g_s \bar q \sigma G q \rangle^2 \over 96 \pi^2
} - { \langle g_s \bar q \sigma G q \rangle \langle g_s \bar s \sigma G s \rangle \over 96 \pi^2
} - { \langle g_s \bar s \sigma G s \rangle^2 \over 96 \pi^2
}
\\ \nonumber && -  { \langle g_s^2 GG \rangle \langle
\bar q q \rangle^2 \over 1152 \pi^2 } - { \langle g_s^2 GG \rangle \langle
\bar q q \rangle \langle \bar s s \rangle \over 192 \pi^2 } - { \langle g_s^2 GG \rangle \langle
\bar s s \rangle^2 \over 1152 \pi^2 } - { 14 m_s
\langle \bar q q \rangle^2 \langle \bar s s \rangle \over 9 }
\\ \nonumber && - { m_s
\langle \bar q q \rangle \langle \bar s s \rangle^2 \over 9 }
+ { m_s \langle g_s^2 GG \rangle \langle g_s \bar q
\sigma G q \rangle \over 1024 \pi^4 } + { m_s \langle g_s^2 GG \rangle \langle g_s \bar s
\sigma G s \rangle \over 3072 \pi^4 }
+ { 5 m_s^2 \langle \bar q q
\rangle \langle g_s \bar q \sigma G q \rangle \over 12 \pi^2 }
\\ \nonumber && + { m_s^2 \langle \bar s s
\rangle \langle g_s \bar q \sigma G q \rangle \over 16 \pi^2 } + { m_s^2 \langle \bar q q
\rangle \langle g_s \bar s \sigma G s \rangle \over 24 \pi^2 } \Big) + {1 \over M_B^2} \Big( {16 g_s^2 \langle
\bar q q \rangle ^2 \langle \bar s s \rangle^2 \over 27 } + { \langle g_s^2 GG \rangle \langle \bar q q \rangle
\langle g_s \bar q \sigma G q \rangle \over 768 \pi^2 }
\\ \nonumber && + { \langle g_s^2 GG \rangle \langle \bar q q \rangle
\langle g_s \bar s \sigma G s \rangle \over 256 \pi^2 } + { \langle g_s^2 GG \rangle \langle \bar s s \rangle
\langle g_s \bar q \sigma G q \rangle \over 256 \pi^2 } + { \langle g_s^2 GG \rangle \langle \bar s s \rangle
\langle g_s \bar s \sigma G s \rangle \over 768 \pi^2 }
\\ \nonumber && + { m_s \langle \bar q q \rangle^2
\langle g_s \bar s \sigma G s \rangle \over 3} + { m_s \langle \bar q q \rangle \langle \bar s s \rangle
\langle g_s \bar q \sigma G q \rangle } + { m_s \langle \bar q q \rangle \langle \bar s s \rangle
\langle g_s \bar s \sigma G s \rangle \over 12}
\\ \nonumber && + { m_s \langle \bar s s \rangle^2
\langle g_s \bar q \sigma G q \rangle \over 12} - { 3 m_s^2 \langle g_s \bar q \sigma G q \rangle^2 \over 32 \pi^2}
- { m_s^2 \langle g_s \bar q \sigma G q \rangle \langle g_s \bar s \sigma G s \rangle \over 32 \pi^2} - { m_s^2 \langle g_s^2 GG \rangle \langle \bar s s \rangle^2 \over 1152 \pi^2 }  \Big)\, .
\end{eqnarray}
%
%
\begin{eqnarray}
\Pi^{++}_{S,1,2}(M_B^2) &=& \int^{s_0}_{0} \Bigg [ {1 \over 18432
\pi^6} s^4 - { \langle
g_s^2 G G \rangle \over 18432 \pi^6 } s^2 - { 5 \langle \bar q q \rangle^2 \over 18 \pi^2 } s - { \langle
\bar q q \rangle \langle g_s \bar q \sigma G q \rangle \over 4
\pi^2 } \Bigg ] e^{-s/M_B^2} ds
\\ \nonumber && + \Big ( - { \langle g_s \bar q \sigma G q \rangle^2 \over 48 \pi^2
} - { \langle g_s^2 GG \rangle \langle
\bar q q \rangle^2 \over 216 \pi^2 }  \Big) + {1 \over M_B^2} \Big( {32 g_s^2 \langle
\bar q q \rangle^4 \over 81 } + { \langle g_s^2 GG \rangle \langle \bar q q \rangle
\langle g_s \bar q \sigma G q \rangle \over 144 \pi^2 } \Big)\, .
\end{eqnarray}
%
%
\begin{eqnarray}
\Pi^{++}_{S,2,2}(M_B^2) &=& \int^{s_0}_{0} \Bigg [ {1 \over 12288
\pi^6} s^4 + { \langle
g_s^2 G G \rangle \over 18432 \pi^6 } s^2 - { 5 \langle \bar q q \rangle^2 \over 12 \pi^2 }
s - { 3 \langle
\bar q q \rangle \langle g_s \bar q \sigma G q \rangle \over 8
\pi^2 }  \Bigg ] e^{-s/M_B^2} ds
\\ \nonumber && + \Big ( - { \langle g_s \bar q \sigma G q \rangle^2 \over 32 \pi^2
} - { \langle g_s^2 GG \rangle \langle
\bar q q \rangle^2 \over 144 \pi^2 }  \Big) + {1 \over M_B^2} \Big( {16 g_s^2 \langle
\bar q q \rangle^4 \over 27 } + { \langle g_s^2 GG \rangle \langle \bar q q \rangle
\langle g_s \bar q \sigma G q \rangle \over 96 \pi^2 } \Big)\, .
\end{eqnarray}
%

\end{document}